\documentclass[natbib]{svjour3}                     
\smartqed  
\usepackage{graphicx}
\usepackage{epsfig}
\usepackage[dvips]{color}
\usepackage{amsmath}
\usepackage{url}

\begin{document}

\title{Scaling relations for galaxy clusters: properties and evolution
}


\author{Giodini, S.     	\and
        		Lovisari, L. 		\and
			Pointecouteau, E. \and
			Ettori, S.			\and
			Reiprich, T.~H.		\and	
			Hoekstra, H. 	
}

\authorrunning{Giodini, Lovisari et al.} 

\institute{S. Giodini \at
	   Leiden Observatory, Leiden University, PO Box 9513, 2300 RA Leiden, the Netherlands  \\
	  TNO, Acoustic and Sonar, Oude Waalsdorperweg 63, 2597AK The Hague, the Netherlands\\ 
             \email{stefania.giodini@tno.nl} \\          
           \and
           L. Lovisari \at
              Argelander-Institut f\"ur Astronomie, Bonn University, Auf dem H\"ugel 71,
	       53121 Bonn, Germany \\
            \and
    	   E. Pointecouteau \at
    	     Universit{\'e} de Toulouse, CNRS, CESR, 9 av. du colonel Roche, BP 44346, 31028, Toulouse Cedex 04, France\\
    	   \and
    	   S. Ettori   \at
    	     INAF-Osservatorio Astronomico, via Ranzani 1, 40127 Bologna, Italy \\
		INFN, Sezione di Bologna, viale Berti Pichat 6/2, 40127 Bologna, Italy
		   \and
		   T.~H. Reiprich \at
		     Argelander-Institut f\"ur Astronomie, Bonn University, Auf dem
		     H\"ugel 71, 53121 Bonn, Germany \\
		   \and
		   H. Hoekstra \at
		     Leiden Observatory, Leiden University, PO Box 9513, 2300 RA Leiden, the Netherlands    	     
}

\date{Received: date / Accepted: date}

\maketitle

\begin{abstract}
  Well-calibrated scaling relations between the observable properties
  and the total masses of clusters of galaxies are important for
  understanding the physical processes that give rise to these
  relations. They are also a critical ingredient for studies that aim
  to constrain cosmological parameters using galaxy clusters. For this
  reason much effort has been spent during the last decade to better
  understand and interpret relations of the properties of the
  intra-cluster medium.  Improved X-ray data have expanded the mass
  range down to galaxy groups, whereas SZ surveys have openened a new
  observational window on the intracluster medium.  In addition,
  continued progress in the performance of cosmological simulations
  has allowed a better understanding of the physical processes and
  selection effects affecting the observed scaling relations. Here we
  review the recent literature on various scaling relations, focussing
  on the latest observational measurements and the progress in our
  understanding of the deviations from self similarity.
\end{abstract}

\section{Introduction}
\label{intro}

\begin{figure}
\centering
\includegraphics[width=0.4\textwidth]{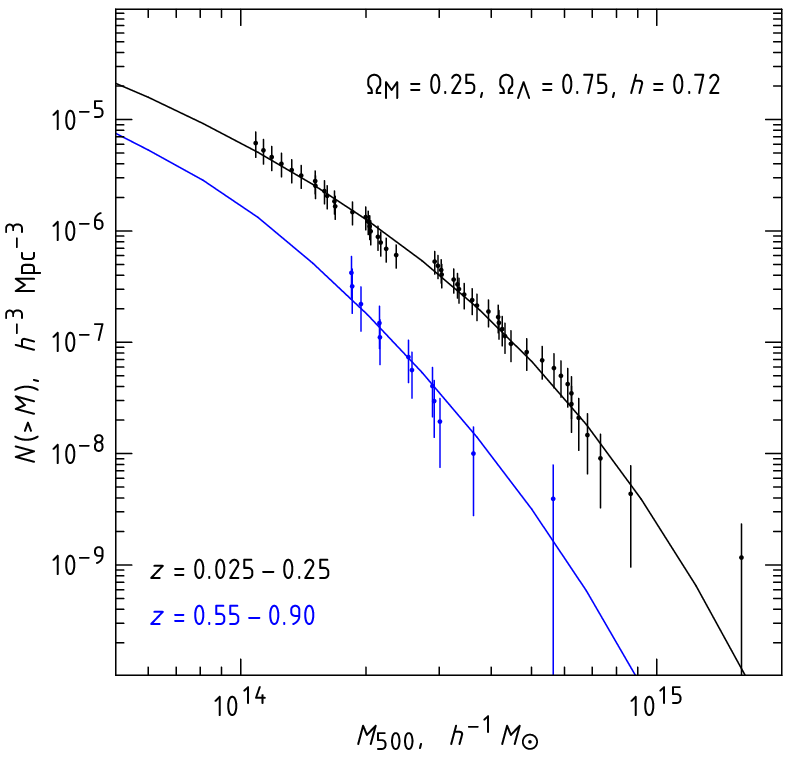}
\includegraphics[width=0.4\textwidth]{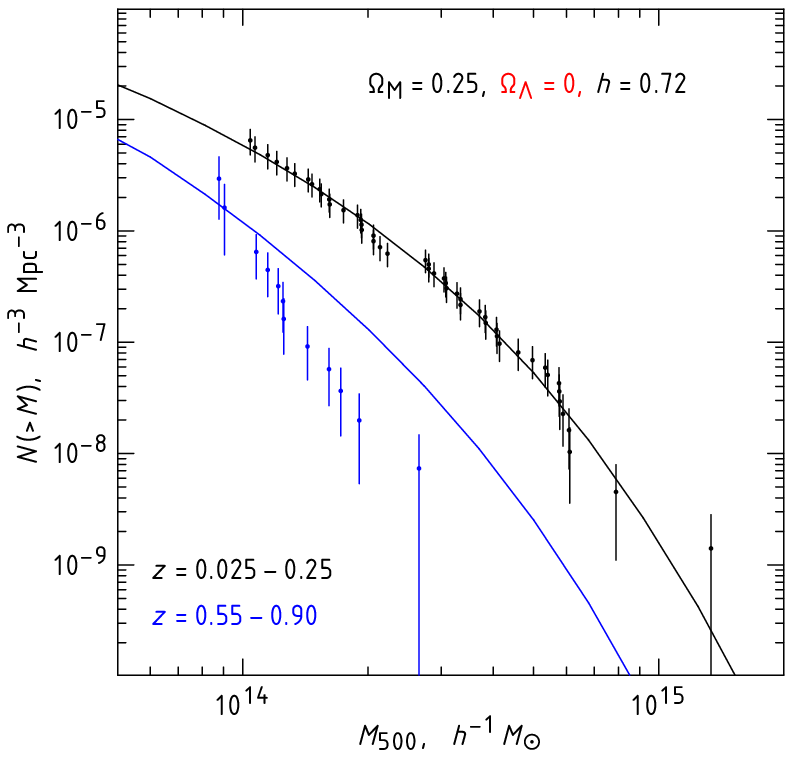}
\caption{Illustration of sensitivity of the cluster mass function to the 
cosmological model \citep[taken from][]{Vikhlinin09a}. In the left
panel the measured mass function and predicted models are shown. In
the right panel, both the data and the models are computed for a
cosmology with $\Omega_{\Lambda}=0$. Both the model and the data at
high redshifts are changed relative to the $\Omega_{\Lambda}=0.75$
case. The measured mass function is changed because it is derived for
a different distance-redshift relation. The model is changed because
the predicted growth of structure and overdensity thresholds
corresponding to $\Delta_{crit}=500$ are different. When the overall
model normalization is adjusted to the low-z mass function, the
predicted number density of $z>0.55$ clusters is in strong
disagreement with the data, and therefore this combination of
$\Omega_{M}$ and $\Omega_{\Lambda}$ can be
rejected.} \label{vikhlinin_MF}
\end{figure}

In our current paradigm of structure formation, tiny
density fluctuations rise and grow in the early Universe under the influence of
gravity, to create the massive, dark matter dominated structures we observe
today. Clusters of galaxies correspond to the densest regions of the
resulting large-scale structure. The spatial distribution and
number density of clusters carries the imprint of the process of structure
formation and, as a consequence, these properties are sensitive to the underlying cosmological parameters.
This strong dependence of the evolution of the halo mass function at the cluster scale on the cosmology is shown in Fig.~\ref{vikhlinin_MF}, which gives a convincing visual example of why clusters of galaxies attractive probes of cosmology. 
and have been suggested as a potential probe of
the dark energy equation of state \citep[e.g.][]{Schuecker03,
Albrecht06, Henry09, Vikhlinin09a, Vikhlinin09b, Mantz10, Allen11}.
Results from these studies are consistent with other observations that
indicate a Universe dominated by dark energy ($\sim73\%$), with
sub-dominant dark matter ($\sim23\%$), and a relatively small amount of
baryonic material ($\sim4.5\%$) \citep{Komatsu11}.

In order to use cluster number counting to constrain cosmological
parameters, accurate knowledge of their total mass is a crucial
ingredient. Masses can be measured directly by means of weak and
strong lensing \citep[see][in this volume]{Hoekstra13} or, under the
assumption of virial equilibrium, through measurements of the velocity
dispersion of the cluster galaxies. Obtaining individual mass
measurements for a large number of system is observationally very
expensive. Instead it is of interest to rely on robust and well
understood scaling relations that are able to relate the total mass to
quantities that are more easily observed. 

These relations are the result of the physics of cluster formation and
evolution. If gravity is the dominant process, the resulting
self-similar models predict simple scaling relations between basic
cluster properties and the total mass \citep{Kaiser86}.  Three
correlations are particularly important, namely the X-ray
luminosity--temperature, mass--temperature and luminosity--mass
relations. In general these are described as power laws in the
average, around which points scatter according to a lognormal
distribution. These relations describe positive correlations, with the
larger systems having on average more of everything. Hence scaling
relations are not merely a tool for cosmology but are also precious
diagnostics to study the thermodynamical history of the intra-cluster
medium (ICM).

With the advent of large, deep surveys of galaxy groups below
temperatures of 4~keV, a number of observational studies have reported
deviations from the self-similar scaling relations for low mass
systems \citep[e.g.][]{Gastaldello07,Sun09,Eckmiller11}.  Such deviations indicate
that non-gravitational processes may be a significant contributor to
the global energy budget in clusters.  These findings have triggered an
interest from the scientific community working on cosmological simulations
to take such processes into account. Nowadays many cosmological
simulations include prescriptions for non-gravitational processes such
as pre-heating during collapse (due to star formation or
shocks), radiative cooling and feedback by super-massive black holes.
There is general agreement that these processes need to be included in
order to reproduce the observed scaling relations. The relative
contributions of the various non-gravitational processes, however, are
still a matter of debate and will remain a major subject of research for
the next decade.

The need for a good mass tracer does not only require an understanding
of the physics of individual galaxy clusters, but also of the cluster
population as a whole. To understand the shape, evolution and
intrinsic scatter in the scaling relations, representative populations
need to be studied. Selecting galaxy clusters using their X-ray
emission is an efficient way of identifying bound, evolved and
virialized systems. In the last decade a large effort has been made to
understand possible biases in this selection. Indeed flux limited
surveys suffer from selection biases (in particular Malmquist bias), and
additional complications have to be taken into account when considering the
scatter or biases in the observables and the total mass determination.

\begin{figure} 
\includegraphics[width=\hsize]{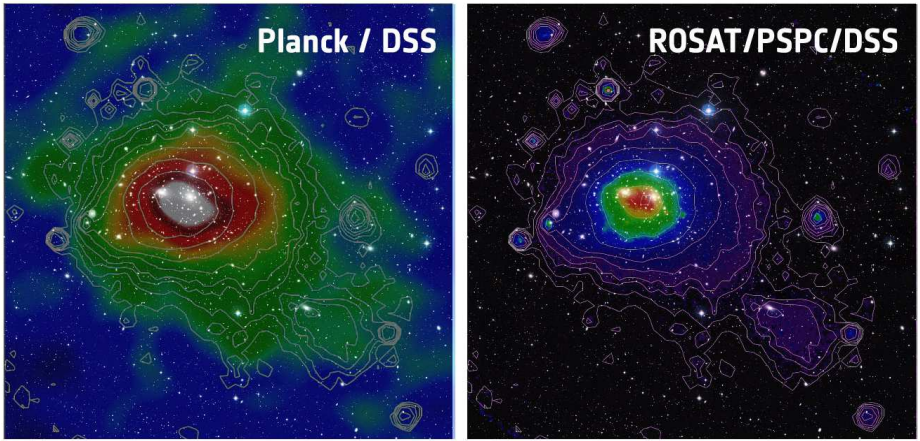}
\caption{The Coma cluster as seen by Planck (left) through the SZ
  effect and ROSAT (right) in X-rays. The images are overlaid on
  visible light images of the cluster obtained by DSS. \textit{Image
    credits: ESA / LFI and HFI Consortia (Planck image); MPI (ROSAT
    image); NASA/ESA/DSS2 (visible image). Acknowledgement: Davide De
    Martin (ESA/Hubble)}}\label{fig:szox}
\end{figure}

A key advantage of the multi-component nature of galaxy clusters is
the fact that they can be observed at different wavelengths (see
Fig.~\ref{fig:szox}). Therefore additional scaling relations that
relate the cluster total mass to properties inferred from optical,
infrared, submillimeter and radio observations, have been derived.
These relations allow a deeper insight into the biases and selection
effects which affect the X-ray based results. In particular samples of
clusters observed in large Sunyaev-Zel'dovich (SZ) surveys are highly
complementary to the X-ray ones because of the different scaling of SZ
and X-ray fluxes with electron density and temperature. Furthermore,
they are less biased towards clusters with cool cores. On the other
hand, because of their current sensitivity limits, SZ samples are
restricted to high mass systems (M$_{\rm tot}>10^{14}$ M$_{\odot}$).  It
is therefore important to complement SZ and X-ray samples with those
obtained from optical surveys. The latter are able to detect the
lowest mass structures, even if they are not virialized, and thus give
a more complete census of the large scale structure.

In this review we present an overview of the studies of scaling
relations of a number of observables with the total mass (or its
proxies), focusing mainly on results from the last decade. We start with
X-ray relations in \S2 and discuss their evolution in \S3. The SZ
results are reviewed in \S4 and optical scaling relations in \S5. The
interpretation of observational results using simulations are
discussed in \S6. General considerations are discussed in \S7 with
conclusions and an outlook presented in \S8.

\section{X-ray relations}
\label{sec:1}

Before discussing observational constraints on X-ray scaling
relations, it is useful to consider first what to expect in the case of
simple models in which only gravity is important. As shown in
\cite{Kaiser86} this leads to a so-called self-similar model, with
power law scaling relations. In general an object is said to be
self-similar when each portion of itself can be considered a
reduced-scale image of the whole \citep{Mandelbrot67}. Mathematically
speaking, a self-similar function is invariant under dilatation, such
that

\begin{equation}
f(x)=f(\alpha x).
\end{equation}

\noindent Power laws (i.e. $f(x)=x^n$) are typical functions for which
self-similarity applies. Exact self-similarity is a typical property
of fractals, such as the Koch curve or the Serpiensky gasket, where
the rescaled system is identical to the original one for each
rescaling length. Nature, instead, exhibits the property of
\emph{statistical} self-similarity, meaning that only statistical
quantities are the same for the rescaled and the original system, and
only for a range of scales; physical systems are considered
self-similar if dimensionless statistics are invariant under
rescaling.

We follow the arguments from \citet{Kaiser86}, considering the case
where the Universe has closure density (i.e., $\Omega=1$). Under this
assumption the initial spectrum of density fluctuations, $P(k)$, is a
power law over some range of wave-number, such that

\begin{equation}
P(k)\propto k^n.
\end{equation}

\noindent The mass variance of the fluctuations $\sigma^2$ scales as

\begin{equation}
\sigma^2(k) \sim k^3 P(k) \propto r^{-(n+3)} \propto M^{-(n+3)/3},
\end{equation}

\noindent where the last two proportionalities follow because
$k\propto \frac{1}{r}$ and $M\propto r^3$.  Therefore, under these
conditions, the amplitude $\sigma$ of the fluctuations is a power law
function of the size or mass. Hence the fluctuations are self-similar.
They grow in time leading to non-linear evolution when $\sigma=1$. The
growth of density fluctuations is described by

\begin{equation}
 \sigma(M,t)\propto a(t) M^{-(n+3)/6},
\end{equation}

\noindent where $a(t)$ is the scale factor. When $\sigma=1$ we obtain
the scaling of the mass-scale of non-linearity, $M_{\rm NL}$, given by

\begin{equation}
M_{\rm NL} \propto a^{\frac{6}{(n+3)}}.
\end{equation}


The transition from the linear to the non-linear regime is the only
scale introduced in the problem and as a results all statistical
quantities of the evolved fluctuation field (i.e. the number density
of halos of a given mass at time $t$), depend on the ratio $M/M_{\rm
  NL}$ only.  In other words, $M_{\rm NL}$ is a characteristic
variable that captures the dependence on the normalization and shape
of the matter power spectrum\footnote{$M_{\rm NL}$ is the variable of
  choice when the power spectrum is a power law of $k$. When this is
  not the case, other choices are more adequate.}. With the power
spectrum specified, the only dependence a function of $M/M_{\rm NL}$
can have is on $a(t)$, which itself is a power law of time
(i.e. $a\propto t^{2/3}$). This implies that the function is
self-similar with respect to time. For example, the halo properties
and halo abundance of two structures which have the same $M/M_{NL}$ at
two different times are the same.

In general, the statistical properties of haloes are expressed as a
function of the density contrast $\Delta(r,t)$ at a given time and
(comoving) scale, with $\Delta\propto a(t)^{-2/3}$. The statistic
$S(\Delta)$ obeys self-similarity, so that

\begin{equation}
S(\Delta(r, t_{1}))=S(\Delta(r, t_{2}))=S(\Delta(\alpha(t_{2})r)) \, .
\end{equation}

\noindent In this sense, a universe starting from a power-law power
spectrum is defined as self-similar. As discussed by \citet{Kaiser86},
this power-law shape cannot be expected at all scales, but it is a
good approximation on the scales of galaxy clusters and groups.

\subsection{Self-similar scaling relations for galaxy clusters}

The argument for self-similarity holds for collisionless particles,
such as dark matter, because gravity is the only force acting on the
particles. Gas in galaxy clusters can be considered "weakly
collisional" since the ion Larmor radius is much smaller than the mean
free Coulomb path ($10^8$ cm versus 10-30 kpc for a typical density of
$n\sim10^{-3}$cm$^{-3}$, e.g. \citealt {Lyutikov07}). Numerical
simulations \citep[e.g.][]{NFW1} have shown that self-similarity holds
also for the gas component if the effects of gravity and shock heating
are included, neglecting any of the dissipative, non-gravitational
effects. This means that when observing collapsed structures such as
galaxy clusters, provided dissipation can be neglected, their
dimensionless properties (e.g. their gas fraction, temperature
distribution, etc.)  can be expected to be self-similar in time and
$M_{{\rm gas},\Delta}\propto M_{{\rm DM},\Delta}$ \citep{Kaiser86}. As a
consequence, in a hierarchical scenario, where small structures form
first and provide the building blocks for larger ones, these small
structures are expected to be scaled down versions of the big ones.

In such a self-similar universe several simple relations between the
X-ray properties of the gas can be predicted. Since structures are
self-similar in time, two haloes that have formed at the same time
must have the same mean density. Hence

\begin{equation}\label{temp1}
\frac{M_{\Delta_z}}{R_{\Delta_z}^3}={\rm constant},
\end{equation}

\noindent where $R_{\Delta_z}$ is the radius where the density
contrast\footnote{The density contrast $\Delta$ is usually expressed
  with respect to the critical density at the cluster redshift.  As
  detailed in \cite{Boehringer12}, the evolution of the background and
  critical density across the cosmic epoch introduces a redshift
  dependence in the definition of $\Delta$. Indeed to relate clusters
  at different epoch and sizes the density contrast should be scaled
  as $\Delta_z=\Delta(z=0)\frac{\Delta_{\rm vir}(z)}{\Delta_{\rm
      vir}(z=0)}$.} is $\Delta_z$.  M$_{\Delta_z}$ is the mass within
a sphere of radius R$_{\Delta_z}$ defined as:

\begin{equation} 
\label{MDELTA}
M_{\Delta_{z}}=\frac{4\pi}{3} \Delta_z \rho_{\rm crit,0} E_{z}^{2} R_{\Delta_z}^{3} ,
\end{equation}

\noindent where
$E_z=H_z/H_0=[(\Omega_m(1+z)^3+(1-\Omega_m-\Omega_{\Lambda})(1+z)^2+\Omega_{\Lambda})]^{1/2}$
describes the evolution of the Hubble parameter with redshift $z$.  A
cluster of galaxies is considered to be in hydrostatic equilibrium
when the pressure gradient balances the gravitational force. If
hydrostatic equilibrium holds, the temperature of the gas provides a
good estimate of the depth of the potential well and thus of the
virial mass of the cluster:

\begin{equation}\label{temp2}
T_{\rm gas}\propto \frac{G M}{R}\propto R_{\rm vir}^2 ,
\end{equation}

\noindent where $R_{\rm vir}$ is the virial radius.  If we substitute
Eqn.~\ref{temp1} into Eqn.~\ref{temp2} it follows that

\begin{equation}\label{scal1}
M_{\Delta_z}\propto T_{\rm gas}^{\frac{3}{2}},
\end{equation}

\noindent which is the expected scaling relation between mass and
ICM temperature.

To relate the X-ray luminosity, which is easier to observe, to the
temperature we need to assume an emission mechanism.  For sufficiently
massive systems the ICM is shock heated to temperatures of
10$^{7}$-10$^{8}$ K and emits mainly by thermal bremsstrahlung.  In
this emission regime (for a plasma with solar metallicity) the total
emissivity $\epsilon$ (luminosity per unit volume) and the temperature
are related as follows

\begin{equation}
\label{bremss}
\epsilon \simeq 3.0 \times 10^{-27 } T_{gas}^{1/2} \rho_{\rm gas}^2
\ erg \ cm^{-3} s^{-1},
\end{equation}

\noindent where we are implicitly assuming thermal equilibrium, such
that the temperature of the electrons is the same as that of the
ions. By means of Eqns.~\ref{scal1} and~\ref{bremss} it is then
possible to relate the X-ray luminosity to the total mass:

\begin{equation}
  L_X \approx \epsilon R_{\Delta_z}^3 \approx T_{\rm gas}^{1/2} \rho_{\rm
    gas}^2 R_{\Delta_z}^3 \approx T_{\rm gas}^{1/2} f_{\rm gas}^2 M_{\rm
    tot}\approx f_{\rm gas}^2 T_{\rm gas}^2,
\end{equation}

\noindent where $f_{\rm gas}$ is the gas fraction defined as M$_{\rm
  gas}$/M$_{\rm tot}$ and where we used the second proportionality in
Eqn.~\ref{temp2} to obtain the last scaling. Since the gas
fraction is predicted to be a constant in the self-similar scenario,
this implies that \citep[e.g.][]{Ponman99}:

\begin{equation}\label{scal2}
L_{X}\propto T_{\rm gas}^{2}.
\end{equation}

Equations \ref{scal1} and \ref{scal2} are the basic scaling relations
between X-ray properties in galaxy clusters predicted by the
self-similar model. These hold for halo masses where dissipative
processes can be ignored. Consequently, a departure from this
prediction can be used to quantify the importance of non-gravitational
processes.

We stress that the so-called \emph{self-similar scenario} results from
a property of the dark matter power spectrum of initial fluctuation
and that it predicts a particular value for the power law exponent (as
that in Eqns.~\ref{scal1} and~\ref{scal2}). Hence, an observed
power law scaling between the X-ray properties different from that
predicted above does not imply the self-similar scenario, even though
a power law relation is self-similar in a mathematical sense.

A very useful quantity to describe the ICM is the entropy $S$: it
determines the structure of the ICM in galaxy clusters, together with
the profile of the potential well. The low entropy gas sinks while the
high entropy gas floats; hence the gas will convect until the
iso-entropic surfaces will coincide with the equipotential surfaces of
the dark matter potential \citep{Voit05}. This naturally leads to a
state of hydrodynamical equilibrium, which is just an expression of
its underlying entropy and potential profiles. Furthermore, since
entropy can only increase, if we consider the cluster as a closed
system within a certain radius, it retains the memory of the
thermodynamical history of the intracluster gas. 

\noindent The entropy $S$ is defined as\footnote{The definition of
  entropy used in astrophysics of galaxy clusters is different from
  the classic thermodynamical entropy $s$, but the two are related through
  $s=k_{B}S+constant$ \citep{Voit05}.}

\begin{equation}
S=\frac{k_{B}T_{\rm gas}}{(n_{e})^{2/3}}\,.
\end{equation}
 
\noindent Therefore, the scaling laws described above imply that the
entropy parameter scales as

\begin{equation}
 S\propto T_{\rm gas} \propto M_{\rm tot}^{2/3}
\end{equation}

\noindent for purely gravitational heating. 

These results can be combined to obtain other scaling relations and we
list the most popular ones. In doing so, it is convenient to combine
the dependence on cosmology and redshift in the factor $F_{z}=E_z
\times \left( \Delta_z / \Delta_{z=0} \right)^{1/2}$:

\begin{center}
\begin{minipage}{3in}
\begin{equation}
\begin{aligned}
L_X&\propto & F_z  & T_{gas}^2 \\
L_X& \propto & F_z^{7/3} & M_{\rm tot}^{4/3} \\
L_X& \propto & F_z^{9/5} & Y_{\rm X}^{4/5} \\
M_{\rm tot} & \propto & F_z^{-1} & T_{gas}^{3/2} \\
M_{\rm tot} & \propto & F_z^{-2/5} & Y_{\rm X}^{3/5}  \\
M_{\rm gas} & \propto & F_z^{-1} & T_{gas}^{3/2} \\
S & \propto & F_z^{-4/3} &  T_{gas}
\label{eq:scalaw}
\end{aligned}
\end{equation}
\end{minipage}
\end{center}

It should be stressed these equations are valid only if the condition
of hydrostatic equilibrium holds. This is true only in the central
part of the clusters, which is the most evolved one, while in the
outskirts both the assumptions of thermal \citep[e.g.][]{Fox97} and
hydrostatic equilibrium \citep[e.g.][]{Nagai07} brake down (for a
review on the physical processes occurring in the clusters outskirts
see \cite{Reiprich13} in this volume). This assumption also breaks
down in disturbed systems undergoing mergers \citep{Poole07}. Also,
the very central core of a galaxy cluster can be out of equilibrium
when there is AGN activity. Therefore one has to take care when
interpreting the cluster profiles at both small and large radii and
for unrelaxed systems using equations that rely on the assumptions of
hydrostatic and thermal equilibrium. 

Furthermore these simple analytic scaling relations employing $F_z$
implicitly assume that clusters formed only recently. The validity of
this assumption was examined in \cite{Boehringer12} who compared
scaling relations obtained from the results from N-body simulations.
They find that modifications are needed, especially for relations that
involve the gas density or gas mass. \cite{Boehringer12} also provide
a comprehensive comparison of their modified scaling relations to a
number of observational studies. Here we limit the comparison to a
number of recent studies and list constraints on the various scaling
relations in Table~\ref{table:laws}. Some of the relations agree
fairly well with the predictions from the self-similar model, whereas
others show significant deviations.

\subsection{Observations and deviation from self-similarity}
\subsubsection{The $L_{X}-T$ and $S-T$ scaling relation}
\label{sec:2}

Among the X-ray scaling relations, the first one to be studied was the
$L_X-T$ relation, and it remains the best studied one
\citep[e.g.][]{Mitchell77,Mitchell79,Mushotzky84,Edge91,David93,
  Markevitch98, Allen01,Ikebe02,Ettori04,Pratt09,Mittal11,Maughan12}.
This is not surprising, since both quantities can be measured easily
and almost independently from X-ray data. The gas temperature is
determined from X-ray spectroscopic data while the luminosity is
obtained from by integrating the surface brightness profile of the
cluster from X-ray imaging data.

Several independent observational studies have shown that the $L_X-T$
relation does not scale self-similarly, as it would do if the heating
is mostly due to gravitational processes (i.e. adiabatic compression
during the infall and shock heating from supersonic accretion).
Already from the earliest X-ray observations of galaxy clusters
performed with ASCA, EXOSAT and ROSAT there has been a general
consensus that the slope of the $L_X-T$ relation is significantly
steeper, with a slope for the bolometric luminosity closer to 3 than
to the theoretically predicted exponent of 2
\citep[e.g.][]{Mushotzky84,Edge91,Markevitch98}. Further studies with
samples of lower mass galaxy groups assessed that the deviation from
the self-similar scaling becomes larger below kT$\sim$3.5 keV, marking
a clear transition between galaxy groups and clusters \citep{Ponman96,
  Balogh99,Maughan12}. 

\begin{figure}
\centering
\includegraphics[width=8cm]{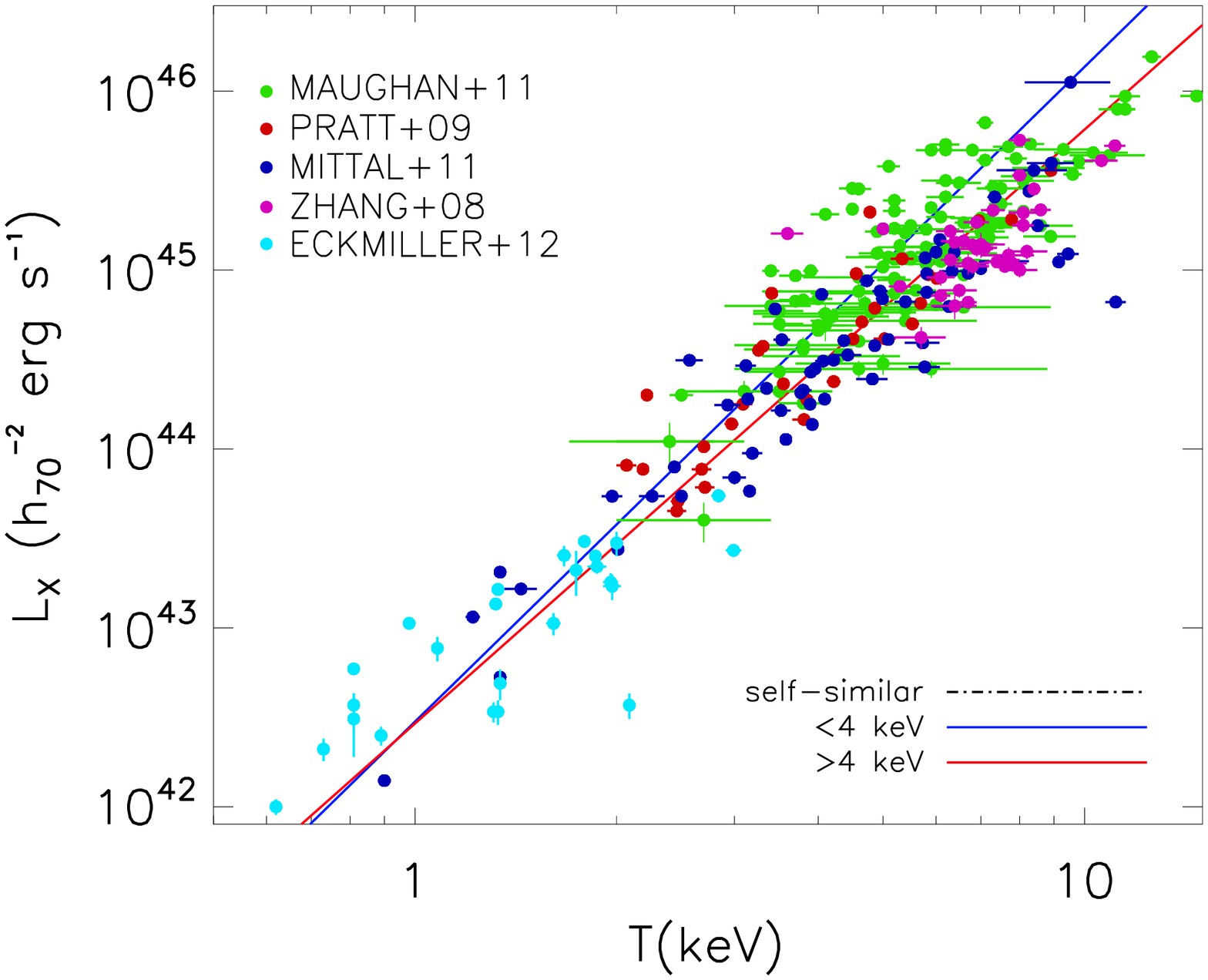}
\includegraphics[width=8cm,angle=270]{multi_egas_01R200.eps}
\caption{\textit{Top Panel:} Recent measurements of the $L_X-T$
relation for different samples of groups and clusters. Cyan circles
mark measurements from the groups sample from \cite{Eckmiller11},
green circles from \cite{Maughan12}. Blue circles show the
HIFLUGCS massive clusters \citep{Mittal11}, red circles mark
the REXCESS clusters \citep{Pratt09} and pink circles are
LoCuSS clusters \citep{Zhang08}. All the parameters are
calculated at $R_{500}$. {\it Bottom panel}: Gas entropy versus
temperature measured for a sample of galaxy groups and clusters.
Observations suggest that gas entropy varies with the mean temperature
to the power 2/3 (solid line), a scaling which is at odds with the
self-similar expectation (dotted line). The flattening of the
relationship is likely due to  the action of non-gravitational
heating/cooling sources that has a greater impact on the least massive
systems. (Figure from \citealt{Ponman03}) } 
\label{fig:groups}
\end{figure}

Figure \ref{fig:groups} ({\it top panel}) shows a compilation of
recent data for the $L_{X}-T$ relation. Table~\ref{table:laws} lists
the best fit slopes from a number of studies. The best fitting
relations obtained for the samples with $T>4$ keV (red line) and
$T<4$ keV (blue line) are also shown separately in
Figure~\ref{fig:groups}. Indeed the two subsamples do not share the same
best fit solution.

Much effort has been spent over the last decade to determine the
processes responsible for the deviation from self-similarity. These
studies are now possible because samples of low mass systems are
becoming available. Because of their fainter X-ray luminosity, galaxy
groups require very deep observations, and this has limited the number
of systems that have been observed.  In the last few years deep X-ray
surveys have provided the first statistically significant samples of
X-ray galaxy groups \citep[e.g.][]{Finoguenov07,Gastaldello07,Sun09,Eckmiller11}.
For example \citet{Pratt09} examined possible causes for the observed
steeper slopes, concluding that it is mostly associated with the
variation of the gas content with mass, while structural variations
play only a minor role. There is currently a general consensus that
the fraction of gas decreases as the mass decreases
\citep{Vikhlinin06, Gastaldello07,Pratt09,Dai10}\footnote{We note that
  \citet{Juett10} claims this may be a selection effect.}.  As L$_{X}$
is proportional to the square of the gas fraction, a change in the gas
content of low mass systems (and as a function of radius) would lead
to a reduction in the observed luminosity and consequently to a
steepening of the relations.  Complications to this very simple
reasoning can be added by the increased importance of line emission
from metals at low temperatures, which implies that the assumption of
pure bremsstrahlung is not fulfilled, resulting in a different
dependence of the luminosity on $f_{\rm gas}$.

The deviation of the $L_{X}-T$ relation from the pure gravitational
prediction can also be interpreted in terms of entropy variation
\citep{Evrard91,Bower97,Tozzi01,Borgani01,Voit02,Younger07,Eckmiller11}.
One way to inhibit the gas from reaching the center of the potential
well, thus changing the gas fraction, is to increase the entropy of
the gas. This implies the existence of an "entropy floor" for low mass
systems that would make the gas more resistive to compression. 
Observations show that the cores of galaxy groups exhibit entropy in
excess to that achievable by pure gravitational collapse
\citep{Ponman99}.  Consequently, the scaling relation between entropy
and temperature is flatter then predicted (\citealt{Ponman03}, see bottom panel of 
Fig. \ref{fig:groups}). Furthermore observations of clusters at larger
radii showed that the excess entropy increases by up to factor of 4 in
the outskirts of galaxy clusters \citep{Finoguenov02}. See
\cite{Reiprich13} for a summary of recent entropy observations in
cluster outskirts.

The key to understanding the deviation from self-similarity is to know
which processes regulate the increase of the entropy in galaxy
clusters and groups. A boost in entropy can be induced either by
heating the gas or by selectively removing gas with low entropy
(i.e. lowering the gas density). This can only be achieved through
non-gravitational processes such as radiative cooling, AGN feedback,
star formation or galactic winds. These will affect low mass groups
more strongly because of the lower gravitational binding energy for
the gas.  Furthermore, if feedback processes are triggered by
galaxies, the combined mass of the member galaxies in groups is at
least equal to that of the gas \citep{Giodini09} making the ratio
source/recipient of excess entropy just about unbalanced.

The ICM heating can be due to processes internal to the clusters, such
as late stellar or AGN heating coming mostly from the central galaxy,
or due to pre-heating, i.e. prior to the cluster infall.  In the last
ten years studies have focused on understanding which processes
contribute most during the thermodynamical history of the ICM.
\citet{Pratt10} examined the entropy profiles for clusters in the
REXCESS sample, revealing that the scaling of gas entropy is shallower
than self-similar in the inner regions, but that it steepens with
radius, becoming consistent with the self-similar prediction at
$R_{500}$. They argue that variations of the gas content with mass and
radius are at the root of the observed departures from self-similarity
of cluster entropy profiles and that results are consistent with a
central heating source. The variation in the gas fraction within
$R_{500}$ as a function of the cluster mass was quantified by
\citet{Pratt09}.  The results are consistent with a scenario in which
AGN feedback combined with merger mixing maintains an elevated central
entropy level in the majority of the clusters.  Similar conclusions
were reached by \citet{Maughan12} using a sample of 114 clusters
observed with XMM. They pointed out, however, that the most massive
cool core systems follow the self similar $L_{X}-T$ scaling relation
(when the core is excluded) and do not exhibit a central entropy
excess. Non-cool core systems, on the other hand, being dynamically
unrelaxed, would never follow self-similar scaling relations because
merger shocks enhance the entropy input. Further evidence supporting
central heating has also been found by \citet{Mantz10b} and
\citet{Mittal11}.

The $L_{X}-T$ relation shows a large scatter about the mean of
$\sigma\sim$0.7~dex \citep{Pratt09}. Using high-resolution imaging it
has now become clear that this scatter reflects the prevalence in a
mass limited sample of clusters exhibiting a boost in the X-ray
surface brightness in their inner 50-100~kpc. Because of the
corresponding temperature drop in this region, these systems have been
dubbed `cool cores', and they are associated with the inflow of gas from
the external regions of the clusters towards the core. 

For a given mass, cool-core clusters are generally more X-ray luminous
than non-cool-cores. As a result they are common in X-ray selected
samples. Since most of the luminosity in cool-cores comes from the
central region of the cluster, the scatter in the $L_{X}-T$ is
strongly reduced (to roughly half) when the X-ray luminosity is
estimated outside the core of the cluster
\citep{Markevitch98,Pratt09,Mittal11,Maughan12}.  Another source of
scatter in the $L_{X}-T$ relation is caused by morphologically
disturbed systems where the assumptions of hydrodynamic equilibrium
and the spherical symmetry are invalid, biasing the estimate of the
luminosity \citep[e.g.][]{Maughan12}.

\subsection{The $M_{\rm tot}-T$ relation}

The total mass-temperature relation ($M_{\rm tot}-T$) is another
important source of information about the cluster physics, because it
provides the link between the properties of the hot gas in the ICM and
the overall mass: in the absence of strong cooling, the temperature of
a cluster is $T$ is only determined by the depth of its potential
well.

There are two approaches to determine the $M_{\rm tot}-T$ relation with an
X-ray survey. The first is to study a small sample of clusters for
which the assumptions required determine its total mass can be trusted
with an high level of confidence (e.g. a massive relaxed
systems). Unfortunately such a limited sample may not be
representative of the cluster population as a whole. Alternatively, a
large sample can be used, but the usual assumption of hydrostatic
equilibrium may be invalid for some of the clusters, introducing
additional scatter in the measured relation. However, since the link
between the ICM temperature and the total mass is determined only by
the condition of hydrostatic equilibrium, the $M_{\rm tot}-T$ relation
should have a small scatter because it is less sensitive to processes
of heating and cooling. As before, any deviations from self-similarity
would indicate that other physical processes are at play in addition
to gravity.

\cite{Boehringer12} have summarized observational constraints from the
literature (also see our compilation in Table~\ref{table:laws}) The
general consensus is that slope of the $M_{\rm tot}-T$ relation is
self-similar for massive clusters
\citep[e.g.][]{Finoguenov01,Arnaud05,Vikhlinin09a}. When low mass
systems are considered, the best fit slope is slightly steeper than
the self similar prediction \citep[the values range from 1.5--1.7;
e.g.][]{Arnaud05,Sun09,Eckmiller11}. The $M-T$ relation seems to be
fairly robust against deviations from self-similarity. Furthermore,
the scatter in the measured $M_{\rm tot}-T$ relation is considerably
smaller than that of the $L_{X}-T$ relation (15$\%$,
\citealt{Mantz10b}), making it very attractive for its use in
cosmological studies with galaxy clusters.

It is important to understand the nature of the scatter in this
relation since the current uncertainty on the determination of the
cosmological parameters $\Omega_m$ and $w$ from X-ray studies of
clusters is dominated by uncertainties in the mass-observable
relation, as shown by \citet{Cunha10}. For instance, underestimating
the scatter in the $M_{\rm tot}-T$ relation can lead to an
overestimate of $\sigma_8$ \citep{Randall02}. An added concern is that
the assumption of hydrostatic equilibrium is incorrect.

Simulations have also shown that the intrinsic scatter in the
$M_{\rm tot}-T$ relation is associated with the presence of substructure
\citep{Hara06,Yang09}. Substructure is mostly associated with merging
systems, where the mass measurement will be biased because the
assumptions of hydrostatic equilibrium and spherical symmetry are
invalid. This will result in systematically underestimated masses up
to 20$\%$ as shown in numerical simulations \citep{Evrard90, Evrard96,
  Rasia06,Nagai07,Shaw10,Rasia12} and observations
\citep{Mahdavi08,Mahdavi12}. Additional sources of scatter are likely
present, as discussed in \cite{Poole07} who showed that the increase
in the dispersion due to mergers it is not enough to account for all
the scatter in the $M_{\rm tot}-T$ relation.

\subsection{$M_{\rm tot}-Y_X$ and $M_{\rm tot}-M_{\rm gas}$ relation}
\label{sec:yx}

\cite{Kravtsov06} proposed the use of the X-ray equivalent of the SZ
signal (see \S\ref{sec:3} for details), defined as

\begin{equation}
 Y_X=M_{gas}\times T.
\end{equation} 

\begin{figure}
	\hspace{0.2cm}
  \includegraphics[width=5.4cm]{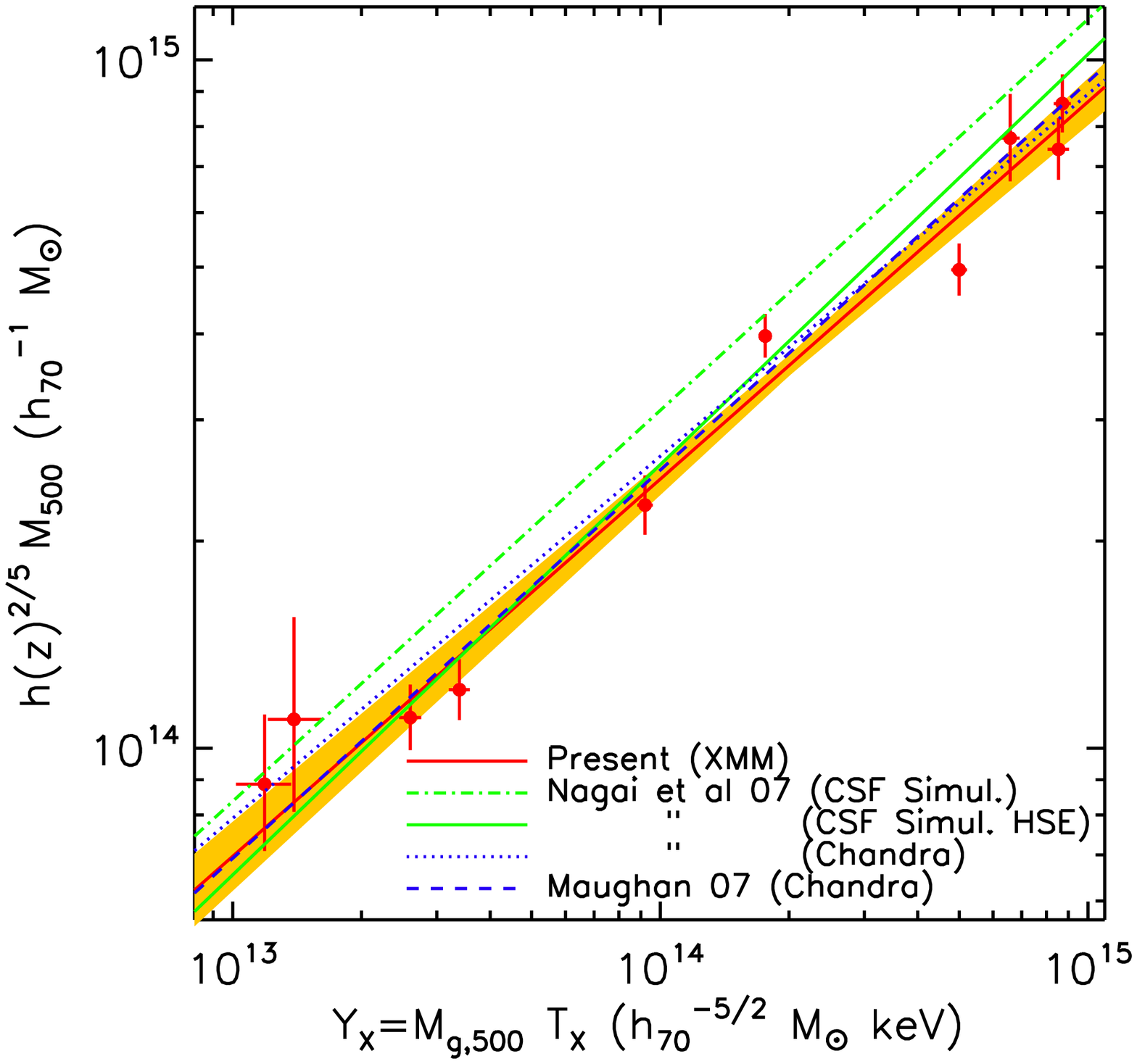}
	\hspace{1cm}
  \includegraphics[width=5cm]{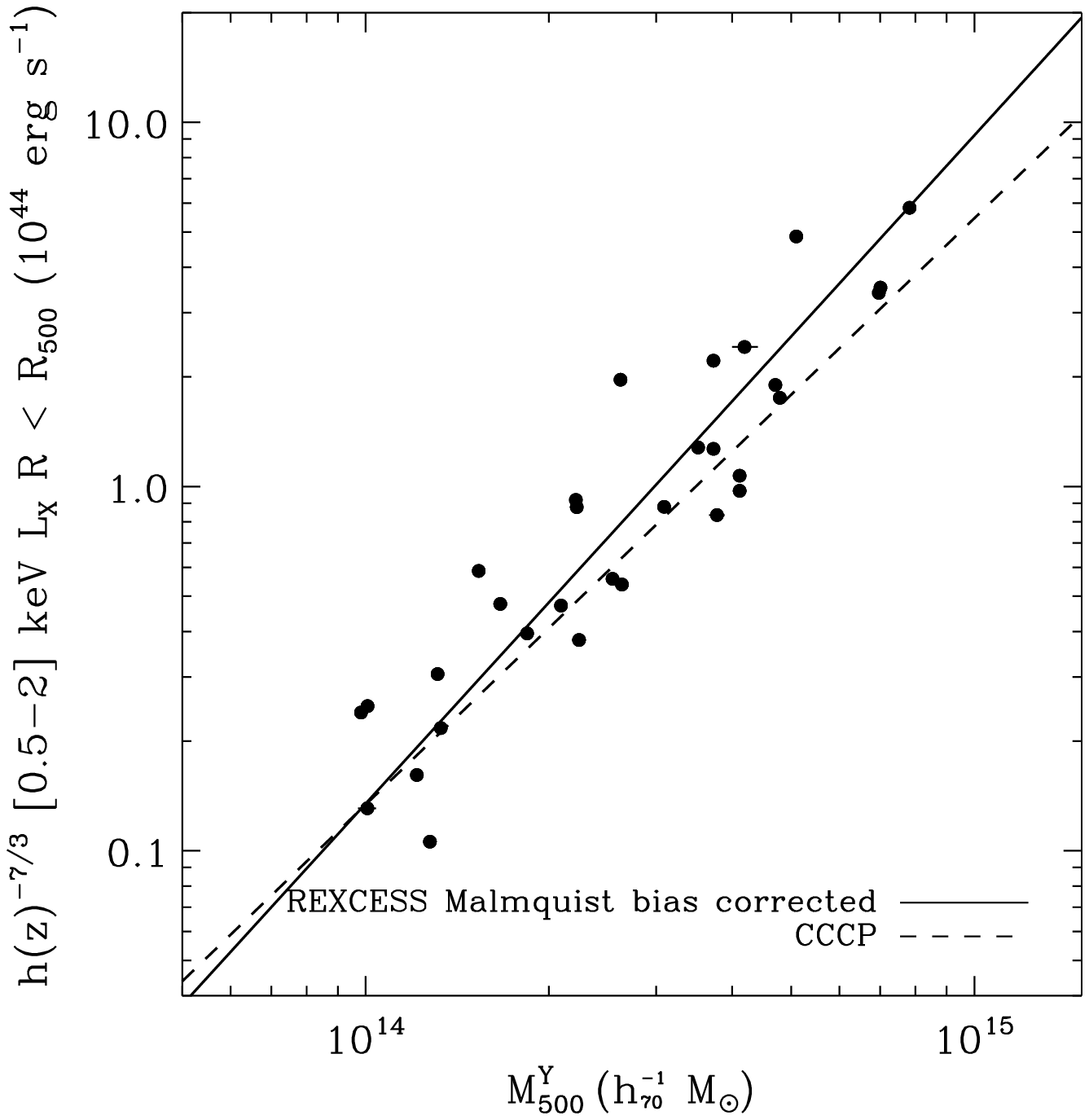}
\caption{{\it left}: $E_z^{2/5} M_{\rm tot}-Y_X$ relation for a sample of 10 local
  relaxed clusters observed with XMM compared with the predicted
  relations from numerical simulations and the observed one with
  Chandra. (Figure from \citealt{Arnaud07a}). {\it right}: Comparison between the Malmquist bias corrected
  $L_X(0.5-2 \ keV)-M_Y$ relations obtained by \cite{Pratt09} and
  \cite{Vikhlinin09a}. The points are the bias-corrected REXCESS
  values. Figure from \citealt{Pratt09}.
  }
\label{fig:bias} 
\end{figure}

\noindent This quantity is related to the total thermal energy of the
ICM and it appears to be a low scatter mass indicator.  The use of
this quantity has been motivated by results from hydrodynamic
numerical simulations, which showed that the temperature deviations
from the $M_{\rm tot}-T$ relation are anti-correlated with the
residuals in M$_{\rm gas}$ from the $M_{\rm tot}-M_{\rm gas}$
relation. This anti-correlation tends to suppress the scatter in the
$M_{\rm tot}-Y_X$ relation (down to 5-7$\%$ for $M_{500}$-$Y_X$)
independently of the dynamical state of the objects. Whether or not
$Y_{X}$ is the lowest scatter estimator in simulations is a matter of
debate. For instance, \citet{Stanek10} used an SPH code, and found a
positive correlation between temperature and gas-mass deviations, thus
contradicting the result by \citet{Kravtsov06},

X-ray observations have shown that the measured $M_{tot}-Y_X$ relation
agrees with the self similar prediction from the simulations
\citep{Arnaud07a,Maughan07,Zhang08,Vikhlinin09a}, albeit with an
offset in the normalization. This could be due to an underestimate of
the gas fraction in simulations or due to deviations from hydrostatic
equilibrium \citep{Arnaud07a}. Figure~\ref{fig:bias} ({\it left panel}) shows the $M_{\rm
  tot}-Y_X$ relation from \cite{Arnaud07a} and the comparison with the
predictions from numerical simulations. Recent observational studies
\citep{Juett10,Okabe10,Mahdavi12} found a larger scatter of this
relation (up to $\sim$20$\%$ against $<$10$\%$ in the
simulations). Interestingly, \cite{Mahdavi12} find that the scatter in
the $M_{\rm tot}-Y_X$ relation is the same for clusters with low and
high central entropies, suggesting that $Y_X$ may be well suited as a
proxy for large cluster surveys.

Interestingly, $M_\mathrm{gas}$ appears to have a very small scatter
with the cluster total mass, with only a mild dependence with redshift
\citep[e.g.][]{Vikhlinin03}. Comparing to weak lensing observations
both \cite{Okabe10} and \cite{Mahdavi12} argue that M$_\mathrm{gas}$
is the lowest scatter mass proxy. \cite{Mahdavi12} found that the
scatter for clusters with low central entropies is particularly low,
suggesting that the gas fractions vary very little for such clusters.
\cite{Zhang08} found a lower gas mass in low mass systems than
expected from a purely gravitational scenario, implying a steepening
with respect to the prediction of the self-similar scenario.

\begin{table}
\caption{Overview of the most recent published scaling relations.} \label{table:laws}
\[
\begin{array}{ccccccc}
	\hline
        \noalign{\smallskip}
{\rm relation}  & {\rm observed} & {\rm predicted} & {\rm comments} & {\rm reference} & {\rm Note}\\
        \noalign{\smallskip}
        \hline
        \noalign{\smallskip}
        & 2.26 \pm 0.29^{\dagger} &       & {\rm 50~clusters,z=0.15-0.55} & {\rm Mahdavi \ et \ al. \ 2012 }    & a,d,e \\
 	& 2.25 \pm 0.21^{\ddagger} &	& {\rm 26~clusters,z=0.01-0.05} & {\rm  Eckmiller \ et \ al. \ 2011 } & c,d,e,f \\
 	& 2.64 \pm 0.20^{\ddagger} & 	& {\rm 64~clusters,z=0.01-0.15} & {\rm Mittal \ et \ al. \ 2011 }     & a,e,h     \\ 
 	& 2.53 \pm 0.15^{\parallel} &	& {\rm 232~clusters,z=0.04-1.46} & {\rm  Reichert \ et \ al. \ 2011 } & a       \\
L_X - T & 2.72 \pm 0.18^{\uparrow} &   2 & {\rm 114~clusters,z=0.10-1.30} & {\rm Maughan \ et \ al. \ 2012 }   & a,d,e   \\
        & 2.70 \pm 0.24^{\parallel} & 	& {\rm 31~clusters,z=0.06-0.17} & {\rm Pratt \ et \ al. \ 2009 }      & a       \\ 
        & 3.35 \pm 0.32^{\uparrow} & 	& {\rm 31~clusters,z=0.06-0.17} & {\rm Pratt \ et \ al. \ 2009 }      & a       \\ 
 	& 2.78 \pm 0.13^{\parallel} & 	& {\rm 31~clusters,z=0.06-0.17} & {\rm Pratt \ et \ al. \ 2009 }      & a,d,e   \\ 
 	& 2.61 \pm 0.32^{\downarrow} & 	& {\rm 37~clusters,z=0.14-0.30} & {\rm Zhang \ et \ al. \ 2008 }      & a,e     \\ 
 
	\noalign{\smallskip}
        \hline
        \noalign{\smallskip}
 	    & 1.68 \pm 0.20^{\ddagger} &     & {\rm 26~clusters,z=0.01-0.05} & {\rm  Eckmiller \ et \ al. \ 2011 }   & e,f \\
	    & 1.76 \pm 0.08^{\updownarrow} & 1.5 & {\rm 232~clusters,z=0.04-1.46} & {\rm  Reichert \ et \ al. \ 2011 }   &  \\ 
M_{\rm tot}-T & 1.53 \pm 0.08^{\ddagger} &     & {\rm 17~clusters,z=0.03-0.05}  & {\rm  Vikhlinin \ et \ al. \ 2009a } & e   \\ 
	    & 1.65 \pm 0.04^{\uparrow} &     & {\rm 43~clusters,z=0.01-0.12}  & {\rm  Sun \ et \ al. \ 2009 } & f   \\ 
 	    & 1.65 \pm 0.26^{\downarrow} &     & {\rm 37~clusters,z=0.14-0.30}  & {\rm Zhang \ et \ al. \ 2008 }       & e   \\ 
	\noalign{\smallskip}
        \hline
        \noalign{\smallskip}
	      & 1.34 \pm 0.18^{\ddagger} &     & {\rm 26~clusters,z=0.01-0.05} & {\rm  Eckmiller \ et \ al. \ 2011 } & c,d,f \\
	      & 1.51 \pm 0.09^{\parallel} &  & {\rm 232~clusters,z=0.04-1.46} & {\rm  Reichert \ et \ al. \ 2011 }  & a	 \\
	      & 1.76 \pm 0.13^{\uparrow} &     & {\rm 31~clusters,z=0.06-0.17} & {\rm  Arnaud \ et \ al. \ 2010 }     & c,g \\
	      & 1.64 \pm 0.12^{\uparrow} &     & {\rm 31~clusters,z=0.06-0.17} & {\rm  Arnaud \ et \ al. \ 2010 }     & c   \\
 	      & 1.90 \pm 0.11^{\parallel} & 1.3 & {\rm 31~clusters,z=0.06-0.17} & {\rm  Pratt \ et \ al. \ 2009 }      & a,g \\
L_X-M_{\rm tot} & 1.62 \pm 0.11^{\parallel} &     & {\rm 31~clusters,z=0.06-0.17} & {\rm  Pratt \ et \ al. \ 2009 }      & c,g \\
	      & 1.83 \pm 0.14^{\uparrow} &     & {\rm 31~clusters,z=0.06-0.17} & {\rm  Pratt \ et \ al. \ 2009 }      & c,g \\
	      & 1.53 \pm 0.10^{\parallel} &     & {\rm 31~clusters,z=0.06-0.17} & {\rm  Pratt \ et \ al. \ 2009 }      & c,d \\
	      & 1.71 \pm 0.12^{\uparrow} &     & {\rm 31~clusters,z=0.06-0.17} & {\rm  Pratt \ et \ al. \ 2009 }      & c,d \\
	      & 1.61 \pm 0.14^{\ddagger} &     & {\rm 17~clusters,z=0.03-0.05} & {\rm  Vikhlinin \ et \ al. \ 2009 }  & b   \\
	      & 2.33 \pm 0.70^{\downarrow} &     & {\rm 37~clusters,z=0.14-0.30} & {\rm  Zhang \ et \ al. \ 2008 }      & a   \\
 	\noalign{\smallskip}
        \hline
        \noalign{\smallskip}
M_{\rm gas}-M_{\rm WL} & 1.04\pm 0.10^{\dagger} & 1 & {\rm 50~clusters,z=0.15-0.55} & {\rm Mahdavi \ et \ al. \ 2012 }    & i \\
M_{\rm WL} - Y_X & 0.56 \pm 0.08^{\dagger} & 0.6 & {\rm 50~clusters,z=0.15-0.55} & {\rm Mahdavi \ et \ al. \ 2012 }    & i \\
M_{\rm tot} - Y_X & 0.53 \pm 0.06^{\ddagger} & 0.6 & {\rm 26~clusters,z=0.01-0.05} & {\rm  Eckmiller \ et \ al. \ 2011 } & e,f \\
M_{\rm tot} - Y_X & 0.57 \pm 0.03^{\ddagger} & 0.6 & {\rm 17~clusters,z=0.03-0.05} & {\rm  Vikhlinin \ et \ al. \ 2009 } & e \\ 
M_{\rm tot} - Y_X & 0.57 \pm 0.01^{\uparrow} & 0.6 & {\rm 43~clusters,z=0.01-0.12} & {\rm  Sun \ et \ al. \ 2009 } & f \\
S - T       & 0.92 \pm 0.24^{\uparrow} & 1   & {\rm 31~clusters,z=0.06-0.17} & {\rm Pratt \ et \ al. \ 2010 }     &     \\
L_X - Y_X   & 1.07 \pm 0.08^{\uparrow} & 0.8 & {\rm 31~clusters,z=0.06-0.17} & {\rm  Arnaud \ et \ al. \ 2010 }   & c,g \\
L_X - Y_X   & 0.82 \pm 0.03^{\parallel} & 0.8 & {\rm 31~clusters,z=0.06-0.17} & {\rm  Pratt \ et \ al. \ 2009 }   & c,d,e \\
M_{\rm gas} - T & 2.12 \pm 0.12^{\uparrow} & 1.5 & {\rm 31~clusters,z=0.06-0.17} & {\rm Croston \ et \ al. \ 2008 }   &     \\
M_{\rm gas} - T & 1.99 \pm 0.11^{\flat} & 1.5 & {\rm 31~clusters,z=0.06-0.17} & {\rm Croston \ et \ al. \ 2008 }   &     \\
M_{\rm gas} - T & 1.86 \pm 0.19^{\downarrow} & 1.5 & {\rm 37~clusters,z=0.14-0.30} & {\rm Zhang \ et \ al. \ 2008 }     & e   \\

         \noalign{\smallskip}
         \hline
\end{array}
\]
\begin{list}{}{}                                                                       
\item[$^{\rm a}$ bolometric luminosity.]
\item[$^{\rm b}$ luminosity in the 0.5-2 keV band.]
\item[$^{\rm c}$ luminosity in the 0.1-2.4 keV band.]
\item[$^{\rm d}$ core excised luminosity.]
\item[$^{\rm e}$ core excised temperature.]
\item[$^{\rm f}$ limited to systems with temperature $\le 3$ keV.]
\item[$^{\rm g}$ corrected for Malmquist bias.]
\item[$^{\rm h}$ individual Malmquist bias corrections for SCC, WCC and NCC clusters.]
\item[$^{\rm i}$ weak lensing mass.]
\item[$^{\rm \dagger}$ bayesian method by \cite{Hogg10}.]
\item[$^{\rm \ddagger}$ BCES \citep{AkritasBershady} bisector.]
\item[$^{\rm \uparrow}$ BCES orthogonal.]
\item[$^{\rm \parallel}$ BCES Y$|$X.]
\item[$^{\rm \updownarrow}$ BCES X$|$Y.]
\item[$^{\rm \downarrow}$ ODRPACK orthogonal.]
\item[$^{\rm \flat}$ WLSS.]
\item[]
\end{list}            
 \end{table}

\subsection{$L_X-M_{tot}$ relation}

Future all-sky X-ray surveys, such as eROSITA, will image hundreds of
thousands of clusters with very shallow observations, collecting too
few photons to extract spectra or mass profiles. On the other hand a
measure of the X-ray luminosity will be always possible if redshift
information is available. Hence the correlation between the X-ray
luminosity and total mass is an important tool for cosmology because
it correlates the total mass of a system with its `cheapest' X-ray
observable. This does, however, require a very accurate determination
of the $L_X-M_{\rm tot}$ relation and its scatter.

If a large range in mass is covered, the degeneracy between $\Omega_m$
and $\sigma_8$ can be broken \citep{Reiprich02}. Hence the calibration
of this relation needs to be extended to low mass groups. A large
number of observations
\citep[e.g.][]{Reiprich02,Ettori02,Maughan06,Maughan07, Chen07,
  Vikhlinin09a, Pratt09,Arnaud10a,Mantz10b,Eckmiller11,Reichert11}
show that the X-ray luminosity is heavily affected by
non-gravitational processes. Observed slopes for the $L_{X}-M_{\rm
  tot}$ relation are $\sim$1.4-1.9, steeper than the self-similar
prediction of $4/3$. Furthermore, both the slope and normalization of
this relation can vary quite significantly depending on the energy
band\footnote{Apart from the bolometric value which requires an
  extrapolation, the luminosity is often derived using the 0.1-2.4 or
  0.5-2 keV bands.}  and method used for the flux extraction.  In
Figure~\ref{fig:bias} ({\it right panel}) we compare the $L_{X}-M_{\rm tot}$ relation
derived by \citet{Pratt09} to the results from \citet{Vikhlinin09a}.
There is a general agreement for the recovered slopes and
normalizations between measurements.

Among the various X-ray scaling relations the scatter of $\sim 40\%$
in the $L_X-M_{\rm tot}$ relation is the largest. This has been attributed
to the presence of cool-cores and the overall dynamical state of 
clusters. Most of the scatter derives from the central part of the
cluster (within $\sim$0.1-0.2 Mpc) where cooling and merging effects
are most pronounced. Excluding the cluster core can reduce the scatter
to less than 10$\%$ in mass \citep{Markevitch98, Mantz10b}.

In an attempt to reduce the scatter between the mass proxies and the
total cluster mass, \cite{Ettori12} introduced a generalized scaling
law, defined as

\begin{equation}
 M_{\rm tot}=10^KA^aB^b\,.
\end{equation}

They found a locus of minimum scatter that relates the
logarithmic slopes of two generalized independent variables, namely
the temperature $T$, which traces the depth of the cluster potential, and
another one accounting for the gas density distribution, such as gas
mass $M_{\rm gas}$ or X-ray luminosity. This minimum scatter locus
corresponds to the plane where $L_X$: $b_M = -3/2 a_M +3/2$ and $b_L =
-2 a_L +3/2$ for $A=M_{\rm gas}$ and $L_X$, respectively, and $B=T$.
Within this approach, all the known scaling relations appear as
particular realizations of generalized scaling relations.  A new
relation is also introduced, $M_{\rm tot} \propto (L_XT)^{1/2}$, which
is analogous to the $M_{\rm tot}-Y_X$ relation, once luminosity is
used instead of gas mass. Although, this approach is still affected by
mass calibration and selection effects, it allows a minimization of
the scatter imposing a new constraint on the slope of the scaling
laws.

\section{Evolution}
\label{sec:2}
The X-ray scaling relations are expected to be redshift-dependent,
even in the simplest case where gravity dominates. This is because of
the cosmological expansion and the corresponding evolution of the
background matter density of the Universe. The evolution is expected
to be stronger when non-gravitational processes are considered, due to
the growing relative importance of such processes to the energy budget
of galaxy clusters as a function of redshift (e.g. the AGN luminosity
function evolves strongly with redshift in both X-ray and radio
bands).

In the self-similar scenario, the scale (in mass or $T$) does not play
any role (i.e. groups and clusters are the same kind of objects) and
as a result only the normalization depends on cosmic
time/redshift. This dependence is generally parametrized by the
relative change in the Hubble parameter $E_z$ (or $F_z$) and one can
write a scaling relation between quantities $X$ and $Y$ as:

\begin{equation}
Y(X,z) = X_0 \times E(z)^\beta X^\alpha.
\end{equation}

One can consider more complicated scenarios in which the slope also
depends on redshift, although this would require some additional
physics. At the moment, however, the paucity of well defined samples
at high redshift (and the narrow range in mass surveyed as a good
sample of galaxy groups at high-z is lacking) strongly limits the present
constraints on the redshift evolution of the scaling relations or any
clear detection of departure from the self-similar predictions.

Understanding the evolution of the scaling relations is nonetheless
crucial in order to use clusters for cosmology, especially for the
determination of the evolution of the mass function with redshift.
While the mass-observable scaling relations are calibrated reasonably
well at low redshift, at least for relaxed clusters, measuring these
relations at high redshift is considerably more challenging, due to
the long observations required to obtain sufficiently deep X-ray data
to constrain the cluster properties. For this reason no clear consensus
has been reached on the evolution of the X-ray scaling relations,
despite a number of observational studies carried out in the past 
decade \citep[e.g.][]{Vikhlinin02, Vikhlinin09a, Ettori04b, Kotov05,
Maughan06,Maughan07, Maughan12,Hara06,Morandi07a,Branchesi07,Pacaud07,
Andreon11,Reichert11}.

\begin{figure}
\centering
\hspace{0.2cm}
\includegraphics[width=0.7\textwidth]{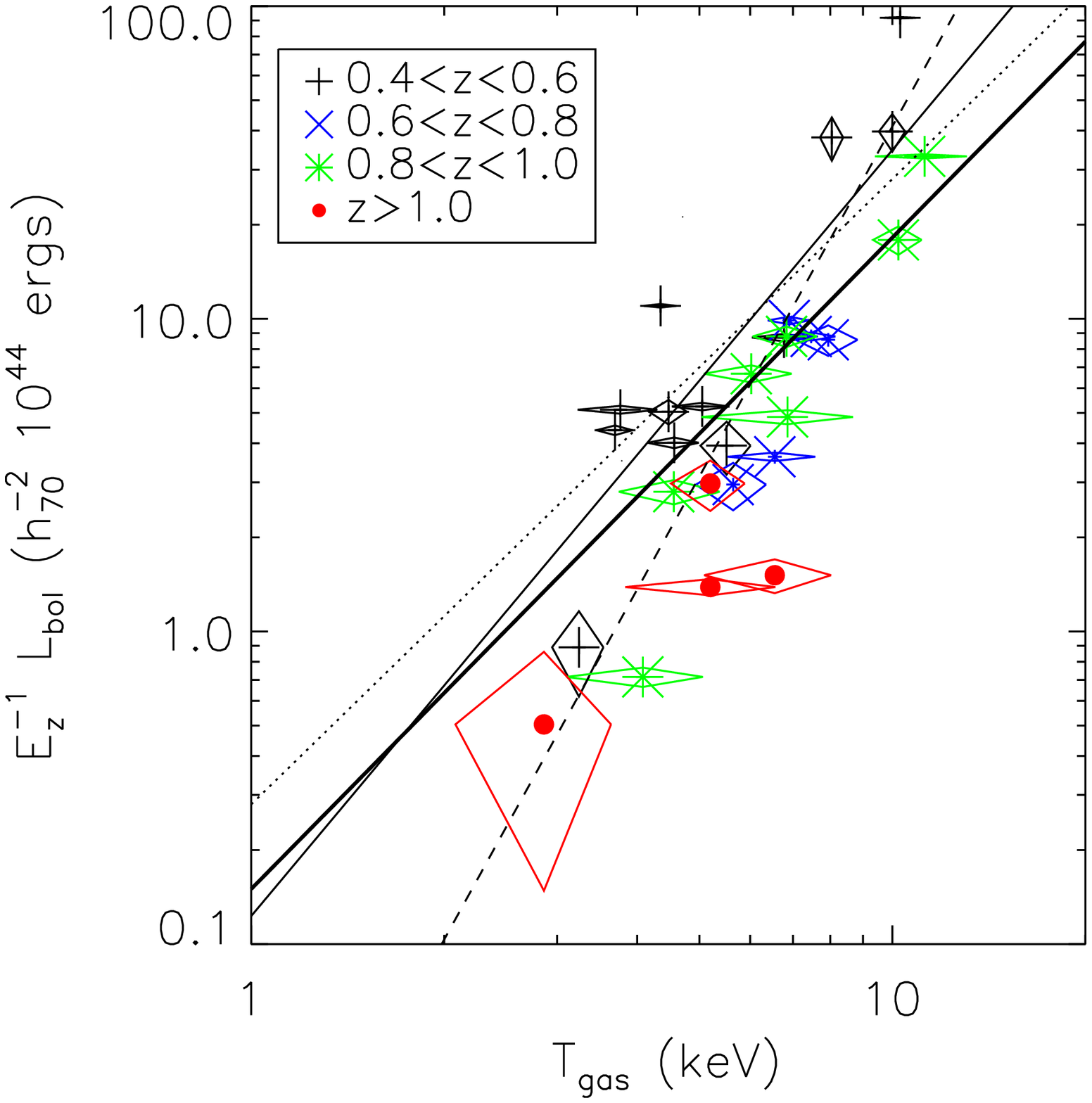}
\hspace{0.5cm}
\includegraphics[width=0.7\textwidth,angle=270]{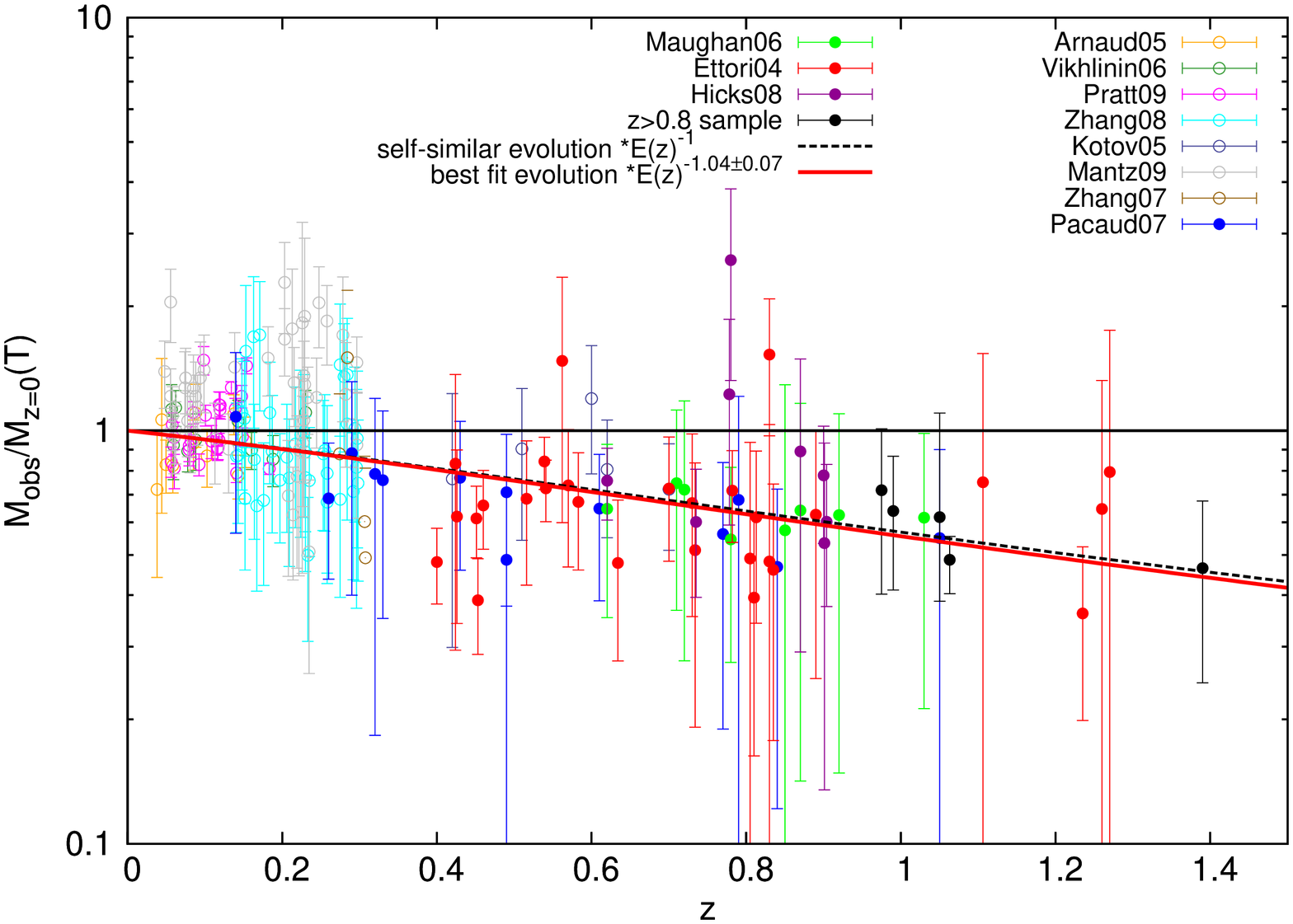}
\caption{\textit{Upper Panel:} $E_z^{-1}$ $L_X-T$ relation for a sample
  of 28 objects observed with Chandra in the z-range 0.4-1.3. Dotted
  line: slope fixed to the self-similar value. Dashed line: slope
  free. The solid lines represent the local best-fit results (from
  thinnest to thickest line): \cite{Markevitch98}, \cite{Arnaud99},
  \cite{Novicki02}. The evolution is evaluated by fitting 
  the relation $\log{Y}=\alpha+A\log{X}+B\log{(1+z)}$ to the data, where
  ($\alpha$, $A$) are the best-fit results obtained from a sample of
  objects observed at lower redshift. (Credit: Ettori et al., A$\&$A, vol.417, pg.13, 2004, reproduced with permission $\copyright$ ESO). {\textit{Bottom Panel:}} Redshift
  evolution of the $M_{tot}-T$ relation. Black-dashed line:
  self-similar prediction ($\propto E_z^{-1}$).  Continuous red line:
  best-fit evolution ($\propto E_z^{-1.04\pm0.07}$).Credit: Reichert et al., A$\&$A, vol.535, pg.A4, 2011, reproduced with permission $\copyright$ ESO.}
\label{fig:evol}
\end{figure}

For example \cite{Ettori04b} (see top panel of Fig. \ref{fig:evol}), \cite{Hara06} and \cite{Reichert11} found a
negative evolution for the $L_{X}-T$ relation, at odds with the result
by \cite{Kotov05} who observed a positive evolution, and
\cite{Pacaud07} who found no significant evolution. In general, all
the relations involving parameters that depend on the gas density show
significant deviations from the predictions; a clear indication
that non-gravitational processes cannot be neglected. In contrast, the
$M_{\rm tot}-T$ relation is generally very close to the self-similar
prediction (see bottom panel of Fig. \ref{fig:evol}), which is not
surprising because it mostly depends on the dark matter potential.  A
compilation of the most recent publications and their main results can
be found in \cite{Reichert11}.

There are several reasons why the results from different studies
appear to be contradictory. One of the main problems in achieving a
consensus is the difficulty in accounting for selection biases caused
by the lack of concordance between different studies in the cluster
selection. The use of miscellaneous archival cluster samples leads to
selection bias corrections that may vary from sample to sample and
alter the measured evolution especially when considering the poor
statistics due to the small sample size. Since clusters do not have a
clear outer boundary \citep[see][for a review on the cluster
outskirts]{Reiprich13} the different choices of defining the fiducial
radius (e.g. redshift-independent or not) within which the cluster
properties are considered, may also play a role, although
\cite{Reichert11} found that this effect should be negligible.
Importantly, the assumed local scaling relation which is the reference
to compare the high-redshift data to, has a direct impact on the
inferred evolution. This is likely to introduce systematic errors.  As
shown in the previous sections, the luminosity is sensitive to the
central gas density such that tighter scaling relations involving the
X-ray luminosity are obtained by excising the core. The lack of photon
collecting power of current instruments makes the cool cores excision
problematic at high redshift and increases the scatter in the scaling
relations making it difficult to disentangle the evolution from the
no-evolution scenario.

Major progress will come from X-ray observations that measure the
thermodynamical properties of high redshift clusters. Among these are
the XMM  Deep Cluster Project (XDCP2; \citealt{Fassbender11}) that has
so far spectroscopically confirmed 22 clusters at $z>0.8$. The study
of these high redshift objects is extremely important. Although
massive clusters are rare at any redshift, they are most sensitive to
cosmology, allowing a more precise study of the evolution.

As discussed in more detail in \S\ref{sec:5}, the expected evolution
of mass-observable scaling relations has also been studied using
samples extracted from large cosmological hydrodynamical
simulations. The predictions depend on the adopted prescriptions for
cooling and feedback. Hence observational constraints have the
potential to constrain the non-gravitational physics. However,
\cite{Short10} point out that a statistically meaningful comparison
with observations is impossible at the moment, because the largest
samples of high-redshift clusters currently available are still
affected by strong selection biases.

\section{SZ scaling relations}
\label{sec:3}
The thermal Sunyaev-Zeldovich effect (SZE) is a distortion of the
black body spectrum of the photons from the cosmic microwave
background (CMB) in the direction of galaxy clusters.  As first
predicted by \cite{SZ70, SZ72}, the low-energy CMB photons can
interact via inverse Compton scattering with the free electrons in the
intra-cluster medium.  This scattering causes a small change of the
mean photon energy as

\begin{equation}
\frac{\Delta\nu}{\nu}\simeq\left(\frac{kT}{m_{e}c^{2}}\right)\sim 10^{-2}\,,
\end{equation}

\noindent where m$_{e}$ is the mass of the electron. The frequency
shift causes an increase in the CMB intensity in the high frequency
(Wien) part of the spectrum and a decrement in the Rayleigh-Jeans
tail. This corresponds to a brightness fluctuation in the CMB of
$\sim$10$^{-4}$, which is roughly an order of magnitude larger than
the cosmological signal from the primary anisotropies.

\begin{figure}
\centering
\includegraphics[width=7cm]{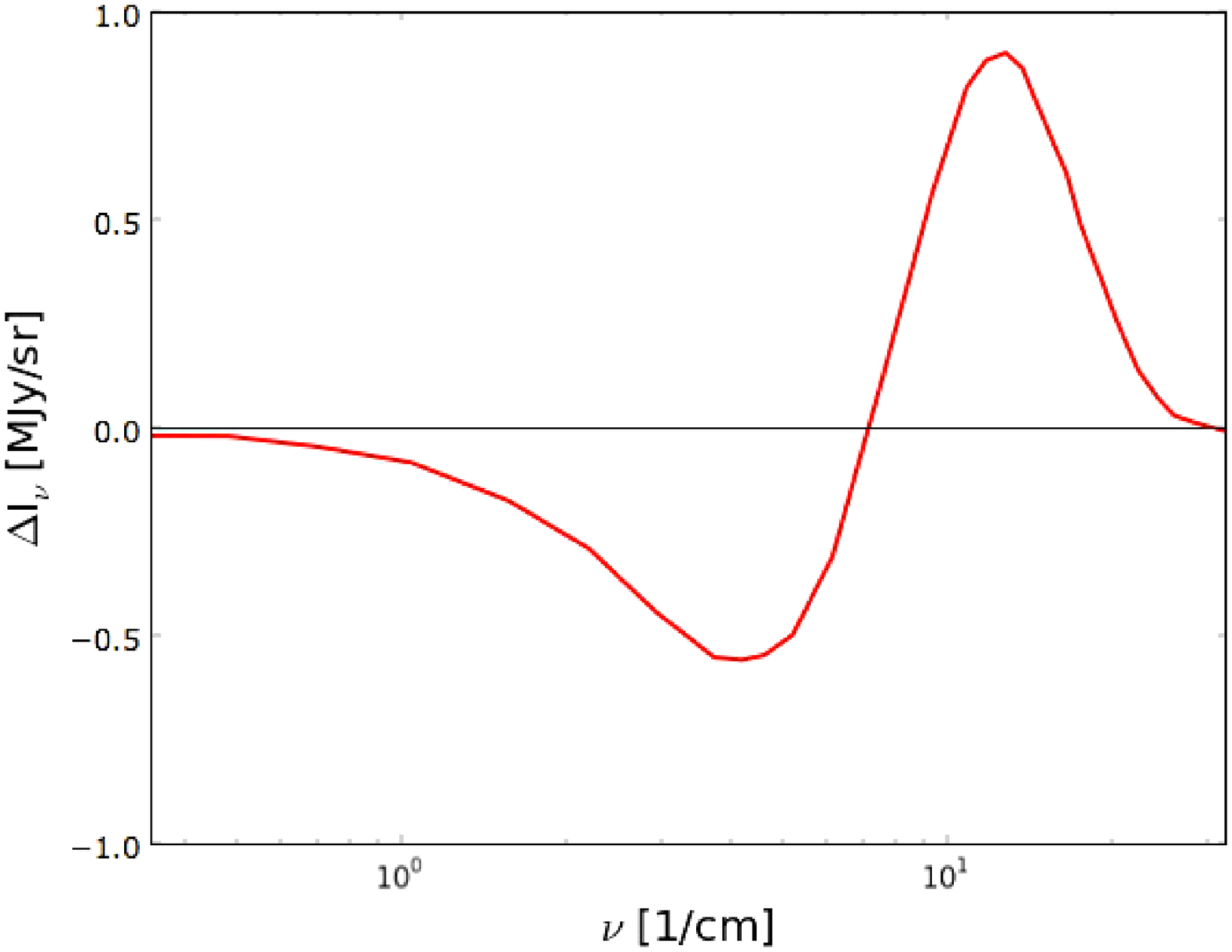}
\caption{Differences in intensity between the on-cluster distorted
  spectrum and the off-cluster black-body spectrum for a massive
  cluster with y=5$\times$10$^{-4}$.}
\label{fig1_sz}
\end{figure}

Figure \ref{fig1_sz} plots the difference in intensity between the
on-cluster distorted spectrum and the off-cluster black-body spectrum
for a massive cluster ($y$=5$\times$10$^{-4}$, where $y$ is defined
below).  Within a non-relativistic Thomson diffusion
limit\footnote{The non-relativistic approximation for the SZ effect
  roughly holds for clusters with ICM temperatures $kT\leq$10 keV.},
the spectral shape for the SZ effect derives from the Kompaneets
equation \citep{Kompaneets57} and is defined
analytically. Implementation of relativistic corrections due to the
weakly relativistic tail of the electron velocity distribution
leads to a weak dependence of this spectral shape with the electron
gas temperature \citep[e.g.,][]{Rephaeli95,Pointecouteau98,Challinor99}.

The magnitude of the decrement in the CMB is proportional to the
line-of-sight integral of the product of gas density ($n_{e}$) and
temperature ($T_{e}$)

\begin{equation}\label{eq1}
\Delta y=-2y I_{\nu}\,,
\end{equation}

\noindent where $y$ is the Comptonization parameter, and is defined as

\begin{equation}
\label{eq2}
y=\frac{\sigma_{T}k_{B}}{m_{e}c^2}\int{T_{e}n_{e}dl} \,,
\end{equation}

\noindent where $k_{B}$ is the Boltzmann constant and $\sigma_{T}$ is
the Thompson cross-section. This equation corresponds to the
integrated thermal pressure of the intra-cluster gas along the line of
sight (since $P=n_{e} k_{B} T_{e}$ in the ideal gas approximation).

Being proportional to the integrated thermal pressure support within
clusters, the magnitude of the SZ-effect is an ideal proxy for the
mass of the gas in a galaxy cluster, $M_{\rm gas}$, and thereby of the
total mass, $M_{\rm tot}$. This can be illustrated through the integrated
Comptonization parameter defined as the integral of $y$ over the solid
angle under which the cluster is seen, i.e., $\Omega$:

\begin{equation}
  Y_{\rm SZ}=\int_{\Omega}{yd\Omega}=\frac{1}{D_{A}^2}\frac{\sigma_{T}k_{B}}{m_{e}c^2}\int_V{n_{e}T_{e}dV}
\label{eq3}
\end{equation}

\noindent where $D_{A}$ is the angular distance to the cluster
and $V$ is the physical volume of the cluster. In the context of an
isothermal model, $Y_{SZ}$ is proportional to the integral of the
electron density $n_e$ over a cylindrical volume, which corresponds to
the gas mass in the same volume. Assuming a gas fraction $f_{\rm gas}=
M_{\rm gas}/M_{\rm tot}$ we thus obtain

\begin{equation}
Y_{\rm SZ}D_{A}^2\propto T_{e}\int{n_{e}dV}=M_{\rm gas}T_{e}=f_{\rm gas}M_{\rm tot}T_{e}\,.
\label{eq4}
\end{equation}

Using Eqn.~\ref{eq4} in combination with the scaling $T_{e}
\propto M_{\rm tot}^{2/3}E(z)^{2/3}$, assuming hydrostatic equilibrium and
an isothermal distribution for both the dark matter and the cluster
gas \citep[e.g][]{Bryan98}, we can obtain the following
scaling relations for the integrated SZ signal and others observables:

\begin{center}
\begin{minipage}{3in}
\begin{equation}
\begin{aligned}
  Y_{\rm SZ}D_{A}^2&\propto & f_{\rm gas} T_{e}^{5/2} E(z)^{-1} \\
  Y_{\rm SZ}D_{A}^2&\propto & f_{\rm gas} M_{\rm tot}^{5/3} E(z)^{2/3} \\
  Y_{\rm SZ}D_{A}^2& \propto & f_{\rm gas}^{-2/3} M_{\rm gas}^{5/3} E(z)^{2/3}
\end{aligned}
\end{equation}
\end{minipage}
\end{center}

Equations~\ref{eq1} and~\ref{eq2} show that the amplitude of the
SZ-effect is independent of redshift. Therefore, in contrast to X-ray
and optical measurements, it does not undergo surface brightness
dimming (i.e., $\propto (1+z)^{-4}$) since this is exactly compensated
by the increase of the CMB intensity as $\propto (1+z)^4$ (at higher
redshift we are probing a younger Universe where the CMB temperature
is higher). It should be stressed that the actual SZ measurements are
flux measurements; they are directly proportional to the integrated
Compton parameter as expressed in Eqn.~\ref{eq3}. Hence they do suffer
from a dimming: in the case of unresolved clusters this is determined
by the ratio of the cluster solid angle over the instrumental beam.

The lack of a dependence with redshift and the direct proportionality
to the total mass of the cluster should make SZ selected samples very
close to mass limited. This makes the SZE an excellent probe for
cluster cosmology.  Large area SZ surveys carried out by large, single
dish ground-based or space telescopes (such as SPT, ACT and {\it
  Planck}) are performing such studies and are delivering samples with
significantly higher median redshifts compared to X-ray selected
cluster catalogues \citep[e.g.,][]{Reichardt13}.

The biggest challenge for blind SZ surveys is the extraction of the SZ
signal and the separation between the fore- and background structures
as well as the astrophysical signal. The diffuse gas that resides in
large-scale filaments provides only a negligible contamination due to
its comparatively low density and temperature. Small halos, on the
other hand, are expected to be present in large number and cannot be
resolved with the current SZ observations; they may provide a
significant contamination as shown by \citet{White02} using
cosmological hydrodynamical simulations. The main signal contamination
is expected to come from other astrophysical emissions such as
infra-red and radio point sources, cosmic infrared background
fluctuations, Galactic emission and CMB contaminations
\citep[e.g.,][]{Aghanim05}. To optimally exploit large SZ samples in
cosmological and astrophysical studies, well calibrated relations
between the total mass and the SZ flux are required
\citep[e.g.,][]{Dasilva04,Aghanim09}.

SZ and X-ray measurements naturally complement each other. In fact,
due to the respective dependence on the gas density profile, i.e.,
$n_{e}$ and $n_e^2$ respectively, density, temperature and therefore
mass profiles can be inferred from joint SZ and X-ray analyses out to
very large radii \citep[e.g., ][]{Kitayama04,Basu10}.  Furthermore
X-ray and SZ measurement can be combined to determine the Hubble
parameter (H$_{0}$; \citealt{SW78}) by measuring the distances to
clusters.

The SZ effect as described here is usually called `thermal' and
largely dominates over the `kinetic' SZ effect that is caused by the
comoving bulk motion of the hot electrons in the intra-cluster medium
\citep{Sunyaev80}. The detection and quantification of the kinetic SZ
effect is an ongoing topic of discussion
\citep[e.g.,][]{Atrio08,Kashlinsky10,Hand12}. We refer to the reviews
by \citet{Rephaeli95}, \citet{Birkinshaw99} and \citet{Carlstrom02} for
a more detailed discussion of the SZE effect and related issues.


\subsection{SZ scaling relations:  pre-survey-era observations}

The first significant detection of the thermal SZE was reported by
\cite{Birkinshaw78}, only six years after the concept was proposed by
\cite{SZ72}.  The history of successful targeted observations of
clusters to detect the SZ effect goes back two decades, when
pioneering observations were made with interferometers such as the
Owen Valley Radio Observatory (OVRO; \citealt{Birkinshaw91},
\citealt{Herbig95}), the OVRO/BIMA interferometers
\citep{Carlstrom96}, or single dish bolometric instruments such as
the Sunyaev-Zel'dovich Imaging Experiment (SuZIE;
\citealt{Holzapfel97}), the Diabolo photometer at the focus of the
IRAM 30m radio telescope \citep{Desert98} and the NOBA instrument
on the 45m NRO telescope \citep{Komatsu99}.

In the last decade these measurements have been expanded to samples of
clusters and the focus has moved to understanding the correlation
between the SZ signal and other clusters observables, especially those
related to the total mass. To this end a number of early studies
targeted small samples of (well-)known clusters.  Due to the intrinsic
limitations of these SZ measurements, as well as the reach of the
X-ray data, most of these studied were intrinsically limited to the
inner regions of the clusters \citep[e.g.][]{Cooray99,McCarthy03,
  Morandi07a}. \citet{Benson04} showed that the
integrated SZ flux is a more robust observable than the central values
of the SZ signal and found a strong correlation with X-ray
temperature using a sample of 15 clusters obtained with SuZIE and
X-ray temperatures from the ASCA experiment.  More recently,
\citet{Bonamente08} examined the scaling relations between $Y_{\rm SZ}$
and total mas, gas mass and gas temperature using 38 clusters observed
with \textit{Chandra} and OVRO/BIMA and found that the slope and the
evolution of the observed relations agree with that predicted by a
self-similar model in which the evolution of cluster is dominated by
gravitational processes \citep[also see][for a similar
result using AMiBA]{Huang10}.

\citet{Marrone09} measured the relation between $Y_{\rm SZ}$ and lensing
mass within 350~kpc, derived from a strong and weak lensing analysis
of HST observations of 14 X-ray luminous clusters of galaxies.  They
found no evidence of segregation in $Y$ between disturbed and
undisturbed clusters, as had been seen with $T_{X}$ on the same
physical scales. This result confirmed that SZE may be less sensitive
to the details of cluster physics in cluster cores compared to X-ray
observations, as suggested by simulations. More recently,
\cite{Marrone12} studied a sample of 18 clusters using weak lensing.
They found an intrinsic scatter of 20\% in the weak lensing mass at
fixed $Y$, with a suggestion of a dependence on
morphology. \cite{Hoekstra12} compared their weak lensing masses to
results from \cite{Bonamente08} and \cite{Planck11}, concluding that
the SZ signal correlates well with weak lensing mass. The intrinsic
scatter that they measure is smaller but consistent with the results
from \cite{Marrone12}.

The SZ signal can be predicted using X-ray observations. As discussed
in \S\ref{sec:yx}, \citet{Kravtsov06} introduced an X-ray analog of
$Y_{\rm SZ}$, $Y_{X}$, which is the product of the gas mass and the
spectroscopic X-ray temperature. The comparison between $Y_{\rm SZ}$ and
$Y_{X}$ provides information on the ICM inner structure and especially
the clumpiness. Indeed, as aforementioned, the quadratic and linear
dependence of the X-ray and SZ signal on $n_e$ for $Y_{X}$ and
$Y_{\rm SZ}$ respectively enable us to equate $Y_{X}$ and $Y_{\rm SZ}$ only if
the gas distribution is completely smooth, i.e. $\langle
n_{e}^{2}\rangle$=$\langle n_{e}\rangle^2$.

From these early studies, no consensus was reached whether predictions
for the SZ signal based on ICM properties from X-ray observations are
in agreement with direct SZ observations. \citet{Lieu06} and
\citet{Bielby07} found evidence for a weaker SZ signal than expected
from X-ray predictions in the WMAP3 data. This case was strengthened
by the WMAP7 data analysis \citep{Komatsu11} which argues for a
deficit of SZ signal, especially at low halo masses. However,
re-analysing the same data, \citet{Afshordi07} and \citet{Melin11}
found a good agreement between the SZ measurements and X-ray
predictions. These conflicting results have demonstrated the need for
more precise SZ measurements for larger samples of clusters from
dedicated and multi-wavelength surveys in order to improve our
understanding of cluster physics and cosmology.

\subsection{SZ scaling relations: first results from large dedicated
  SZ surveys}

Thanks to the start of wide-area SZ surveys, performed with dedicated
instruments, there has been a lot of progress in recent years.
Indeed, from the current generation of high sensitivity, high
resolution and large coverage microwave telescopes (such as {\it
  Planck}, ACT, and SPT), new cluster surveys are producing catalogues
of hundreds of SZ-detected clusters including new high-redshift
objects. These numbers will continue to increase in the next few years.

The first clusters discovered in a blind SZ survey were reported by
\citet{Staniszewski09} who used data from the South Pole Telescope
\citep[SPT;][]{Carlstrom11} and demonstrated the capability of the SZ signal
for cluster-finding. \citet{Hincks10}, \citet{Vanderlinde10} and
\citet{Marriage11} reported more blind cluster detections (each
$\sim$20 candidates), setting the stage for SZ selected cluster
catalogs.

SPT and ACT have been able to realize blind detections from ground
based facilities because they combine three essential design features:
resolution matched to the size of the cluster, degree-scale field of
view for efficient surveying and the unprecedented sensitivity of
bolometric detector arrays with 1000 elements. These observations have
also demonstrated the need for multi-wavelength observations for a
blind SZ survey in order to reduce the contamination from
astrophysical foreground and background, as well as from primary CMB
anisotropies. For instance the aforementioned SPT results were
obtained with a single band survey (150 GHz) and the contamination
hampered the determination of the aperture size required to integrate
the SZ flux. Since the uncertainty on the scale aperture greatly
affects the relation between $Y_{\rm SZ}$ and total mass, the SPT
detection significance has been used as a mass proxy rather than
$Y_{\rm SZ}$ in these initial studies.

The {\it Planck} satellite, launched in 2009, will soon provide the
first multi-band, all-sky catalog of blind SZ detections. This catalog
will be nearly mass selected and less affected by the systematics of
X-ray selection. This makes the SZ signal a very attractive
alternative for an unbiased proxy of the cluster mass
\citep{Vanderlinde10,Williamson11}. The early release from the {\it
  Planck} Collaboration consists of 189 SZ extended sources at
low/intermediate redshift \citep[i.e., the ESZ
sample;][]{Planck11}. Although most of the clusters in this catalog
were previously detected (either in the optical or in the X-ray band),
the sample also contains 20 clusters discovered by {\it
  Planck}. Twelve of these have been confirmed with XMM-Newton
\citep{Planck11_IX}, and 8 remained unconfirmed cluster candidates.
Seven were further confirmed by targeted observations with SPT
\citep{Story11}, AMI \citep{PlanckAMI13}, Bolocam \citep{Sayers12} and
CARMA \citep{Muchovej12}.

Recently the SPT collaboration released a new extended sample of 224
SZ clusters detected at 150~GHz in an area of 750~sq.~deg.,
drastically increasing the number of SZ detected clusters
\citep{Reichardt13}. More than half (117) of the systems in this
catalog are new detections. Part of the sample was used, together with
other probes, to constrain cosmology \citep{Benson13}. More recently
the ACT collaboration published a sample of 68 clusters out of which
19 were new discoveries \citep{Hasselfield13}.

{\it Planck} released some early results on scaling relations derived
from three main studies, each using a different approach.  The first
is a statistical study of scaling relations by \cite{Planck11_VII}
starting from a X-ray selected sample of about 1600 galaxy clusters
drawn from the X-ray meta catalog (MCXC) by \citet{Piffaretti11}. This
study combined the accuracy of the {\it Planck} measurements with the
statistical size of the sample to overcome the dispersion within
individual measurements and recovered X-ray--SZ scaling relations
consistent with the predictions from X-ray constraints. The intrinsic
scatter in the D$^{2}_{A} Y_{500} -L_{X,500}$ relation amounts to
$\sim$40$\%$ and it is likely due to variation in the dynamical states
of the clusters. \citet{Planck11_XI} studied a subsample of 62 local
($z\leq 0.4$) clusters from the ESZ sample for which high quality
XMM-Newton archive data were available. This study confirmed the
agreement between the SZ and X-ray scaling relations. A remarkably
small logarithmic intrinsic scatter (10$\%$) in the
$D_A^{2}Y_{500}-Y_{X,500}$ relation was derived, consistent with the
idea that both quantities are low-scatter mass proxies.

\begin{figure}
\centering
\includegraphics[width=5.3cm,height=4.9cm]{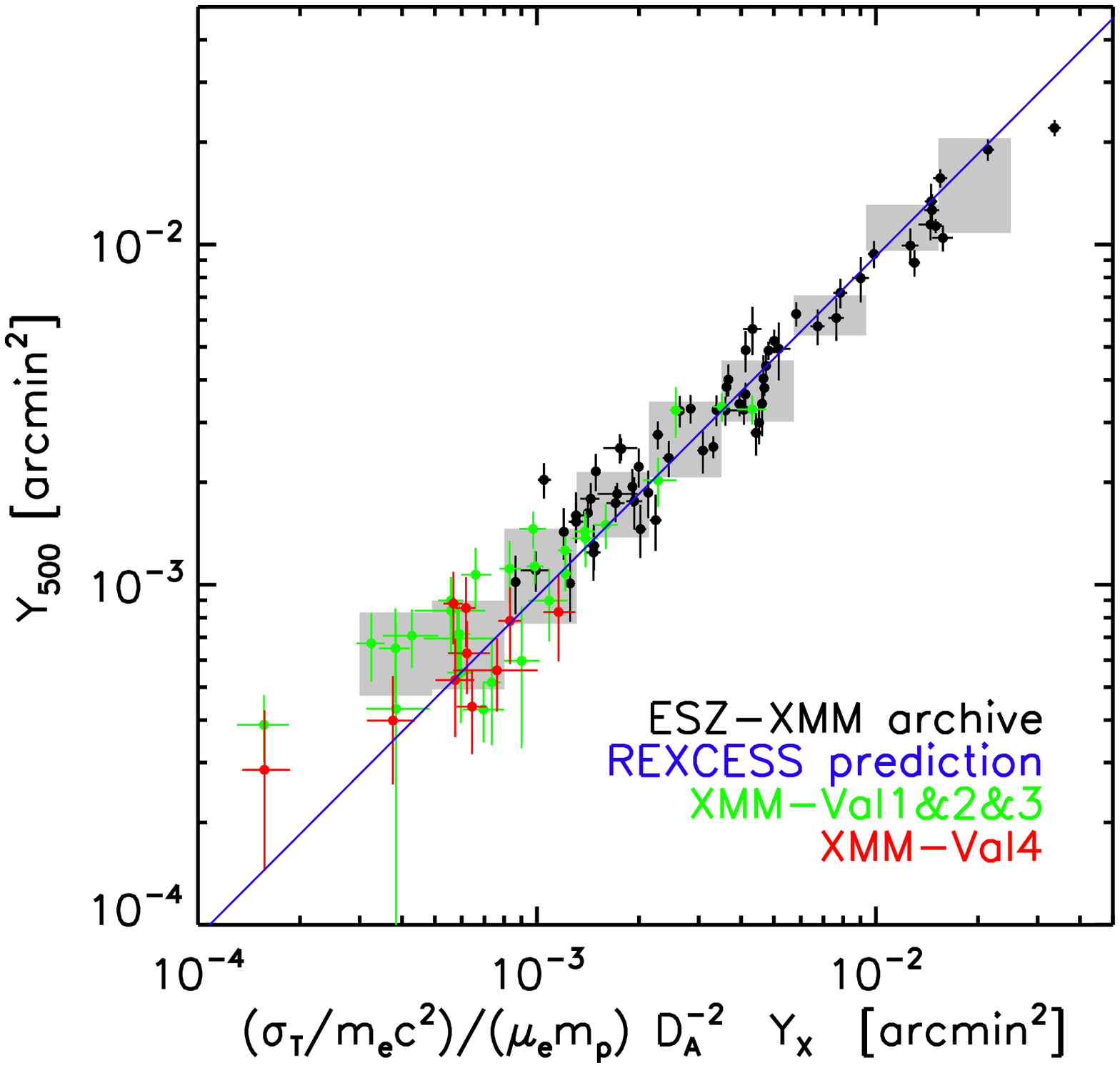}
\includegraphics[width=5.3cm]{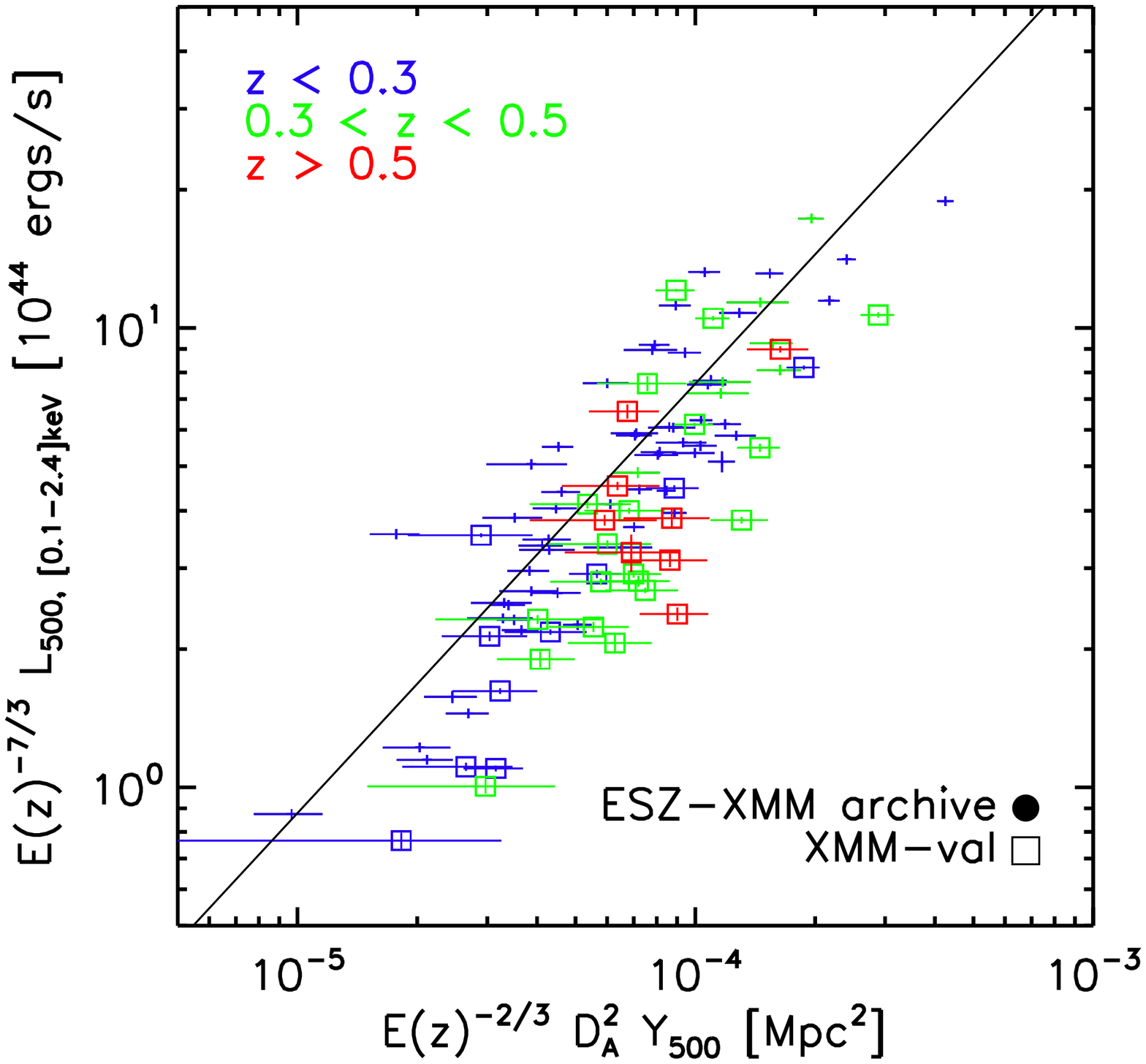}
\caption{ Scaling properties from {\it Planck} studies. {\it (left)}
  Relation between $Y_\mathrm{X}$ and $Y_\mathrm{500}$ for the new
  detected clusters confirmed by XMM-Newton (red and green dots) and
  of the ESZ clusters with XMM-Newton archive data (black dots). The
  blue solid line shows the prediction from the REXCESS sample
  measurements. Credit:Planck Collaboration, A$\&$A,Vol. 550, pg. 130,2013, reproduced with permission $\copyright$ ESO.  {\it (Right)} Scaling relations between the X-ray
  luminosity and $Y_\mathrm{500}$ for new {\it Planck} clusters
  confirmed by XMM-Newton (squares) and for the ESZ clusters with
  XMM-Newton archive data (dots). Each quantity is scaled with
  redshift, as expected from standard self-similar evolution. The
  solid black lines denote the predicted $Y_\mathrm{500}$ scaling
  relations from the REXCESS X-ray observations
  \citep{Arnaud10a}. Credit:Planck Collaboration, A$\&$A,Vol. 550, pg. 130,2013, reproduced with permission $\copyright$ ESO.}
\label{planck}
\end{figure}

Finally \citet{Planck11_IX}, \citet{Planck12_I} and
\citet{Planck13_IV} analysed a sample of 37 newly detected clusters by
{\it Planck} that were confirmed with XMM-Newton as single systems.
This study revealed a non-negligible population of massive dynamically
perturbed objects with low X-ray surface brightness, lying around
or below the flux limit of X-ray surveys such as REFLEX and NORAS
\citep{Boehringer00, Boehringer04}. In this {\it Planck} sample the
proportion of objects with a perturbed dynamical state tops $\sim
70$\%, which is to be compared to the $\sim 30$\% observed in the
X-ray selected representative REXCESS sample
\citep{Boehringer07}. These clusters have much flatter density
profiles, lower X-ray luminosity and a more disturbed morphology when
compared to X-ray selected samples. These SZ selected clusters show a
larger scatter in the plane of the scaling relations involving $Y_X$
or $L_{X}$ with respect to the X-ray selected ones (see 
Fig.~\ref{planck}).

The consistency of these three approaches and results highlights the
very good agreement between the SZ and X-ray measurements of the
intra-cluster medium, at least within a radius of $R_{500}$. Similar
results drawn from smaller samples of clusters spreading over a wider
range of redshifts derived by SPT \citep{Andersson11} and ACT
\citep{Marriage11} also agree with the {\it Planck} findings.  It is,
however, important to test the effects of clumping or assumptions of
hydrostatic equilibrium by comparing to observations that are not
related to the ICM. Comparison to weak lensing masses were discussed
above. Alternatively one can compare to dynamical masses, which was
done in \cite{Rines10} and \cite{Sifon12}.

\begin{figure}
  \includegraphics[width=6.cm,height=5.1cm]{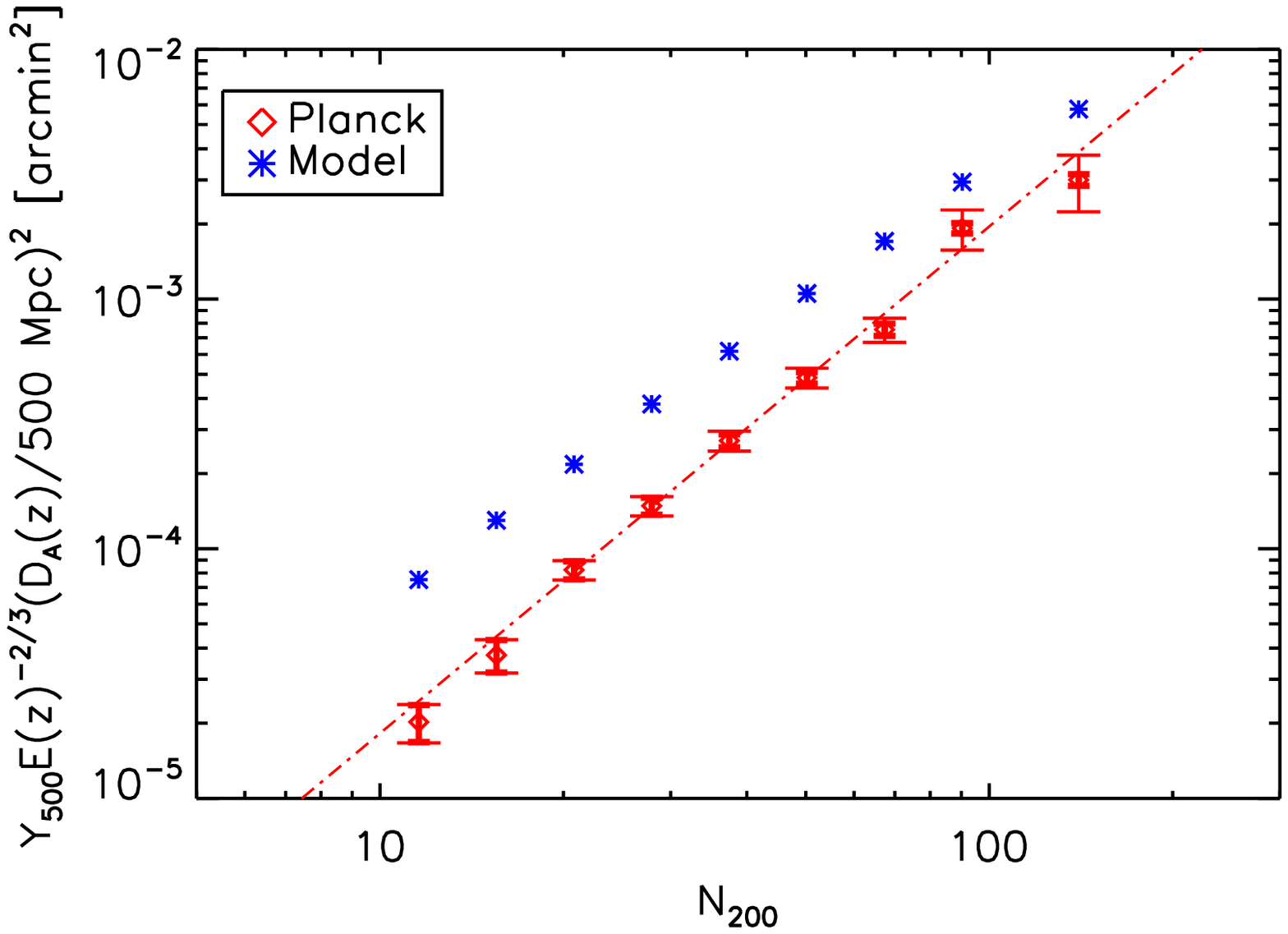}%
  \includegraphics[width=5.4cm]{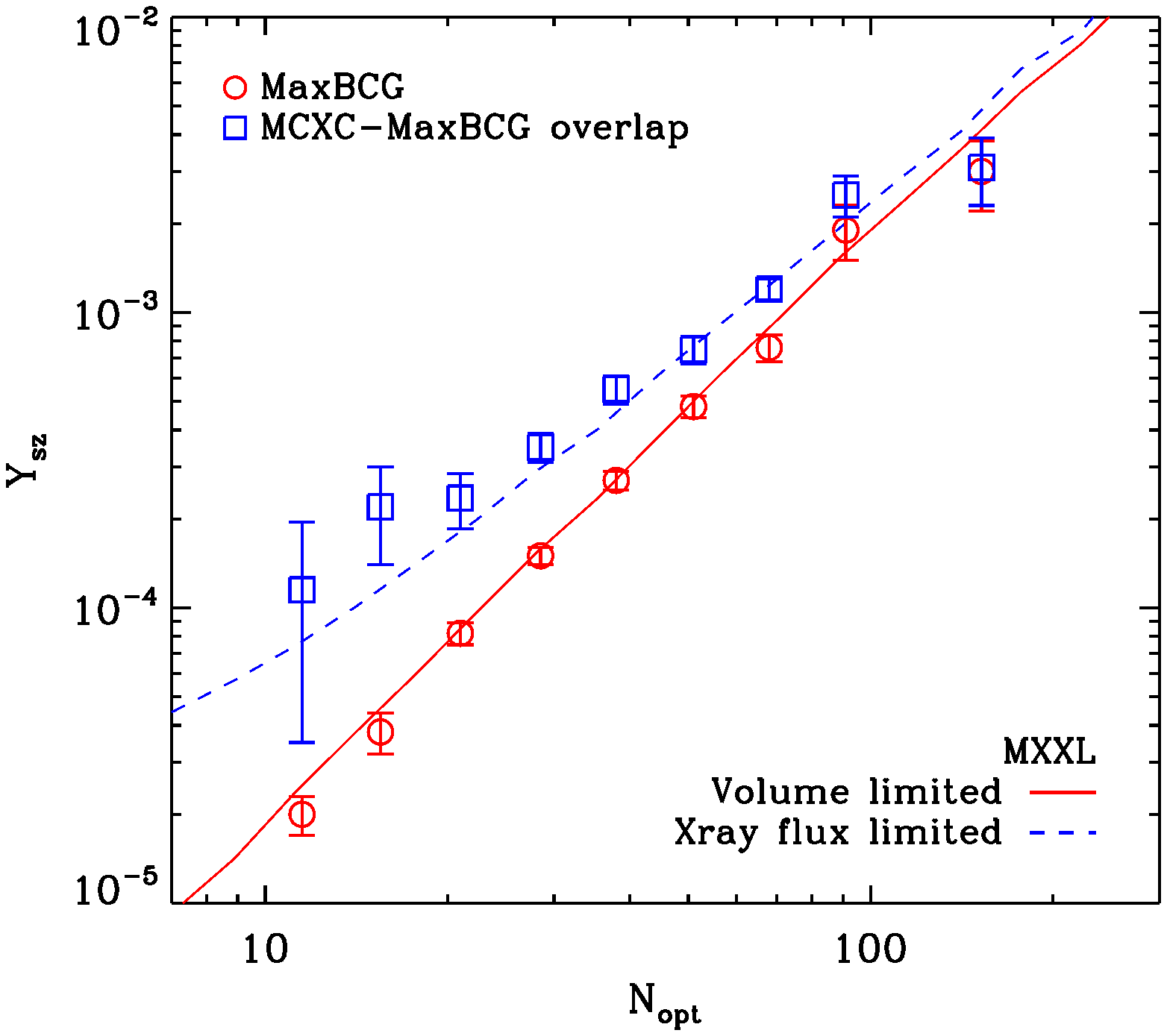}
  \caption{ The optical richness versus the SZ flux scaling
    relation. {\it Left}: Measured from stacking the signal at the
    location of the maxBCG clusters in the {\it Planck} survey
    \citep[reprinted from Fig~2, right panel
    of][]{Planck11_XII}. Red diamonds and blue triangles
    mark the data and predicted signal from the X-ray constraints
    respectively. {\it Right} Obtained from mock samples built from
    the XXL simulations for an optical maxBCG like catalogue (red
    squares) and the overlap between a maxBCG and MCXC like catalogues
    (blue squares) showing the differences obtained from the {\it Planck}
    measurements could be an effect of combined selection biases
    \citep[reprinted from Fig~11, right panel of][]{Angulo12}.  }
\label{sz2}       
\end{figure}

\citet{Planck11_XII} used an optically selected sample
\citep[maxBCG;][]{Koester07} to investigate the stacked relation
between SZ signal and weak lensing mass. The derived amplitude is
found to be significantly lower than when X-ray masses are used. The
two relations are shown in the left panel of Figure \ref{sz2}.
To date a conclusive explanation of this discrepancy has not been
presented, although it is expected to arise from the cumulative effect
of various biases. For instance, a bias in the weak-lensing mass
measurements and/or a high contamination of the optical catalogue have
been proposed as possible explanations; a bias in the hydrostatic
X-ray masses relative to the weak-lensing based ones can also cause a
different normalization \citep[e.g.][]{Nagai07, Mahdavi08,Mahdavi12},
although the required level of bias would be much larger than is
expected from simulations and observations. \citet{Angulo12} used
simulated clusters from the new Millennium-XXL simulations to show
that the discrepancy in the amplitude can result from the Malmquist
bias in flux limited samples. In this case the discrepancy would
arise from the propagation of the Malmquist bias from the X-ray
luminosities to the SZ signal through covariance in their scatter at
fixed cluster mass (Fig.~\ref{sz2}, right panel).  Clearly a better
understanding of the link between the X-ray, lensing and SZ
constraints on the cluster mass should help to achieve a better
definition and understanding of the mass proxies for galaxy clusters.

\section{Optical scaling relations}
\label{sec:4}
The total mass of a galaxy cluster can be directly estimated using
spectroscopic measurements of the projected velocity dispersion of the
member galaxies by applying the virial theorem under the assumption of
dynamical equilibrium \citep{Zwicky33}. The mass enclosed within a
radius, $r$, is given\footnote{As was the case for X-ray observations:
  T$_{X}$ is now replaced by the velocity dispersion $\sigma$ which
  can be interpreted as the ``temperature'' of the galaxy distribution.}
by the Jeans equation (e.g. \citealt{BT87}):

\begin{equation}
  M(r)=-\frac{r\sigma^2_r(r)}{G}\left[\frac{d\ln{\rho}}{d\ln{r}}+\frac{d\ln{\sigma^2_r(r)}}{d\ln{r}}+2\beta(r)\right],
\end{equation}

\noindent where $\rho(r)$ is the galaxy number density, $\sigma_r$ the
radial component of the velocity dispersion and
$\beta(r)=1-\sigma^2_t(r)/\sigma^2_r(r)$ the isotropy parameter, which
characterizes the ratio of the tangential to the radial dispersion.

If we consider for simplicity an isothermal system with an isotropic
velocity field the second and third terms vanish. In this case it
becomes clear that the velocity dispersion as a function of radius and
the radial distribution of a galaxy population are not independent
variables, but must be balanced to provide the correct mass of the
system.

By applying the virial theorem (which is an integration of the Jeans
equation; \citealt{BT87}), we obtain that the total virial mass of the
cluster ($M_{V}$) depends on the global velocity dispersion ($\sigma$)
and the spatial distribution of the galaxy population. In the
approximation that the galaxies trace the matter perfectly, we obtain

\begin{equation}
\label{MV}
M_{V}=\frac{\sigma^2R_V}{G}=\frac{3\pi \sigma_P^2 R_{V,P}}{2G},
\end{equation}

\noindent where $R_{V}$ is the virial radius. In the spherical
approximation $\sigma^2=3\sigma_{P}^2$ and it is possible to express
the virial mass in terms of the projected radius and line-of-sight
velocity dispersion (respectively $R_{V,P}$ and $\sigma_{P}$) as in
the second part of Eqn.~\ref{MV}. This estimator has the advantage
over the Jeans equation that it uses the integrated quantity
$\sigma_{P}$ rather than the dispersion profile. For this reason
Eqn.~\ref{MV} is the most commonly used to estimate the virial
mass \citep{Girardi98}.

Compared to X-ray observations, an advantage of using galaxies as
tracers is that the galaxy population can be observed with good
accuracy out to large radii. In the cluster outskirts, where the
virial equilibrium assumption does not hold, the caustic method has
yielded promising results for the total cluster masses
\citep[e.g.,][]{Rines03,Diaferio05,Rines13}. The accuracy of dynamical
methods was studied in detail by \cite{Biviano06}, who found that
the virial estimator can recover the virial mass for a galaxy cluster
within 10$\%$ for samples of at least 60 cluster members. In this
sense, the dynamical approach is expensive in terms of telescope time.

For this reason, especially considering large cluster samples, it is
interesting to consider inexpensive proxies based on the global
optical properties of clusters. Given that the gas fraction $f_{\rm
  gas}$ appears to be a low-scatter mass proxy (see \S\ref{sec:yx}),
one might expect observations of the stellar content to yield good
proxies. This reasoning suggests that the total optical luminosity
($L_{\rm op}$) or richness ($N_{\rm gal}$) of a galaxy cluster
provides a direct indication of its mass. Such optical mass proxies
are relatively inexpensive to measure, requiring only direct images of
moderate depth, even for high redshift clusters. Furthermore, these
estimators are applicable to low mass groups that typically lack a
sufficient number of member galaxies for a robust dynamical mass
estimation.

$N_{\rm gal}$ and $L_{\rm op}$ are simply evaluated by counting
galaxies or summing their luminosities in an aperture down to a
certain magnitude. If the sample is incomplete, a correction must be
applied.  Both $N_{\rm gal}$ and $L_{\rm op}$ must be corrected for
the expected contamination by field galaxies. The latter can be
estimated from a comparison with the surrounding field where no
cluster was detected or from the number counts of blank-field galaxy
surveys.

The relation between optical and X-ray observables was studied by
\cite{Yee03} and \cite{Popesso04}. The latter combined observations of
114 clusters of galaxies in the SDSS and RASS. They found that the
luminosity in the red optical bands ($i$ and $z$), which are more
sensitive to the light of the old stellar population and therefore to
the stellar mass of cluster galaxies, have tight correlations with the
X-ray properties.  Furthermore \cite{Popesso04} found that by using
$L_{\rm op}$ (in the $z$-band) it is possible to predict the
temperature of the cluster (and thus the mass) with an precision of
60$\%$. More recently \citet{Lopes09} found that the scatter between
optical luminosity and X-ray temperature for a sample of massive
clusters amounts to 40$\%$, which is comparable to that of the
corresponding relation based on X-ray data alone.

Lack of multi-colour data results in potentially large corrections
for the background, thus increasing the uncertainties in the richness
estimates. Nowadays, optical cluster surveys employ multiple bands,
which improves the purity of the samples and provides photometric
redshift estimates, thanks to the well-known observation that
early-type galaxies form a tight ridgeline in color--magnitude
space. With the advent of large optical surveys aimed at constraining
cosmology, more effort is being spent on refining these mass proxies.

Current large optically selected cluster samples, such as the maxBCG
sample \citep{Koester07}, have been used to study the relation between
total mass and optical observables. Mass estimates of the clusters in
the maxBCG sample were derived via a stacked weak lensing analysis
\citep{Johnston07} by binning the clusters in richness. For this
sample \cite{Rykoff08} studied the relation between the X-ray
luminosity and weak lensing mass. The measurements indicate a
power-law relation between mass and richness.  \cite{Rozo09} measured
a logarithmic scatter in mass at fixed richness of $\sigma_{\ln
  M|N_{200}}=0.45^{+0.20}_{-0.18}$ (with 95\% confidence) at
$N_{200}\approx 40$, where $N_{200}$ is the number of red-sequence
galaxies within $r_{200}$.  \cite{Sheldon09} measured the optical
luminosities of the clusters in this sample. They found that the
signal within a given richness bin depends on luminosity, which
suggests that the luminosity is more closely correlated with mass than
$N_{\rm gal}$. These studies have drawn attention to optical scaling
relations as a very effective way to obtain mass estimates for a large
number of systems.

Optical scaling relations are generally more difficult to interpret
because their behaviour cannot be predicted from simple physics scaling
arguments (with the possible exception of the stellar mass).
This is because the observed galaxy properties are the end result of
the complicated non-linear process of galaxy formation and evolution.
The M$_{\rm tot}-L_{\rm op}$ relations have power law slopes close to
unity, but not quite so, as most studies indicate an increase of the
mass-to-light ratio $M/L_{\rm op}$ with cluster mass. This is a direct
consequence of the variation of the fraction of stars in galaxies
\citep{Lin03,Giodini09}, suggesting that the efficiency of star
formation or galaxy evolution processes depend on the total mass. We
note that these same processes may also affect the ICM properties in
low mass systems.

The mass-richness relation shows a large intrinsic scatter
\citep{Gladders07, Rozo09}, which is mostly caused by the large
Poisson noise due to the low number of galaxies. It is possible to
reduce the scatter using more optimal estimators.  For example
\citep{Rozo09b,Rozo11} improved on the estimation of the richness
parameter, obtaining a significant reduction of the scatter in mass at
fixed richness for maxBCG clusters, by using a matched filter and an
optimized iterative measure of the cluster extent \citep[also
see][]{Rykoff12}. \cite{Andreon10} investigated the mass-richness
relation using caustic mass measurements for a sample of local X-ray
detected clusters. They stressed that once a careful statistical
analysis is performed, the richness has a similar performance as the
X-ray luminosity in predicting the total mass of a cluster.

\section{Interpretation of scaling relations with simulations}
\label{sec:5}
Only under certain (ideal) conditions we can derive scaling relations
between the baryonic properties and the total mass. However,
observations indicate the real situation is more complicated and we
need to rely on numerical simulations to gain further insights. The
simulation box represents a controlled laboratory where the models can
be tested and compared directly with the constraints derived from
observations.  Simulations start from a set of initial conditions
which consist of a realization of a density field with statistical
properties (e.g. the power spectrum) appropriate for the adopted
background cosmological model. The evolution of these initially small
density fluctuations is followed by advancing the density and velocity
fields forward by numerically integrating the equations governing the
dynamics of dark matter and baryons. The evolution of the
collision-less dark matter is relatively easily implemented, but
including the effects of baryons has proven to be more
complicated. For a recent review on cosmological simulations of galaxy
clusters we refer the interested reader to \cite{Borgani11}.

The best way to test if simulations correctly model the various
physical processes is to examine whether they faithfully reproduce the
statistical properties of a cluster sample, such as the observed
scaling relations.  While a simple gravity-only simulation naturally
reproduces the scaling relations for massive galaxy clusters (as they
are mostly self-similar), more physics needs to be included to
reproduce the observed deviations from the purely gravitational
scenario.

Radiative cooling was one of the first processes to be explored in
simulations. As discussed by \citet{Bryan00}, it can cause a selective
removal of the low entropy gas from the hot phase. Simulations show
that including radiative cooling leads to somewhat steeper scalings of
the X-ray luminosity relations by reducing the fraction of hot gas in
a mass dependent fashion. However, the effect is not sufficient to
reproduce the observed steepening \citep{Dave02}. Furthermore, a
``cooling only scenario" suffers from an excessive conversion of gas into
stars in the densest cluster regions (``overcooling";
\citealt{Blanchard92}, \citealt{Balogh01}), leading to an unreasonably
high predicted baryon fraction in the cores of galaxy clusters
\citep{Kravtsov05}. Given the short gas cooling time this should lead
to significant star formation, which is not observed; nor is the cold
gas \citep[e.g.][]{Kaastra01,Peterson03}.

\begin{figure}
\includegraphics[width=0.5\hsize]{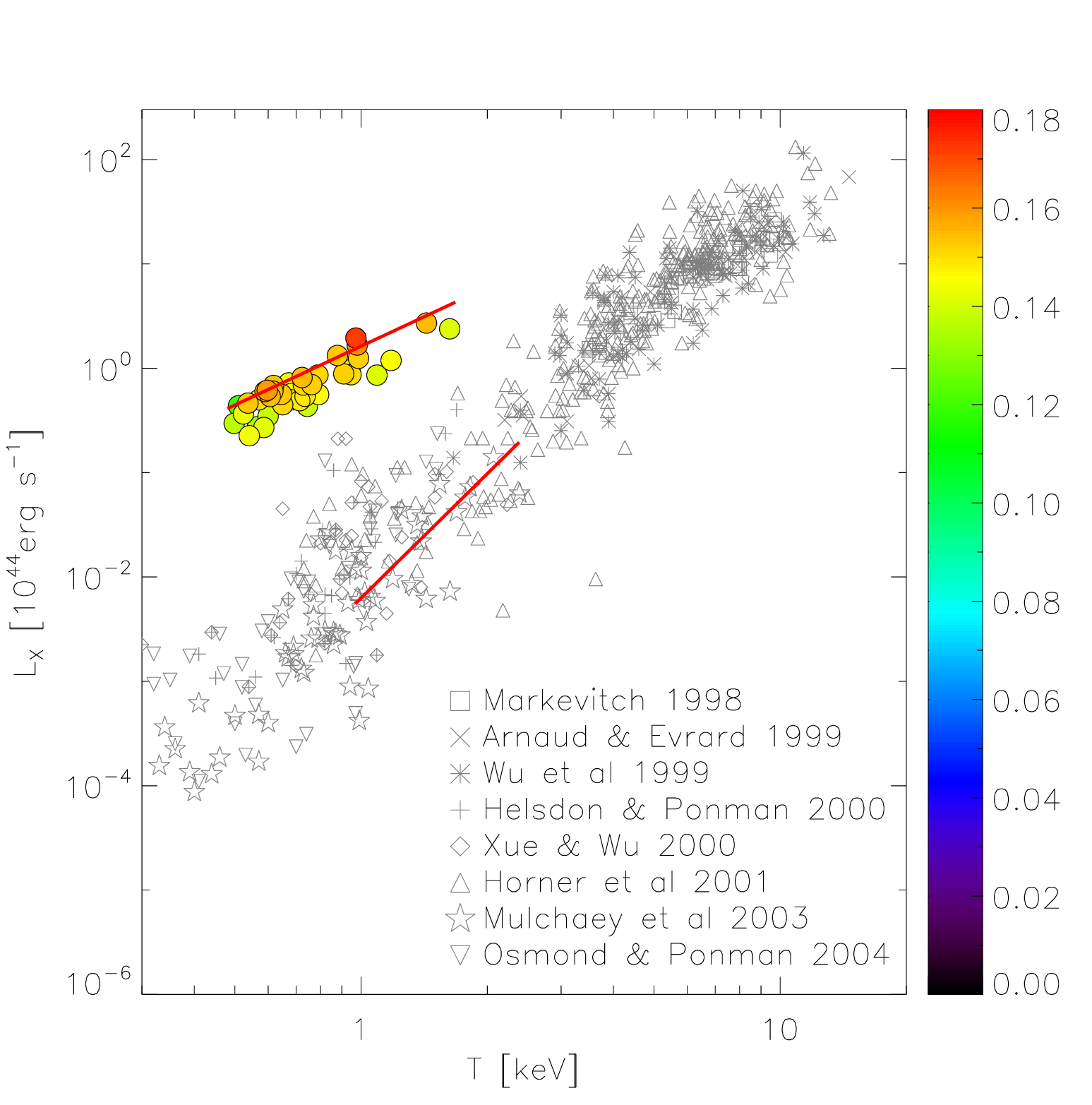}
\includegraphics[width=0.5\hsize]{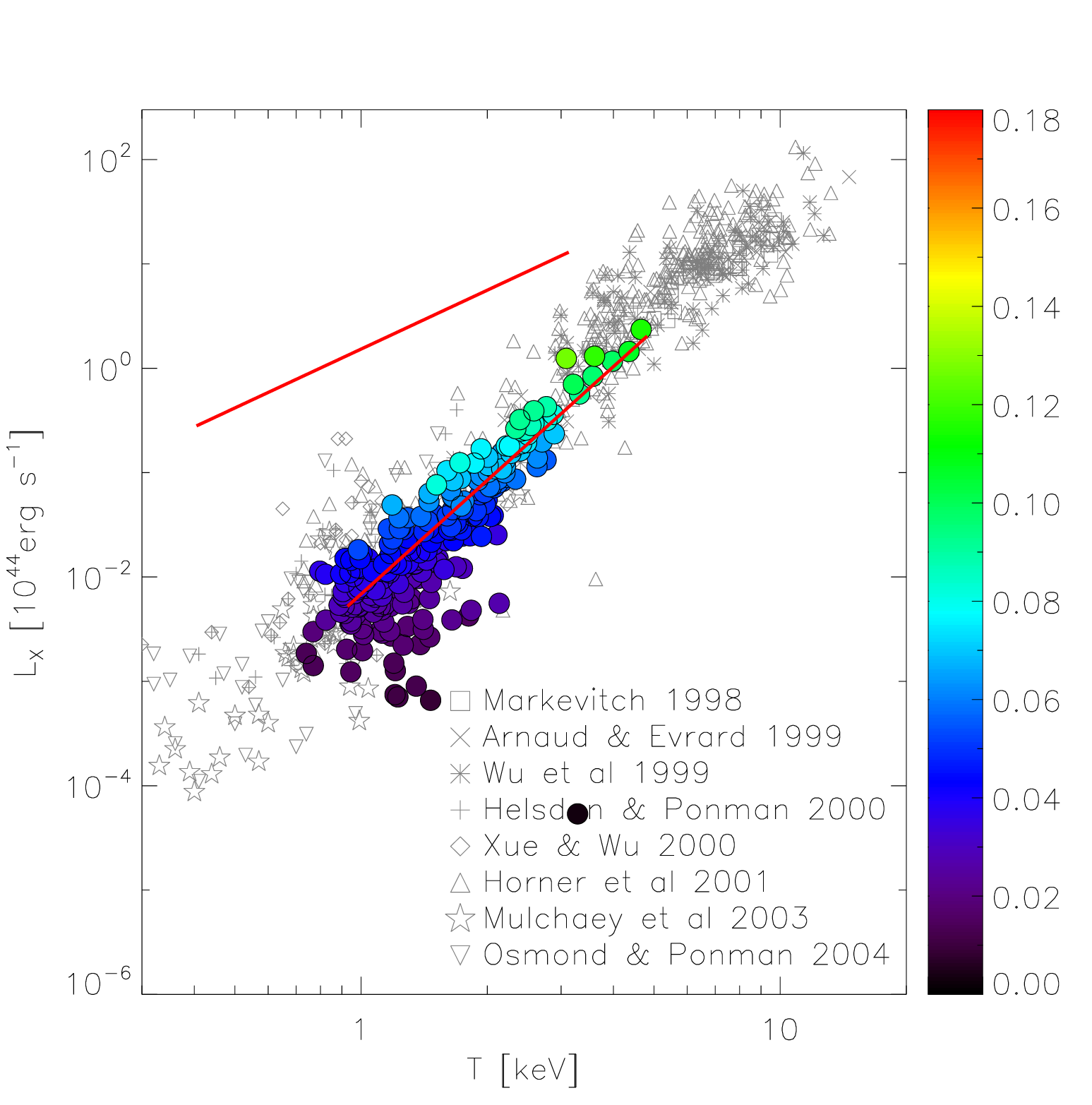}
\caption{Effect of feedback on simulated scaling relations (colored
  circles) between X-ray luminosity and temperature from
  \protect{\citet{Short09}} plotted on top of observed data-points (in
  gray). On the right the simulation include only supernova feedback,
  while on the right AGN feedback is included, where the energy is
  strongly coupled with the gas. It is clear that an energetic event
  similar to the latter is needed to reproduce the observed scaling. (Reproduced by permission of the AAS)}
\label{short}
\end{figure}

A solution can be provided by a suitable scheme of gas heating that
compensates for the radiative losses, pressurizes the gas in the core
regions, and regulates star formation. Feedback from supernovae is in
principle a good candidate to regulate gas cooling in cluster cores.
The energy injected by the supernova explosions can be used to keep
the relatively low-entropy gas in the hot phase despite its short
cooling time. Although this is a plausible mechanism, stellar feedback
is generally not considered the complete solution because the heating
is thought to be insufficient to reproduce the observed $L_{X}-T$
relation (see Fig.~\ref{short}). Another observation is that the
brightest cluster galaxies contain mostly old stellar populations that
are not capable of providing the needed amount of energy to offset the
cooling (although star formation is observed in cool core clusters
\citep[e.g.][]{Crawford99, Edge01, Edwards07, Bildfell08}.

Therefore another (powerful) feedback process is needed. Furthermore,
it should not be related directly to star formation activity. The most
popular candidate that appears to fit the bill is AGN feedback.  The
energy output of the AGN is provided by the accretion energy released by
the gas surrounding the supermassive black hole at the center of the
central cluster galaxy. There is convincing observational evidence
suggesting that AGN feedback has a large impact on the surrounding
gas. At radio wavelengths observations show large bubbles of
relativistic gas being spewed from the central galaxies and their
locations overlap with large cavities in the X-ray luminous gas
\citep[e.g.][]{Fabian06}. The amount of energy that is injected is
sufficiently large to offset the cooling and affect the cluster gas in
the core and beyond. Furthermore its effect will be even more dramatic
in low mass groups, where the injected energy can be comparable to the
binding energy of the gas \citep[e.g.][]{Giodini10}. AGN heating is
nowadays considered the most likely mechanism, also because is is
thought to play an important role in quenching the star formation in
the brightest cluster galaxy, which otherwise ends up too luminous in
the simulations. Finally, trends of the hot gas fraction and entropy
with cluster mass also support AGN heating.

The main challenge for simulators is that the details of the heating
mechanism are still poorly understood. First of all it is unknown how
the coupling between the AGN energy injection and the ICM
occurs. Furthermore, AGN heating works through episodic jets, but it
is not clear how the cooling and the heating episodes may be tuned.
In the past years the first attempts to include AGN heating in full
cosmological simulations have been carried out. \citet{Sijacki06}
developed a model for AGN heating via hot thermal bubbles in a
cosmological simulation of cluster formation. \citet{Puchwein08} have
shown that the observed $L_{X}-T$ relation is successfully reproduced
without invoking excessive cooling in the central region of the
simulated clusters. However this model does fail to reproduce the
observed entropy profiles. \citet{McCarthy10} used simulated groups
from the OverWhelmingly Large Simulations project \citep{Schaye10}. In
these simulations AGN feedback is included to match the observed
relation between black hole mass and the galaxy bulge mass
\citep{Booth10}. Encouragingly, they find entropy profiles and an
$L_{X}-T$ relation similar to the observations. \citet{Short09}
reproduce the observed scaling relations with simulations where
feedback from galaxies is incorporated via a hybrid approach: the
energy imparted to the ICM by SNe and AGN is computed from a
semi-analytic model of galaxy formation.

The observed deviation from self-similarity can also be explained by
pre-heating of the gas at early times. This is mostly motivated by
observational results that indicate the existence of an universal
entropy floor for clusters \citep{Evrard91,Kaiser91}. In this
scenario, the energy injection into the ICM from non–gravitational
processes such as supernovae, star formation, and galactic winds heats
the gas at high redshift, before the gas collapses in the deep
cluster/group potential well, causing a high-redshift entropy
modification.  Simulations show that the increase in entropy arises
from the shift from clumpy to smooth accretion in the cluster
outskirts due to the heating \citep{Borgani05}. The extra entropy
would inhibit the gas from falling into the potential well. The effect
would be larger for low mass systems which have shallower potentials,
alleviating the discrepancy between simulations and
observations. However, simulations have shown that the resulting
entropy profiles of the simulated groups are much too flat compared to
observations \citep{Borgani05, Younger07}. Therefore the observed lack
of iso-entropic core entropy profiles in groups and poor clusters has
shown that simple preheating is unlikely to be the sole explanation of
the observations \citep{Ponman03, Pratt03}.

\citet{Ettori04b} using a simulation that included
radiative cooling, star formation and supernova feedback, found a
significant negative evolution in the normalization of the $L_X-T$ and
$S-T$ relations in objects selected in the range $0.5<z<1$. This
result suggests either that the hot X-ray-emitting plasma measured in
the central regions of simulated systems is smaller than the observed
one or that systematically higher values of gas temperatures are
recovered in the simulated dataset. \citet{Muanwong06} and
\citet{Kay07} using different prescriptions for cooling and stellar
feedback found qualitatively similar results. 

\citet{Muanwong06} used three hydrodynamic simulations in the
$\Lambda$CDM cosmology. They considered a ``radiative-cooling",
``pre-heating" and ``AGN-feedback" scenario. They concluded that all
the models could reproduce the observed local $L_{X}-T$ scaling
relation but substantial differences between the models are predicted
at $z=1.5$. According to these simulations, if the evolution of the
scaling relation is parametrized as

\begin{equation}
L_{\rm bol}=C_{0}\times T_{\rm bol}^{\alpha}\times(1+z)^{A},
\end{equation}

\noindent the value of $A$ at $z=1.5$ is predicted to be $\sim -0.6$
(mildly negative evolution) for the AGN feedback model,
$\sim 0.7$ (mildly positive evolution) in a pre-heating scenario 
and $\sim 1.9$ (strong positive evolution) for a radiative cooling model. 

Current observational constraints (see \S\ref{sec:2}) would support a
mildly positive evolution, pointing towards early and widespread
preheating of the ICM, to be preferred over an extended period of
preheating. \citet{Short10}, using the Millennium Gas Simulations
which include AGN feedback following \citet{Short09}, confirms the
different model predictions but concludes that the feedback model is
favoured for $z\le0.5$ while the preheating model is preferred at
higher redshift, when comparing simulations to recent observations by
\citet{Pratt09} and \citet{Maughan08}. There remain, however, concerns
about strong selection biases in the current samples of high-redshift
clusters.

\section{Some general considerations}
\label{sec:6}
Any survey of galaxy clusters provides catalogs of systems that are
somewhat biased and incomplete. Biases and incompleteness arise
because of the chosen survey strategy or simply due to the finite
sensitivity. Furthermore the distribution of cluster properties is not
uniform in the observable-mass plane, but there is segregation
(e.g. cool-cores segregate at high luminosity).  Therefore the
determination of correct scaling relations relies heavily on
understanding the statistical properties of the underlying population
and any bias in the observables. In this section we list some of the
issues that need to be considered when determining and interpreting
mass-observable scaling relations.

\begin{itemize}
\item \textit{Malmquist bias:} the empirical determination of scaling
  relations is complicated by selection effects in the observations
  due to the presence of scatter. For instance, in an X-ray flux
  limited survey the intrinsically brighter sources for a given mass
  will appear to be more numerous than the fainter sources at that same
  mass because they can be seen in a larger volume (brighter sources are
  seen out to larger distances). This bias is commonly known as
  Malmquist bias and should be taken into account in both the scaling
  relation calibration and the cosmological analysis based on such
  relations. Some works in which this bias has been properly taken
  into account in the interpretation of scaling relations are
  \cite{Ikebe02},\cite{Stanek06},\cite{Pacaud07},\cite{Pratt09},\cite{Vikhlinin09a}.
  \cite{Mittal11} even applied individual Malmquist bias corrections
  for SCC, WCC, and
  NCC clusters.

\item \textit{Eddington bias:} this is the bias caused by the
  uncertainty in the observables in the sample \citep{Eddington13}. In
  general, sources of a given X-ray luminosity, for instance, will
  follow a distribution associated with the uncertainty in the
  measurement. Because the X-ray luminosity function is non-uniform
  (there are more objects with low luminosity) a larger fraction of
  systems will scatter from low luminosity to high than vice-versa,
  flattening the distribution. Another complication arises if there is
  a detection threshold, e.g. a flux limit. In this case the full
  range in scatter is not well represented at low luminosities and the
  inferred average luminosity will be overestimated. For examples in
  the context of X-ray scaling relations see for example
  \citet{Allen11} or \citet{Maughan12}.

\item \textit{Hidden priors:} the scatter of the points around an
  mass-observable scaling relation depends on the underlying
  cosmology, because it is directly linked to the shape of the halo
  mass function \citep{Stanek06}. Some authors
  \citep[e.g.][]{Mantz10b,Allen11} have recently started performing
  joint fits of scaling relations and cosmological parameters.

\item \textit{Binning of noisy data:} scaling relations are often
  estimated from binned data, instead of individual clusters. However
  particular care must be taken when the observable chosen for the
  binning is very noisy. With a large scatter, the mean values used to
  compute the ensemble averaged values may be biased with respect to
  the median relation, which is more robust. This is because the
  scatter about the mean is typically described by a lognormal
  distribution, and as a result the mean values will be dominated by
  the most luminous clusters. Considering the $L_X-N_{\rm gal}$
  relation as an example, the stacked normalization overestimates the
  median by a factor $\exp(\sigma_{\ln L_X}/2)$, as discussed in
  \cite{Rykoff08b}. As the scatter may depend on the mass, this can
  also impact the recovered slope.  Hence, the intrinsic scatter as a
  function of the various cluster properties needs to be known to
  account for this. Also covariance between observables needs to be
  accounted for. Note that this is also true for unbinned noisy data,
  although such analyses are often restricted to higher masses.

\item \textit{Cluster type bias:} flux-limited samples preferentially
  select certain types of clusters. For instance, \cite{Hudson10} and
  \cite{Mittal11} showed that due to the enhanced $L_{X}$ for a given
  $T_{X}$ (or $M$), clusters with cool cores are overrepresented in an
  X-ray survey. Similarly, \cite{Eckert11} showed that the detection
  efficiency of X-ray instruments is not the same for centrally peaked
  (CC) and flat (NCC) objects; they quantified this dependence on the
  surface brightness and corrected the corresponding fractions. Note
  that both effects should be accounted for, which has not been done
  simultaneously in any study up to now.

\item \textit{Archive bias:} corrections for Malmquist, Eddington, and
  cluster type bias can only be applied to samples that are complete
  according to (relatively) simple selection criteria, e.g., X-ray
  flux-limited samples. Samples constructed from public archives are,
  in general, not complete in any sense; their selection functions are
  often unknown and, therefore, such samples cannot be reproduced by
  mock simulations. Also, certain types of clusters may be preferred
  for proposals by observers and time allocation committees, e.g.,
  strong cool core clusters or major mergers as opposed to more
  ``boring'' weak cool core clusters. Therefore, any scaling relation
  study aiming for high accuracy should be based on a complete sample,
  with appropriate corrections applied.

\item \textit{Halo shape:} the assumption of spherical symmetry of the
  ICM, whereas clusters are known to be triaxial structures, can
  affect the observed scaling relations at the level of $\sim10\%$
  \citep[e.g.][]{Buote12} and introduces scatter. Hence knowledge of
  the intrinsic shape and orientation of halos is crucial for an
  unbiased determination of their masses. As reviewed in
  \cite{Limousin12} multi-wavelength observations of the ICM and mass can be
  used to quantify and account for triaxiality.

\end{itemize}

\section{Conclusions and Future Outlook}
\label{sec:7}
The next generation of X-ray observatories will provide powerful tools
to probe the structure and mass-energy content of the Universe. Such
probes will be complementary to the other planned cosmological
experiments, such as {\it Planck}, {\it
Euclid}\footnote{\url{http://www.euclid-ec.org}} \citep{redbook} and
LSST. They have the potential of placing very tight constraints on
different classes of dark-energy models, possibly finding signatures
of departures from the standard $\Lambda$CDM predictions. eROSITA
\citep{Predehl10, Pillepich12, Merloni12} will produce cluster
catalogs with $\sim 10^5$ objects out to redshift $\sim$1, increasing
the current statistical samples by 1-2 orders of magnitude and
extending the redshift range over which the growth of cosmic
structures can be traced. 

A number of infrared and optical surveys will provide the required
complementary photometric and spectroscopic redshifts, some of which
have already started collecting data. The photometric data will allow
for the identification of the cluster members and will be used in
combination with X-ray data to classify the eROSITA
clusters. Furthermore, these data provide galaxy targets for
additional spectroscopy if needed, and will also provide important
shear information for background galaxies, enabling the calibration of
the galaxy cluster masses through weak lensing analyses
\citep[see][for a review]{Hoekstra13}. Importantly, these surveys
can themselves be used to search for clusters, resulting in large
multi-wavelength databases of clusters.

A few thousand clusters will also have their temperatures determined
directly from the eROSITA survey data.  This will help to reduce the
scatter in the mass measurements for individual clusters providing
tighter constraints on the scaling relations. Thanks to these numbers,
eROSITA will permit to tackle several crucial astrophysical issues,
such as: 

\begin{itemize} \item the cluster mass function and its evolution,
$N(M,z)$, that provide constraints on the matter density, the
amplitude of the primordial power spectrum and dark energy
\citep[e.g.][]{Vikhlinin09b};

\item the angular clustering as a function of redshift
\citep[e.g.][]{Valageas12};

\item the cluster baryon fraction as function of
the redshift, which constrains the dark matter and energy densities
\citep[e.g.][]{Allen08,Ettori09};

\item  the baryonic wiggles due to acoustic oscillations at recombination,
which will give tight constraints on the space curvature and
cosmological parameters \citep[e.g.][]{Amendola12};

\item the spatially-resolved baryonic and total mass distribution over
the entire virial region for a subset of the X-ray bright systems with
complementary SZ and lensing data.

\end{itemize}

To reduce both the statistical and systematic uncertainties in these
measurements further, an X-ray telescope with the specifications of
the concepts such as {\it Athena} (e.g. \citealt{Barcons12}) or {\it
  Wide Field X-ray Telescope} (e.g. \citealt{Rosati10}) is
required. For instance, to improve the characterization of the
thermodynamical properties of X-ray emitting galaxy clusters as well
as the mass modeling, spatially-resolved temperatures of the ICM out
to $z\sim 1$ and beyond are required. To evaluate the thermal
structure of the ICM and how the scaling relations among integrated
quantities depend on the energy feedback from, e.g., mergers, AGNs and
supernovae, high resolution spectroscopy, hard X-ray imaging and
follow-up observations in radio, optical and infrared bands are
needed. In the near future, NuSTAR (launched in 2012;
http://www.nustar.caltech.edu/), ASTRO-H (to be launched in 2015;
http://astro-h.isas.jaxa.jp/), LOFAR (e.g. \citealt{vanWeeren12}) and
the optical/IR telescopes mentioned above will provide invaluable
resources to deepen our knowledge on the ICM physical
properties. Hence, despite the tremendous progress we reviewed here,
much more is yet to be studied.

\begin{acknowledgements}
We would like to thank ISSI for their hospitality.
SG \& HH acknowedge support from NWO Vidi grant 639.042.814.
LL acknowledges support by the German Research Association (DFG)
through grant RE 1462/6 and by the German Aerospace Agency
(DLR) with funds from the Ministry of Economy and Technology
(BMWi) through grant 50 OR 1102. EP acknowledges the support from grant ANR- 11-BD56-015. SE acknowledge the financial contribution from contracts ASI-INAF I/023/05/0 and I/088/06/0. THR acknowledges support
by the DFG through Heisenberg grant RE 1462/5 and grant RE 1462/6.
\end{acknowledgements}

\bibliographystyle{apj_new}
\bibliography{paper}   

\hyphenation{Post-Script Sprin-ger}
\begin{thebibliography}{225}
\expandafter\ifx\csname natexlab\endcsname\relax\def\natexlab#1{#1}\fi

\bibitem[{{Afshordi} {et~al.}(2007){Afshordi}, {Lin}, {Nagai}, \&
  {Sanderson}}]{Afshordi07}
{Afshordi}, N., {Lin}, Y.-T., {Nagai}, D., \& {Sanderson}, A.~J.~R. 2007,
  \mnras, 378, 293

\bibitem[{{Aghanim} {et~al.}(2009){Aghanim}, {da Silva}, \&
  {Nunes}}]{Aghanim09}
{Aghanim}, N., {da Silva}, A.~C., \& {Nunes}, N.~J. 2009, \aap, 496, 637

\bibitem[{{Aghanim} {et~al.}(2005){Aghanim}, {Hansen}, \&
  {Lagache}}]{Aghanim05}
{Aghanim}, N., {Hansen}, S.~H., \& {Lagache}, G. 2005, \aap, 439, 901

\bibitem[{{Akritas} \& {Bershady}(1996)}]{AkritasBershady}
{Akritas}, M.~G., \& {Bershady}, M.~A. 1996, \apj, 470, 706

\bibitem[{{Albrecht} {et~al.}(2006){Albrecht}, {Bernstein}, {Cahn}, {Freedman},
  {Hewitt}, {Hu}, {Huth}, {Kamionkowski}, {Kolb}, {Knox}, {Mather}, {Staggs},
  \& {Suntzeff}}]{Albrecht06}
{Albrecht}, A., {et~al.} 2006, ArXiv Astrophysics e-prints

\bibitem[{{Allen} {et~al.}(2011){Allen}, {Evrard}, \& {Mantz}}]{Allen11}
{Allen}, S.~W., {Evrard}, A.~E., \& {Mantz}, A.~B. 2011, \araa, 49, 409

\bibitem[{{Allen} {et~al.}(2008){Allen}, {Rapetti}, {Schmidt}, {Ebeling},
  {Morris}, \& {Fabian}}]{Allen08}
{Allen}, S.~W., {Rapetti}, D.~A., {Schmidt}, R.~W., {Ebeling}, H., {Morris},
  R.~G., \& {Fabian}, A.~C. 2008, \mnras, 383, 879

\bibitem[{{Allen} {et~al.}(2001){Allen}, {Schmidt}, \& {Fabian}}]{Allen01}
{Allen}, S.~W., {Schmidt}, R.~W., \& {Fabian}, A.~C. 2001, \mnras, 328, L37

\bibitem[{{Amendola} {et~al.}(2012){Amendola}, {Appleby}, {Bacon}, {Baker},
  {Baldi}, \& et~al.}]{Amendola12}
{Amendola}, L., {Appleby}, S., {Bacon}, D., {Baker}, T., {Baldi}, M., \& et~al.
  2012, ArXiv e-prints

\bibitem[{{Andersson} {et~al.}(2011){Andersson}, {Benson}, {Ade}, {Aird},
  {Armstrong}, {Bautz}, {Bleem}, {Brodwin}, {Carlstrom}, \&
  et~al.}]{Andersson11}
{Andersson}, K., {et~al.} 2011, \apj, 738, 48

\bibitem[{{Andreon} \& {Hurn}(2010)}]{Andreon10}
{Andreon}, S., \& {Hurn}, M.~A. 2010, \mnras, 404, 1922

\bibitem[{{Andreon} {et~al.}(2011){Andreon}, {Trinchieri}, \&
  {Pizzolato}}]{Andreon11}
{Andreon}, S., {Trinchieri}, G., \& {Pizzolato}, F. 2011, \mnras, 412, 2391

\bibitem[{{Angulo} {et~al.}(2012){Angulo}, {Springel}, {White}, {Jenkins},
  {Baugh}, \& {Frenk}}]{Angulo12}
{Angulo}, R.~E., {Springel}, V., {White}, S.~D.~M., {Jenkins}, A., {Baugh},
  C.~M., \& {Frenk}, C.~S. 2012, \mnras, 426, 2046

\bibitem[{{Arnaud} \& {Evrard}(1999)}]{Arnaud99}
{Arnaud}, M., \& {Evrard}, A.~E. 1999, \mnras, 305, 631

\bibitem[{{Arnaud} {et~al.}(2005){Arnaud}, {Pointecouteau}, \&
  {Pratt}}]{Arnaud05}
{Arnaud}, M., {Pointecouteau}, E., \& {Pratt}, G.~W. 2005, \aap, 441, 893

\bibitem[{{Arnaud} {et~al.}(2007){Arnaud}, {Pointecouteau}, \&
  {Pratt}}]{Arnaud07a}
---. 2007, \aap, 474, L37

\bibitem[{{Arnaud} {et~al.}(2010){Arnaud}, {Pratt}, {Piffaretti},
  {B{\"o}hringer}, {Croston}, \& {Pointecouteau}}]{Arnaud10a}
{Arnaud}, M., {Pratt}, G.~W., {Piffaretti}, R., {B{\"o}hringer}, H., {Croston},
  J.~H., \& {Pointecouteau}, E. 2010, \aap, 517, A92

\bibitem[{{Atrio-Barandela} {et~al.}(2008){Atrio-Barandela}, {M{\"u}cket}, \&
  {G{\'e}nova-Santos}}]{Atrio08}
{Atrio-Barandela}, F., {M{\"u}cket}, J.~P., \& {G{\'e}nova-Santos}, R. 2008,
  \apjl, 674, L61

\bibitem[{{Balogh} {et~al.}(1999){Balogh}, {Babul}, \& {Patton}}]{Balogh99}
{Balogh}, M.~L., {Babul}, A., \& {Patton}, D.~R. 1999, \mnras, 307, 463

\bibitem[{{Balogh} {et~al.}(2001){Balogh}, {Pearce}, {Bower}, \&
  {Kay}}]{Balogh01}
{Balogh}, M.~L., {Pearce}, F.~R., {Bower}, R.~G., \& {Kay}, S.~T. 2001, \mnras,
  326, 1228

\bibitem[{{Barcons} {et~al.}(2012){Barcons}, {Barret}, {Decourchelle}, {den
  Herder}, {Dotani}, \& et~al.}]{Barcons12}
{Barcons}, X., {Barret}, D., {Decourchelle}, A., {den Herder}, J.-W., {Dotani},
  T., \& et~al. 2012, ArXiv e-prints

\bibitem[{{Basu} {et~al.}(2010){Basu}, {Zhang}, {Sommer}, {Bender}, {Bertoldi},
  \& et~al.}]{Basu10}
{Basu}, K., {Zhang}, Y.-Y., {Sommer}, M.~W., {Bender}, A.~N., {Bertoldi}, F.,
  \& et~al. 2010, \aap, 519,

\bibitem[{{Benson} {et~al.}(2004){Benson}, {Church}, {Ade}, {Bock}, {Ganga},
  {Henson}, \& {Thompson}}]{Benson04}
{Benson}, B.~A., {Church}, S.~E., {Ade}, P.~A.~R., {Bock}, J.~J., {Ganga},
  K.~M., {Henson}, C.~N., \& {Thompson}, K.~L. 2004, \apj, 617, 829

\bibitem[{{Benson} {et~al.}(2013){Benson}, {de Haan}, {Dudley}, {Reichardt},
  {Aird}, {Andersson}, {Armstrong}, {Ashby}, {Bautz}, \& {Bayliss}}]{Benson13}
{Benson}, B.~A., {et~al.} 2013, \apj, 763, 147

\bibitem[{{Bielby} \& {Shanks}(2007)}]{Bielby07}
{Bielby}, R.~M., \& {Shanks}, T. 2007, \mnras, 382, 1196

\bibitem[{{Bildfell} {et~al.}(2008){Bildfell}, {Hoekstra}, {Babul}, \&
  {Mahdavi}}]{Bildfell08}
{Bildfell}, C., {Hoekstra}, H., {Babul}, A., \& {Mahdavi}, A. 2008, \mnras,
  389, 1637

\bibitem[{{Binney} \& {Tremaine}(1987)}]{BT87}
{Binney}, J., \& {Tremaine}, S. 1987

\bibitem[{{Birkinshaw}(1999)}]{Birkinshaw99}
{Birkinshaw}, M. 1999, \physrep, 310, 97

\bibitem[{{Birkinshaw} {et~al.}(1978){Birkinshaw}, {Gull}, \&
  {Northover}}]{Birkinshaw78}
{Birkinshaw}, M., {Gull}, S.~F., \& {Northover}, K.~J.~E. 1978, \mnras, 185,
  245

\bibitem[{{Birkinshaw} {et~al.}(1991){Birkinshaw}, {Hughes}, \&
  {Arnaud}}]{Birkinshaw91}
{Birkinshaw}, M., {Hughes}, J.~P., \& {Arnaud}, K.~A. 1991, \apj, 379, 466

\bibitem[{{Biviano} {et~al.}(2006){Biviano}, {Murante}, {Borgani}, {Diaferio},
  {Dolag}, \& {Girardi}}]{Biviano06}
{Biviano}, A., {Murante}, G., {Borgani}, S., {Diaferio}, A., {Dolag}, K., \&
  {Girardi}, M. 2006, \aap, 456, 23

\bibitem[{{Blanchard} {et~al.}(1992){Blanchard}, {Valls-Gabaud}, \&
  {Mamon}}]{Blanchard92}
{Blanchard}, A., {Valls-Gabaud}, D., \& {Mamon}, G.~A. 1992, \aap, 264, 365

\bibitem[{{B{\"o}hringer} {et~al.}(2012){B{\"o}hringer}, {Dolag}, \&
  {Chon}}]{Boehringer12}
{B{\"o}hringer}, H., {Dolag}, K., \& {Chon}, G. 2012, \aap, 539, A120

\bibitem[{{B{\"o}hringer} {et~al.}(2007){B{\"o}hringer}, {Schuecker}, {Pratt},
  {Arnaud}, {Ponman}, \& et~al.}]{Boehringer07}
{B{\"o}hringer}, H., {Schuecker}, P., {Pratt}, G.~W., {Arnaud}, M., {Ponman},
  T.~J., \& et~al. 2007, \aap, 469, 363

\bibitem[{{B{\"o}hringer} {et~al.}(2000){B{\"o}hringer}, {Voges}, {Huchra},
  {McLean}, {Giacconi}, {Rosati}, {Burg}, {Mader}, {Schuecker}, {Simi{\c c}},
  {Komossa}, {Reiprich}, {Retzlaff}, \& {Tr{\"u}mper}}]{Boehringer00}
{B{\"o}hringer}, H., {et~al.} 2000, \apjs, 129, 435

\bibitem[{{B{\"o}hringer} {et~al.}(2004){B{\"o}hringer}, {Schuecker}, {Guzzo},
  {Collins}, {Voges}, {Cruddace}, {Ortiz-Gil}, {Chincarini}, {De Grandi},
  {Edge}, {MacGillivray}, {Neumann}, {Schindler}, \& {Shaver}}]{Boehringer04}
---. 2004, \aap, 425, 367

\bibitem[{{Bonamente} {et~al.}(2008){Bonamente}, {Joy}, {LaRoque}, {Carlstrom},
  {Nagai}, \& {Marrone}}]{Bonamente08}
{Bonamente}, M., {Joy}, M., {LaRoque}, S.~J., {Carlstrom}, J.~E., {Nagai}, D.,
  \& {Marrone}, D.~P. 2008, \apj, 675, 106

\bibitem[{{Booth} \& {Schaye}(2010)}]{Booth10}
{Booth}, C.~M., \& {Schaye}, J. 2010, \mnras, 405, L1

\bibitem[{{Borgani} {et~al.}(2005){Borgani}, {Finoguenov}, {Kay}, {Ponman},
  {Springel}, {Tozzi}, \& {Voit}}]{Borgani05}
{Borgani}, S., {Finoguenov}, A., {Kay}, S.~T., {Ponman}, T.~J., {Springel}, V.,
  {Tozzi}, P., \& {Voit}, G.~M. 2005, \mnras, 361, 233

\bibitem[{{Borgani} {et~al.}(2001){Borgani}, {Governato}, {Wadsley}, {Menci},
  {Tozzi}, \& et~al.}]{Borgani01}
{Borgani}, S., {Governato}, F., {Wadsley}, J., {Menci}, N., {Tozzi}, P., \&
  et~al. 2001, \apjl, 559, L71

\bibitem[{{Borgani} \& {Kravtsov}(2011)}]{Borgani11}
{Borgani}, S., \& {Kravtsov}, A. 2011, Advanced Science Letters, 4, 204

\bibitem[{{Bower}(1997)}]{Bower97}
{Bower}, R.~G. 1997, \mnras, 288, 355

\bibitem[{{Branchesi} {et~al.}(2007){Branchesi}, {Gioia}, {Fanti}, \&
  {Fanti}}]{Branchesi07}
{Branchesi}, M., {Gioia}, I.~M., {Fanti}, C., \& {Fanti}, R. 2007, \aap, 472,
  739

\bibitem[{{Bryan}(2000)}]{Bryan00}
{Bryan}, G.~L. 2000, \apjl, 544, L1

\bibitem[{{Bryan} \& {Norman}(1998)}]{Bryan98}
{Bryan}, G.~L., \& {Norman}, M.~L. 1998, \apj, 495, 80

\bibitem[{{Buote} \& {Humphrey}(2012)}]{Buote12}
{Buote}, D.~A., \& {Humphrey}, P.~J. 2012, \mnras, 421, 1399

\bibitem[{{Carlstrom} {et~al.}(2002){Carlstrom}, {Holder}, \&
  {Reese}}]{Carlstrom02}
{Carlstrom}, J.~E., {Holder}, G.~P., \& {Reese}, E.~D. 2002, \araa, 40, 643

\bibitem[{{Carlstrom} {et~al.}(1996){Carlstrom}, {Joy}, \&
  {Grego}}]{Carlstrom96}
{Carlstrom}, J.~E., {Joy}, M., \& {Grego}, L. 1996, \apjl, 456,

\bibitem[{{Carlstrom} {et~al.}(2011){Carlstrom}, {Ade}, {Aird}, {Benson},
  {Bleem}, {Busetti}, {Chang}, {Chauvin}, {Cho}, \& et~al.}]{Carlstrom11}
{Carlstrom}, J.~E., {et~al.} 2011, PASP, 123, 568

\bibitem[{{Challinor} \& {Lasenby}(1999)}]{Challinor99}
{Challinor}, A., \& {Lasenby}, A. 1999, \apj, 510, 930

\bibitem[{{Chen} {et~al.}(2007){Chen}, {Reiprich}, {B{\"o}hringer}, {Ikebe}, \&
  {Zhang}}]{Chen07}
{Chen}, Y., {Reiprich}, T.~H., {B{\"o}hringer}, H., {Ikebe}, Y., \& {Zhang},
  Y.-Y. 2007, \aap, 466, 805

\bibitem[{{Cooray}(1999)}]{Cooray99}
{Cooray}, A.~R. 1999, \mnras, 307, 841

\bibitem[{{Crawford} {et~al.}(1999){Crawford}, {Allen}, {Ebeling}, {Edge}, \&
  {Fabian}}]{Crawford99}
{Crawford}, C.~S., {Allen}, S.~W., {Ebeling}, H., {Edge}, A.~C., \& {Fabian},
  A.~C. 1999, \mnras, 306, 857

\bibitem[{{Cunha} \& {Evrard}(2010)}]{Cunha10}
{Cunha}, C.~E., \& {Evrard}, A.~E. 2010, \prd, 81, 083509

\bibitem[{{da Silva}(2004)}]{Dasilva04}
{da Silva}, A.~J.~C. 2004, \apss, 290, 167

\bibitem[{{Dai} {et~al.}(2010){Dai}, {Bregman}, {Kochanek}, \& {Rasia}}]{Dai10}
{Dai}, X., {Bregman}, J.~N., {Kochanek}, C.~S., \& {Rasia}, E. 2010, \apj, 719,
  119

\bibitem[{{Dav{\'e}} {et~al.}(2002){Dav{\'e}}, {Katz}, \& {Weinberg}}]{Dave02}
{Dav{\'e}}, R., {Katz}, N., \& {Weinberg}, D.~H. 2002, \apj, 579, 23

\bibitem[{{David} {et~al.}(1993){David}, {Slyz}, {Jones}, {Forman}, {Vrtilek},
  \& {Arnaud}}]{David93}
{David}, L.~P., {Slyz}, A., {Jones}, C., {Forman}, W., {Vrtilek}, S.~D., \&
  {Arnaud}, K.~A. 1993, \apj, 412, 479

\bibitem[{{D{\'e}sert} {et~al.}(1998){D{\'e}sert}, {Benoit}, {Gaertner},
  {Bernard}, {Coron}, {Delabrouille}, {de Marcillac}, {Giard}, {Lamarre},
  {Lefloch}, {Puget}, \& {Sirbi}}]{Desert98}
{D{\'e}sert}, F.-X., {et~al.} 1998, \na, 3, 655

\bibitem[{{Diaferio} {et~al.}(2005){Diaferio}, {Geller}, \&
  {Rines}}]{Diaferio05}
{Diaferio}, A., {Geller}, M.~J., \& {Rines}, K.~J. 2005, \apjl, 628, L97

\bibitem[{{Eckert} {et~al.}(2011){Eckert}, {Molendi}, \& {Paltani}}]{Eckert11}
{Eckert}, D., {Molendi}, S., \& {Paltani}, S. 2011, \aap, 526, A79

\bibitem[{{Eckmiller} {et~al.}(2011){Eckmiller}, {Hudson}, \&
  {Reiprich}}]{Eckmiller11}
{Eckmiller}, H.~J., {Hudson}, D.~S., \& {Reiprich}, T.~H. 2011, \aap, 535, A105

\bibitem[{{Eddington}(1913)}]{Eddington13}
{Eddington}, A.~S. 1913, \mnras, 73, 359

\bibitem[{{Edge}(2001)}]{Edge01}
{Edge}, A.~C. 2001, \mnras, 328, 762

\bibitem[{{Edge} \& {Stewart}(1991)}]{Edge91}
{Edge}, A.~C., \& {Stewart}, G.~C. 1991, \mnras, 252, 414

\bibitem[{{Edwards} {et~al.}(2007){Edwards}, {Hudson}, {Balogh}, \&
  {Smith}}]{Edwards07}
{Edwards}, L.~O.~V., {Hudson}, M.~J., {Balogh}, M.~L., \& {Smith}, R.~J. 2007,
  \mnras, 379, 100

\bibitem[{{Ettori} {et~al.}(2002){Ettori}, {De Grandi}, \&
  {Molendi}}]{Ettori02}
{Ettori}, S., {De Grandi}, S., \& {Molendi}, S. 2002, \aap, 391, 841

\bibitem[{{Ettori} {et~al.}(2009){Ettori}, {Morandi}, {Tozzi}, {Balestra},
  {Borgani}, \& et~al.}]{Ettori09}
{Ettori}, S., {Morandi}, A., {Tozzi}, P., {Balestra}, I., {Borgani}, S., \&
  et~al. 2009, \aap, 501, 61

\bibitem[{{Ettori} {et~al.}(2012){Ettori}, {Rasia}, {Fabjan}, {Borgani}, \&
  {Dolag}}]{Ettori12}
{Ettori}, S., {Rasia}, E., {Fabjan}, D., {Borgani}, S., \& {Dolag}, K. 2012,
  \mnras, 420, 2058

\bibitem[{{Ettori} {et~al.}(2004{\natexlab{a}}){Ettori}, {Tozzi}, {Borgani}, \&
  {Rosati}}]{Ettori04}
{Ettori}, S., {Tozzi}, P., {Borgani}, S., \& {Rosati}, P. 2004{\natexlab{a}},
  \aap, 417, 13

\bibitem[{{Ettori} {et~al.}(2004{\natexlab{b}}){Ettori}, {Borgani},
  {Moscardini}, {Murante}, {Tozzi}, {Diaferio}, {Dolag}, {Springel}, {Tormen},
  \& {Tornatore}}]{Ettori04b}
{Ettori}, S., {et~al.} 2004{\natexlab{b}}, \mnras, 354, 111

\bibitem[{{Evrard}(1990)}]{Evrard90}
{Evrard}, A.~E. 1990, \apj, 363, 349

\bibitem[{{Evrard} \& {Henry}(1991)}]{Evrard91}
{Evrard}, A.~E., \& {Henry}, J.~P. 1991, \apj, 383, 95

\bibitem[{{Evrard} {et~al.}(1996){Evrard}, {Metzler}, \& {Navarro}}]{Evrard96}
{Evrard}, A.~E., {Metzler}, C.~A., \& {Navarro}, J.~F. 1996, \apj, 469, 494

\bibitem[{{Fabian} {et~al.}(2006){Fabian}, {Sanders}, {Taylor}, {Allen},
  {Crawford}, {Johnstone}, \& {Iwasawa}}]{Fabian06}
{Fabian}, A.~C., {Sanders}, J.~S., {Taylor}, G.~B., {Allen}, S.~W., {Crawford},
  C.~S., {Johnstone}, R.~M., \& {Iwasawa}, K. 2006, \mnras, 366, 417

\bibitem[{{Fassbender} {et~al.}(2011){Fassbender}, {B{\"o}hringer}, {Nastasi},
  {{\v S}uhada}, {M{\"u}hlegger}, {de Hoon}, {Kohnert}, {Lamer}, {Mohr},
  {Pierini}, {Pratt}, {Quintana}, {Rosati}, {Santos}, \&
  {Schwope}}]{Fassbender11}
{Fassbender}, R., {et~al.} 2011, New Journal of Physics, 13, 125014

\bibitem[{{Finoguenov} {et~al.}(2007){Finoguenov}, {Guzzo}, {Hasinger},
  {Scoville}, {Aussel}, \& et~al.}]{Finoguenov07}
{Finoguenov}, A., {Guzzo}, L., {Hasinger}, G., {Scoville}, N.~Z., {Aussel}, H.,
  \& et~al. 2007, \apjs, 172, 182

\bibitem[{{Finoguenov} {et~al.}(2002){Finoguenov}, {Jones}, {B{\"o}hringer}, \&
  {Ponman}}]{Finoguenov02}
{Finoguenov}, A., {Jones}, C., {B{\"o}hringer}, H., \& {Ponman}, T.~J. 2002,
  \apj, 578, 74

\bibitem[{{Finoguenov} {et~al.}(2001){Finoguenov}, {Reiprich}, \&
  {B{\"o}hringer}}]{Finoguenov01}
{Finoguenov}, A., {Reiprich}, T.~H., \& {B{\"o}hringer}, H. 2001, \aap, 368,
  749

\bibitem[{{Fox} \& {Loeb}(1997)}]{Fox97}
{Fox}, D.~C., \& {Loeb}, A. 1997, \apj, 491, 459

\bibitem[{{Gastaldello} {et~al.}(2007){Gastaldello}, {Buote}, {Humphrey},
  {Zappacosta}, {Bullock}, {Brighenti}, \& {Mathews}}]{Gastaldello07}
{Gastaldello}, F., {Buote}, D.~A., {Humphrey}, P.~J., {Zappacosta}, L.,
  {Bullock}, J.~S., {Brighenti}, F., \& {Mathews}, W.~G. 2007, \apj, 669, 158

\bibitem[{{Giodini} {et~al.}(2009){Giodini}, {Pierini}, {Finoguenov}, {Pratt},
  {Boehringer}, \& et~al.}]{Giodini09}
{Giodini}, S., {Pierini}, D., {Finoguenov}, A., {Pratt}, G.~W., {Boehringer},
  H., \& et~al. 2009, \apj, 703, 982

\bibitem[{{Giodini} {et~al.}(2010){Giodini}, {Smol{\v c}i{\'c}}, {Finoguenov},
  {Boehringer}, {B{\^i}rzan}, {Zamorani}, {Oklop{\v c}i{\'c}}, {Pierini},
  {Pratt}, {Schinnerer}, {Massey}, {Koekemoer}, {Salvato}, {Sanders},
  {Kartaltepe}, \& {Thompson}}]{Giodini10}
{Giodini}, S., {et~al.} 2010, \apj, 714, 218

\bibitem[{{Girardi} {et~al.}(1998){Girardi}, {Giuricin}, {Mardirossian},
  {Mezzetti}, \& {Boschin}}]{Girardi98}
{Girardi}, M., {Giuricin}, G., {Mardirossian}, F., {Mezzetti}, M., \&
  {Boschin}, W. 1998, \apj, 505, 74

\bibitem[{{Gladders} {et~al.}(2007){Gladders}, {Yee}, {Majumdar}, {Barrientos},
  {Hoekstra}, {Hall}, \& {Infante}}]{Gladders07}
{Gladders}, M.~D., {Yee}, H.~K.~C., {Majumdar}, S., {Barrientos}, L.~F.,
  {Hoekstra}, H., {Hall}, P.~B., \& {Infante}, L. 2007, \apj, 655, 128

\bibitem[{{Hand} {et~al.}(2012){Hand}, {Addison}, {Aubourg}, {Battaglia},
  {Battistelli}, \& et~al.}]{Hand12}
{Hand}, N., {Addison}, G.~E., {Aubourg}, E., {Battaglia}, N., {Battistelli},
  E.~S., \& et~al. 2012, ArXiv e-prints

\bibitem[{{Hasselfield} {et~al.}(2013){Hasselfield}, {Hilton}, {Marriage},
  {Addison}, {Barrientos}, {Battaglia}, {Battistelli}, {Bond}, {Crichton}, \&
  et~al.}]{Hasselfield13}
{Hasselfield}, M., {et~al.} 2013, ArXiv e-prints

\bibitem[{{Henry} {et~al.}(2009){Henry}, {Evrard}, {Hoekstra}, {Babul}, \&
  {Mahdavi}}]{Henry09}
{Henry}, J.~P., {Evrard}, A.~E., {Hoekstra}, H., {Babul}, A., \& {Mahdavi}, A.
  2009, \apj, 691, 1307

\bibitem[{{Herbig} {et~al.}(1995){Herbig}, {Lawrence}, {Readhead}, \&
  {Gulkis}}]{Herbig95}
{Herbig}, T., {Lawrence}, C.~R., {Readhead}, A.~C.~S., \& {Gulkis}, S. 1995,
  \apjl, 449,

\bibitem[{{Hincks} {et~al.}(2010){Hincks}, {Acquaviva}, {Ade}, {Aguirre},
  {Amiri}, {Appel}, {Barrientos}, {Battistelli}, \& et~al.}]{Hincks10}
{Hincks}, A.~D., {et~al.} 2010, \apjs, 191, 423

\bibitem[{{Hoekstra} {et~al.}(2013){Hoekstra}, {Bartelmann}, {Dahle}, {Israel},
  {Limousin}, \& {Meneghetti}}]{Hoekstra13}
{Hoekstra}, H., {Bartelmann}, M., {Dahle}, H., {Israel}, H., {Limousin}, M., \&
  {Meneghetti}, M. 2013, ArXiv e-prints

\bibitem[{{Hoekstra} {et~al.}(2012){Hoekstra}, {Mahdavi}, {Babul}, \&
  {Bildfell}}]{Hoekstra12}
{Hoekstra}, H., {Mahdavi}, A., {Babul}, A., \& {Bildfell}, C. 2012, \mnras,
  427, 1298

\bibitem[{{Hogg} {et~al.}(2010){Hogg}, {Bovy}, \& {Lang}}]{Hogg10}
{Hogg}, D.~W., {Bovy}, J., \& {Lang}, D. 2010, ArXiv e-prints

\bibitem[{{Holzapfel} {et~al.}(1997){Holzapfel}, {Wilbanks}, {Ade}, {Church},
  {Fischer}, {Mauskopf}, {Osgood}, \& {Lange}}]{Holzapfel97}
{Holzapfel}, W.~L., {Wilbanks}, T.~M., {Ade}, P.~A.~R., {Church}, S.~E.,
  {Fischer}, M.~L., {Mauskopf}, P.~D., {Osgood}, D.~E., \& {Lange}, A.~E. 1997,
  \apj, 479,

\bibitem[{{Huang} {et~al.}(2010){Huang}, {Wu}, {Ho}, {Koch}, \&
  {Liao}}]{Huang10}
{Huang}, C.-W.~L., {Wu}, J.-H.~P., {Ho}, P.~T.~P., {Koch}, P.~M., \& {Liao},
  e.~a. 2010, \apj, 716, 758

\bibitem[{{Hudson} {et~al.}(2010){Hudson}, {Mittal}, {Reiprich}, {Nulsen},
  {Andernach}, \& {Sarazin}}]{Hudson10}
{Hudson}, D.~S., {Mittal}, R., {Reiprich}, T.~H., {Nulsen}, P.~E.~J.,
  {Andernach}, H., \& {Sarazin}, C.~L. 2010, \aap, 513, A37

\bibitem[{{Ikebe} {et~al.}(2002){Ikebe}, {Reiprich}, {B{\"o}hringer}, {Tanaka},
  \& {Kitayama}}]{Ikebe02}
{Ikebe}, Y., {Reiprich}, T.~H., {B{\"o}hringer}, H., {Tanaka}, Y., \&
  {Kitayama}, T. 2002, \aap, 383, 773

\bibitem[{{Johnston} {et~al.}(2007){Johnston}, {Sheldon}, {Wechsler}, {Rozo},
  {Koester}, {Frieman}, {McKay}, {Evrard}, {Becker}, \& {Annis}}]{Johnston07}
{Johnston}, D.~E., {et~al.} 2007, ArXiv e-prints

\bibitem[{{Juett} {et~al.}(2010){Juett}, {Davis}, \& {Mushotzky}}]{Juett10}
{Juett}, A.~M., {Davis}, D.~S., \& {Mushotzky}, R. 2010, \apjl, 709, L103

\bibitem[{{Kaastra} {et~al.}(2001){Kaastra}, {Ferrigno}, {Tamura}, {Paerels},
  {Peterson}, \& {Mittaz}}]{Kaastra01}
{Kaastra}, J.~S., {Ferrigno}, C., {Tamura}, T., {Paerels}, F.~B.~S.,
  {Peterson}, J.~R., \& {Mittaz}, J.~P.~D. 2001, \aap, 365, L99

\bibitem[{{Kaiser}(1986)}]{Kaiser86}
{Kaiser}, N. 1986, \mnras, 222, 323

\bibitem[{{Kaiser}(1991)}]{Kaiser91}
---. 1991, \apj, 383, 104

\bibitem[{{Kashlinsky} {et~al.}(2010){Kashlinsky}, {Atrio-Barandela},
  {Ebeling}, {Edge}, \& {Kocevski}}]{Kashlinsky10}
{Kashlinsky}, A., {Atrio-Barandela}, F., {Ebeling}, H., {Edge}, A., \&
  {Kocevski}, D. 2010, \apjl, 712, L81

\bibitem[{{Kay} {et~al.}(2007){Kay}, {da Silva}, {Aghanim}, {Blanchard},
  {Liddle}, {Puget}, {Sadat}, \& {Thomas}}]{Kay07}
{Kay}, S.~T., {da Silva}, A.~C., {Aghanim}, N., {Blanchard}, A., {Liddle},
  A.~R., {Puget}, J.-L., {Sadat}, R., \& {Thomas}, P.~A. 2007, \mnras, 377, 317

\bibitem[{{Kitayama} {et~al.}(2004){Kitayama}, {Komatsu}, {Ota}, {Kuwabara},
  {Suto}, {Yoshikawa}, {Hattori}, \& {Matsuo}}]{Kitayama04}
{Kitayama}, T., {Komatsu}, E., {Ota}, N., {Kuwabara}, T., {Suto}, Y.,
  {Yoshikawa}, K., {Hattori}, M., \& {Matsuo}, H. 2004, \pasj, 56, 17

\bibitem[{{Koester} {et~al.}(2007){Koester}, {McKay}, {Annis}, {Wechsler},
  {Evrard}, {Bleem}, {Becker}, {Johnston}, {Sheldon}, {Nichol}, {Miller},
  {Scranton}, {Bahcall}, {Barentine}, {Brewington}, {Brinkmann}, {Harvanek},
  {Kleinman}, {Krzesinski}, {Long}, {Nitta}, {Schneider}, {Sneddin}, {Voges},
  \& {York}}]{Koester07}
{Koester}, B.~P., {et~al.} 2007, \apj, 660, 239

\bibitem[{{Komatsu} {et~al.}(1999){Komatsu}, {Kitayama}, {Suto}, {Hattori},
  {Kawabe}, {Matsuo}, {Schindler}, \& {Yoshikawa}}]{Komatsu99}
{Komatsu}, E., {Kitayama}, T., {Suto}, Y., {Hattori}, M., {Kawabe}, R.,
  {Matsuo}, H., {Schindler}, S., \& {Yoshikawa}, K. 1999, \apjl, 516, L1

\bibitem[{{Komatsu} {et~al.}(2011){Komatsu}, {Smith}, {Dunkley}, {Bennett},
  {Gold}, {Hinshaw}, {Jarosik}, {Larson}, {Nolta}, {Page}, {Spergel},
  {Halpern}, {Hill}, {Kogut}, {Limon}, {Meyer}, {Odegard}, {Tucker}, {Weiland},
  {Wollack}, \& {Wright}}]{Komatsu11}
{Komatsu}, E., {et~al.} 2011, \apjs, 192, 18

\bibitem[{{Kompaneets}(1957)}]{Kompaneets57}
{Kompaneets}, A.~S. 1957, Soviet Phys. --- JETP Lett., 4, 730

\bibitem[{{Kotov} \& {Vikhlinin}(2005)}]{Kotov05}
{Kotov}, O., \& {Vikhlinin}, A. 2005, \apj, 633, 781

\bibitem[{{Kravtsov} {et~al.}(2005){Kravtsov}, {Nagai}, \&
  {Vikhlinin}}]{Kravtsov05}
{Kravtsov}, A.~V., {Nagai}, D., \& {Vikhlinin}, A.~A. 2005, \apj, 625, 588

\bibitem[{{Kravtsov} {et~al.}(2006){Kravtsov}, {Vikhlinin}, \&
  {Nagai}}]{Kravtsov06}
{Kravtsov}, A.~V., {Vikhlinin}, A., \& {Nagai}, D. 2006, \apj, 650, 128

\bibitem[{{Laureijs} {et~al.}(2011){Laureijs}, {Amiaux}, {Arduini},
  {Augu{\`e}res}, {Brinchmann}, {Cole}, {Cropper}, {Dabin}, {Duvet}, {Ealet},
  \& et~al.}]{redbook}
{Laureijs}, R., {et~al.} 2011, ArXiv e-prints

\bibitem[{{Lieu} {et~al.}(2006){Lieu}, {Mittaz}, \& {Zhang}}]{Lieu06}
{Lieu}, R., {Mittaz}, J.~P.~D., \& {Zhang}, S.-N. 2006, \apj, 648, 176

\bibitem[{{Limousin} {et~al.}(2012){Limousin}, {Morandi}, {Sereno},
  {Meneghetti}, {Ettori}, {Bartelmann}, \& {Verdugo}}]{Limousin12}
{Limousin}, M., {Morandi}, A., {Sereno}, M., {Meneghetti}, M., {Ettori}, S.,
  {Bartelmann}, M., \& {Verdugo}, T. 2012, ArXiv e-prints

\bibitem[{{Lin} {et~al.}(2003){Lin}, {Mohr}, \& {Stanford}}]{Lin03}
{Lin}, Y.-T., {Mohr}, J.~J., \& {Stanford}, S.~A. 2003, \apj, 591, 749

\bibitem[{{Lopes} {et~al.}(2009){Lopes}, {de Carvalho}, {Kohl-Moreira}, \&
  {Jones}}]{Lopes09}
{Lopes}, P.~A.~A., {de Carvalho}, R.~R., {Kohl-Moreira}, J.~L., \& {Jones}, C.
  2009, \mnras, 399, 2201

\bibitem[{{Lyutikov}(2007)}]{Lyutikov07}
{Lyutikov}, M. 2007, \apjl, 668, L1

\bibitem[{{Mahdavi} {et~al.}(2012){Mahdavi}, {Hoekstra}, {Babul}, {Bildfell},
  {Jeltema}, \& {Henry}}]{Mahdavi12}
{Mahdavi}, A., {Hoekstra}, H., {Babul}, A., {Bildfell}, C., {Jeltema}, T., \&
  {Henry}, J.~P. 2012, ArXiv e-prints

\bibitem[{{Mahdavi} {et~al.}(2008){Mahdavi}, {Hoekstra}, {Babul}, \&
  {Henry}}]{Mahdavi08}
{Mahdavi}, A., {Hoekstra}, H., {Babul}, A., \& {Henry}, J.~P. 2008, \mnras,
  384, 1567

\bibitem[{{Mandelbrot}(1967)}]{Mandelbrot67}
{Mandelbrot}, B. 1967, Science, 156, 636

\bibitem[{{Mantz} {et~al.}(2010{\natexlab{a}}){Mantz}, {Allen}, {Ebeling},
  {Rapetti}, \& {Drlica-Wagner}}]{Mantz10b}
{Mantz}, A., {Allen}, S.~W., {Ebeling}, H., {Rapetti}, D., \& {Drlica-Wagner},
  A. 2010{\natexlab{a}}, \mnras, 406, 1773

\bibitem[{{Mantz} {et~al.}(2010{\natexlab{b}}){Mantz}, {Allen}, {Rapetti}, \&
  {Ebeling}}]{Mantz10}
{Mantz}, A., {Allen}, S.~W., {Rapetti}, D., \& {Ebeling}, H.
  2010{\natexlab{b}}, \mnras, 406, 1759

\bibitem[{{Markevitch}(1998)}]{Markevitch98}
{Markevitch}, M. 1998, \apj, 504, 27

\bibitem[{{Marriage} {et~al.}(2011){Marriage}, {Acquaviva}, {Ade}, {Aguirre},
  {Amiri}, {Appel}, {Barrientos}, {Battistelli}, \& et~al.}]{Marriage11}
{Marriage}, T.~A., {et~al.} 2011, \apj, 737, 61

\bibitem[{{Marrone} {et~al.}(2009){Marrone}, {Smith}, {Richard}, {Joy}, \&
  {Bonamente}}]{Marrone09}
{Marrone}, D.~P., {Smith}, G.~P., {Richard}, J., {Joy}, M., \& {Bonamente},
  e.~a. 2009, \apjl, 701, L114

\bibitem[{{Marrone} {et~al.}(2012){Marrone}, {Smith}, {Okabe}, {Bonamente},
  {Carlstrom}, {Culverhouse}, {Gralla}, {Greer}, {Hasler}, {Hawkins},
  {Hennessy}, {Joy}, {Lamb}, {Leitch}, {Martino}, {Mazzotta}, {Miller},
  {Mroczkowski}, {Muchovej}, {Plagge}, {Pryke}, {Sanderson}, {Takada}, {Woody},
  \& {Zhang}}]{Marrone12}
{Marrone}, D.~P., {et~al.} 2012, \apj, 754, 119

\bibitem[{{Maughan}(2007)}]{Maughan07}
{Maughan}, B.~J. 2007, \apj, 668, 772

\bibitem[{{Maughan} {et~al.}(2012){Maughan}, {Giles}, {Randall}, {Jones}, \&
  {Forman}}]{Maughan12}
{Maughan}, B.~J., {Giles}, P.~A., {Randall}, S.~W., {Jones}, C., \& {Forman},
  W.~R. 2012, \mnras, 421, 1583

\bibitem[{{Maughan} {et~al.}(2008){Maughan}, {Jones}, {Forman}, \& {Van
  Speybroeck}}]{Maughan08}
{Maughan}, B.~J., {Jones}, C., {Forman}, W., \& {Van Speybroeck}, L. 2008,
  \apjs, 174, 117

\bibitem[{{Maughan} {et~al.}(2006){Maughan}, {Jones}, {Ebeling}, \&
  {Scharf}}]{Maughan06}
{Maughan}, B.~J., {Jones}, L.~R., {Ebeling}, H., \& {Scharf}, C. 2006, \mnras,
  365, 509

\bibitem[{{McCarthy} {et~al.}(2003){McCarthy}, {Babul}, {Holder}, \&
  {Balogh}}]{McCarthy03}
{McCarthy}, I.~G., {Babul}, A., {Holder}, G.~P., \& {Balogh}, M.~L. 2003, \apj,
  591, 515

\bibitem[{{McCarthy} {et~al.}(2010){McCarthy}, {Schaye}, {Ponman}, {Bower},
  {Booth}, {Dalla Vecchia}, {Crain}, {Springel}, {Theuns}, \&
  {Wiersma}}]{McCarthy10}
{McCarthy}, I.~G., {et~al.} 2010, \mnras, 406, 822

\bibitem[{{Melin} {et~al.}(2011){Melin}, {Bartlett}, {Delabrouille}, {Arnaud},
  {Piffaretti}, \& {Pratt}}]{Melin11}
{Melin}, J.-B., {Bartlett}, J.~G., {Delabrouille}, J., {Arnaud}, M.,
  {Piffaretti}, R., \& {Pratt}, G.~W. 2011, \aap, 525,

\bibitem[{{Merloni} {et~al.}(2012){Merloni}, {Predehl}, {Becker},
  {B{\"o}hringer}, {Boller}, {Brunner}, {Brusa}, {Dennerl}, {Freyberg},
  {Friedrich}, {Georgakakis}, {Haberl}, {Hasinger}, {Meidinger}, {Mohr},
  {Nandra}, {Rau}, {Reiprich}, {Robrade}, {Salvato}, {Santangelo}, {Sasaki},
  {Schwope}, {Wilms}, \& {German eROSITA Consortium}}]{Merloni12}
{Merloni}, A., {et~al.} 2012, ArXiv e-prints

\bibitem[{{Mitchell} {et~al.}(1979){Mitchell}, {Dickens}, {Burnell}, \&
  {Culhane}}]{Mitchell79}
{Mitchell}, R.~J., {Dickens}, R.~J., {Burnell}, S.~J.~B., \& {Culhane}, J.~L.
  1979, \mnras, 189, 329

\bibitem[{{Mitchell} {et~al.}(1977){Mitchell}, {Ives}, \&
  {Culhane}}]{Mitchell77}
{Mitchell}, R.~J., {Ives}, J.~C., \& {Culhane}, J.~L. 1977, \mnras, 181, 25P

\bibitem[{{Mittal} {et~al.}(2011){Mittal}, {Hicks}, {Reiprich}, \&
  {Jaritz}}]{Mittal11}
{Mittal}, R., {Hicks}, A., {Reiprich}, T.~H., \& {Jaritz}, V. 2011, \aap, 532,
  A133

\bibitem[{{Mohr}(2002)}]{Mohr02}
{Mohr}, J.~J. 2002, in Astronomical Society of the Pacific Conference Series,
  Vol. 257, AMiBA 2001: High-Z Clusters, Missing Baryons, and CMB Polarization,
  ed. L.-W. {Chen}, C.-P. {Ma}, K.-W. {Ng}, \& U.-L. {Pen}, 49

\bibitem[{{Morandi} {et~al.}(2007){Morandi}, {Ettori}, \&
  {Moscardini}}]{Morandi07a}
{Morandi}, A., {Ettori}, S., \& {Moscardini}, L. 2007, \mnras, 379, 518

\bibitem[{{Muanwong} {et~al.}(2006){Muanwong}, {Kay}, \& {Thomas}}]{Muanwong06}
{Muanwong}, O., {Kay}, S.~T., \& {Thomas}, P.~A. 2006, \apj, 649, 640

\bibitem[{{Muchovej} {et~al.}(2012){Muchovej}, {Leitch}, {Culverhouse},
  {Carpenter}, \& {Sievers}}]{Muchovej12}
{Muchovej}, S., {Leitch}, E., {Culverhouse}, T., {Carpenter}, J., \& {Sievers},
  J. 2012, \apj, 749, 46

\bibitem[{{Mushotzky}(1984)}]{Mushotzky84}
{Mushotzky}, R.~F. 1984, Physica Scripta Volume T, 7, 157

\bibitem[{{Nagai} {et~al.}(2007){Nagai}, {Vikhlinin}, \& {Kravtsov}}]{Nagai07}
{Nagai}, D., {Vikhlinin}, A., \& {Kravtsov}, A.~V. 2007, \apj, 655, 98

\bibitem[{{Navarro} {et~al.}(1995){Navarro}, {Frenk}, \& {White}}]{NFW1}
{Navarro}, J.~F., {Frenk}, C.~S., \& {White}, S.~D.~M. 1995, \mnras, 275, 720

\bibitem[{{Novicki} {et~al.}(2002){Novicki}, {Sornig}, \& {Henry}}]{Novicki02}
{Novicki}, M.~C., {Sornig}, M., \& {Henry}, J.~P. 2002, AJ, 124, 2413

\bibitem[{{O'Hara} {et~al.}(2006){O'Hara}, {Mohr}, {Bialek}, \&
  {Evrard}}]{Hara06}
{O'Hara}, T.~B., {Mohr}, J.~J., {Bialek}, J.~J., \& {Evrard}, A.~E. 2006, \apj,
  639, 64

\bibitem[{{Okabe} {et~al.}(2010){Okabe}, {Zhang}, {Finoguenov}, {Takada},
  {Smith}, {Umetsu}, \& {Futamase}}]{Okabe10}
{Okabe}, N., {Zhang}, Y.-Y., {Finoguenov}, A., {Takada}, M., {Smith}, G.~P.,
  {Umetsu}, K., \& {Futamase}, T. 2010, \apj, 721, 875

\bibitem[{{Pacaud} {et~al.}(2007){Pacaud}, {Pierre}, {Adami}, {Altieri},
  {Andreon}, \& et~al.}]{Pacaud07}
{Pacaud}, F., {Pierre}, M., {Adami}, C., {Altieri}, B., {Andreon}, S., \&
  et~al. 2007, \mnras, 382, 1289

\bibitem[{{Peterson} {et~al.}(2003){Peterson}, {Kahn}, {Paerels}, {Kaastra},
  {Tamura}, {Bleeker}, {Ferrigno}, \& {Jernigan}}]{Peterson03}
{Peterson}, J.~R., {Kahn}, S.~M., {Paerels}, F.~B.~S., {Kaastra}, J.~S.,
  {Tamura}, T., {Bleeker}, J.~A.~M., {Ferrigno}, C., \& {Jernigan}, J.~G. 2003,
  \apj, 590, 207

\bibitem[{{Piffaretti} {et~al.}(2011){Piffaretti}, {Arnaud}, {Pratt},
  {Pointecouteau}, \& {Melin}}]{Piffaretti11}
{Piffaretti}, R., {Arnaud}, M., {Pratt}, G.~W., {Pointecouteau}, E., \&
  {Melin}, J.-B. 2011, \aap, 534,

\bibitem[{{Pillepich} {et~al.}(2012){Pillepich}, {Porciani}, \&
  {Reiprich}}]{Pillepich12}
{Pillepich}, A., {Porciani}, C., \& {Reiprich}, T.~H. 2012, \mnras, 422, 44

\bibitem[{{Planck Collaboration} {et~al.}(2011{\natexlab{a}}){Planck
  Collaboration}, {Aghanim}, {Arnaud}, {Ashdown}, {Aumont}, {Baccigalupi},
  {Balbi}, {Banday}, {Barreiro}, {Bartelmann}, \& et~al.}]{Planck11_IX}
{Planck Collaboration} {et~al.} 2011{\natexlab{a}}, \aap, 536, A9

\bibitem[{{Planck Collaboration} {et~al.}(2011{\natexlab{b}}){Planck
  Collaboration}, {Ade}, {Aghanim}, {Arnaud}, {Ashdown}, {Aumont},
  {Baccigalupi}, {Balbi}, {Banday}, {Barreiro}, \& et~al.}]{Planck11_VII}
---. 2011{\natexlab{b}}, \aap, 536, A7

\bibitem[{{Planck Collaboration} {et~al.}(2011{\natexlab{c}}){Planck
  Collaboration}, {Ade}, {Aghanim}, {Arnaud}, {Ashdown}, {Aumont},
  {Baccigalupi}, {Balbi}, {Banday}, {Barreiro}, \& et~al.}]{Planck11}
---. 2011{\natexlab{c}}, \aap, 536, A8

\bibitem[{{Planck Collaboration} {et~al.}(2011{\natexlab{d}}){Planck
  Collaboration}, {Ade}, {Aghanim}, {Arnaud}, {Ashdown}, {Aumont},
  {Baccigalupi}, {Balbi}, {Banday}, {Barreiro}, \& et~al.}]{Planck11_XI}
---. 2011{\natexlab{d}}, \aap, 536, A11

\bibitem[{{Planck Collaboration} {et~al.}(2011{\natexlab{e}}){Planck
  Collaboration}, {Aghanim}, {Arnaud}, {Ashdown}, {Aumont}, {Baccigalupi},
  {Balbi}, {Banday}, {Barreiro}, {Bartelmann}, \& et~al.}]{Planck11_XII}
---. 2011{\natexlab{e}}, \aap, 536, A12

\bibitem[{{Planck Collaboration} {et~al.}(2012){Planck Collaboration},
  {Aghanim}, {Arnaud}, {Ashdown}, {Atrio-Barandela}, {Aumont}, {Baccigalupi},
  {Balbi}, {Banday}, {Barreiro}, {Bartlett}, {Battaner}, {Benabed}, {Bernard},
  {Bersanelli}, {B{\"o}hringer}, {Bonaldi}, {Bond}, {Borrill}, {Bouchet},
  {Bourdin}, {Brown}, {Burigana}, {Butler}, {Cabella}, {Cardoso}, {Carvalho},
  {Catalano}, {Cay{\'o}n}, {Chamballu}, {Chary}, {Chiang}, {Chon},
  {Christensen}, {Clements}, {Colafrancesco}, {Colombi}, {Coulais}, {Crill},
  {Cuttaia}, {Da Silva}, {Dahle}, {Davis}, {de Bernardis}, {de Gasperis}, {de
  Zotti}, {Delabrouille}, {D{\'e}mocl{\`e}s}, {D{\'e}sert}, {Diego}, {Dolag},
  {Dole}, {Donzelli}, {Dor{\'e}}, {Douspis}, {Dupac}, {En{\ss}lin}, {Eriksen},
  {Finelli}, {Flores-Cacho}, {Forni}, {Fosalba}, {Frailis}, {Fromenteau},
  {Galeotta}, {Ganga}, {G{\'e}nova-Santos}, {Giard}, {Gonz{\'a}lez-Nuevo},
  {Gonz{\'a}lez-Riestra}, {G{\'o}rski}, {Gregorio}, {Gruppuso}, {Hansen},
  {Harrison}, {Hempel}, {Hern{\'a}ndez-Monteagudo}, {Herranz}, {Hildebrandt},
  {Hornstrup}, {Huffenberger}, {Hurier}, {Jagemann}, {Jasche}, {Juvela},
  {Keih{\"a}nen}, {Keskitalo}, {Kisner}, {Kneissl}, {Knoche}, {Knox},
  {Kurki-Suonio}, {Lagache}, {L{\"a}hteenm{\"a}ki}, {Lamarre}, {Lasenby},
  {Lawrence}, {Leach}, {Leonardi}, {Liddle}, {Lilje}, {L{\'o}pez-Caniego},
  {Luzzi}, {Mac{\'{\i}}as-P{\'e}rez}, {Maino}, {Mandolesi}, {Mann}, {Marleau},
  {Marshall}, {Mart{\'{\i}}nez-Gonz{\'a}lez}, {Masi}, {Massardi}, {Matarrese},
  {Matthai}, {Mazzotta}, {Meinhold}, {Melchiorri}, {Melin}, {Mendes},
  {Mennella}, {Miville-Desch{\^e}nes}, {Moneti}, {Montier}, {Morgante},
  {Mortlock}, {Munshi}, {Naselsky}, {Natoli}, {N{\o}rgaard-Nielsen},
  {Noviello}, {Osborne}, {Pasian}, {Patanchon}, {Perdereau}, {Perrotta},
  {Piacentini}, {Pierpaoli}, {Plaszczynski}, {Platania}, {Pointecouteau},
  {Polenta}, {Ponthieu}, {Popa}, {Poutanen}, {Pratt}, {Puget}, {Rachen},
  {Rebolo}, {Reinecke}, {Remazeilles}, {Renault}, {Ricciardi}, {Riller},
  {Ristorcelli}, {Rocha}, {Rosset}, {Rossetti}, {Rubi{\~n}o-Mart{\'{\i}}n},
  {Rusholme}, {Sandri}, {Savini}, {Schaefer}, {Scott}, {Smoot}, {Starck},
  {Stivoli}, {Sunyaev}, {Sutton}, {Sygnet}, {Tauber}, {Terenzi}, {Toffolatti},
  {Tomasi}, {Tristram}, {Valenziano}, {Van Tent}, {Vielva}, {Villa},
  {Vittorio}, {Wandelt}, {Weller}, {White}, {Yvon}, {Zacchei}, \&
  {Zonca}}]{Planck12_I}
---. 2012, \aap, 543, A102

\bibitem[{{Planck Collaboration} {et~al.}(2013{\natexlab{a}}){Planck
  Collaboration}, {AMI Collaboration}, {Ade}, {Aghanim}, {Arnaud}, {Ashdown},
  {Aumont}, {Baccigalupi}, {Balbi}, {Banday}, \& et~al.}]{PlanckAMI13}
---. 2013{\natexlab{a}}, \aap, 550, A128

\bibitem[{{Planck Collaboration} {et~al.}(2013{\natexlab{b}}){Planck
  Collaboration}, {Ade}, {Aghanim}, {Arnaud}, {Ashdown}, {Aumont},
  {Baccigalupi}, {Balbi}, {Banday}, {Barreiro}, {Bartlett}, {Battaner},
  {Benabed}, {Beno{\^i}t}, {Bernard}, {Bersanelli}, {Bikmaev}, {B{\"o}hringer},
  {Bonaldi}, {Bond}, {Borgani}, {Borrill}, {Bouchet}, {Brown}, {Burigana},
  {Butler}, {Cabella}, {Carvalho}, {Catalano}, {Cay{\'o}n}, {Chamballu},
  {Chary}, {Chiang}, {Chon}, {Christensen}, {Clements}, {Colafrancesco},
  {Colombi}, {Coulais}, {Crill}, {Cuttaia}, {Da Silva}, {Dahle}, {Davis}, {de
  Bernardis}, {de Gasperis}, {de Zotti}, {Delabrouille}, {D{\'e}mocl{\`e}s},
  {D{\'e}sert}, {Diego}, {Dolag}, {Dole}, {Donzelli}, {Dor{\'e}}, {Douspis},
  {Dupac}, {En{\ss}lin}, {Eriksen}, {Finelli}, {Flores-Cacho}, {Forni},
  {Frailis}, {Franceschi}, {Frommert}, {Galeotta}, {Ganga},
  {G{\'e}nova-Santos}, {Giraud-H{\'e}raud}, {Gonz{\'a}lez-Nuevo},
  {Gonz{\'a}lez-Riestra}, {G{\'o}rski}, {Gregorio}, {Gruppuso}, {Hansen},
  {Harrison}, {Hempel}, {Henrot-Versill{\'e}}, {Hern{\'a}ndez-Monteagudo},
  {Herranz}, {Hildebrandt}, {Hivon}, {Hobson}, {Holmes}, {Hornstrup}, {Hovest},
  {Huffenberger}, {Hurier}, {Jaffe}, {Jagemann}, {Jones}, {Juvela}, {Kneissl},
  {Knoche}, {Knox}, {Kunz}, {Kurki-Suonio}, {Lagache}, {Lamarre}, {Lasenby},
  {Lawrence}, {Le Jeune}, {Leach}, {Leonardi}, {Liddle}, {Lilje},
  {Linden-V{\o}rnle}, {L{\'o}pez-Caniego}, {Luzzi}, {Mac{\'{\i}}as-P{\'e}rez},
  {Maino}, {Mandolesi}, {Mann}, {Maris}, {Marleau}, {Marshall},
  {Mart{\'{\i}}nez-Gonz{\'a}lez}, {Masi}, {Massardi}, {Matarrese}, {Mazzotta},
  {Mei}, {Meinhold}, {Melchiorri}, {Melin}, {Mendes}, {Mennella}, {Mitra},
  {Miville-Desch{\^e}nes}, {Moneti}, {Morgante}, {Mortlock}, {Munshi},
  {Naselsky}, {Nati}, {Natoli}, {N{\o}rgaard-Nielsen}, {Noviello}, {Osborne},
  {Pajot}, {Paoletti}, {Perdereau}, {Perrotta}, {Piacentini}, {Piat},
  {Pierpaoli}, {Piffaretti}, {Plaszczynski}, {Platania}, {Pointecouteau},
  {Polenta}, {Popa}, {Poutanen}, {Pratt}, {Prunet}, {Puget}, {Reinecke},
  {Remazeilles}, {Renault}, {Ricciardi}, {Rocha}, {Rosset}, {Rossetti},
  {Rubi{\~n}o-Mart{\'{\i}}n}, {Rusholme}, {Sandri}, {Savini}, {Scott}, {Smoot},
  {Stanford}, {Stivoli}, {Sudiwala}, {Sunyaev}, {Sutton}, {Suur-Uski},
  {Sygnet}, {Tauber}, {Terenzi}, {Toffolatti}, {Tomasi}, {Tristram},
  {Valenziano}, {Van Tent}, {Vielva}, {Villa}, {Vittorio}, {Wade}, {Wandelt},
  {Welikala}, {Weller}, {White}, {Yvon}, {Zacchei}, \& {Zonca}}]{Planck13_IV}
---. 2013{\natexlab{b}}, \aap, 550, A130

\bibitem[{{Pointecouteau} {et~al.}(1998){Pointecouteau}, {Giard}, \&
  {Barret}}]{Pointecouteau98}
{Pointecouteau}, E., {Giard}, M., \& {Barret}, D. 1998, \aap, 336, 44

\bibitem[{{Ponman} {et~al.}(1996){Ponman}, {Bourner}, {Ebeling}, \&
  {B{\"o}hringer}}]{Ponman96}
{Ponman}, T.~J., {Bourner}, P.~D.~J., {Ebeling}, H., \& {B{\"o}hringer}, H.
  1996, \mnras, 283, 690

\bibitem[{{Ponman} {et~al.}(1999){Ponman}, {Cannon}, \& {Navarro}}]{Ponman99}
{Ponman}, T.~J., {Cannon}, D.~B., \& {Navarro}, J.~F. 1999, \nat, 397, 135

\bibitem[{{Ponman} {et~al.}(2003){Ponman}, {Sanderson}, \&
  {Finoguenov}}]{Ponman03}
{Ponman}, T.~J., {Sanderson}, A.~J.~R., \& {Finoguenov}, A. 2003, \mnras, 343,
  331

\bibitem[{{Poole} {et~al.}(2007){Poole}, {Babul}, {McCarthy}, {Fardal},
  {Bildfell}, {Quinn}, \& {Mahdavi}}]{Poole07}
{Poole}, G.~B., {Babul}, A., {McCarthy}, I.~G., {Fardal}, M.~A., {Bildfell},
  C.~J., {Quinn}, T., \& {Mahdavi}, A. 2007, \mnras, 380, 437

\bibitem[{{Popesso} {et~al.}(2004){Popesso}, {B{\"o}hringer}, {Brinkmann},
  {Voges}, \& {York}}]{Popesso04}
{Popesso}, P., {B{\"o}hringer}, H., {Brinkmann}, J., {Voges}, W., \& {York},
  D.~G. 2004, \aap, 423, 449

\bibitem[{{Pratt} \& {Arnaud}(2003)}]{Pratt03}
{Pratt}, G.~W., \& {Arnaud}, M. 2003, \aap, 408, 1

\bibitem[{{Pratt} {et~al.}(2009){Pratt}, {Croston}, {Arnaud}, \&
  {B{\"o}hringer}}]{Pratt09}
{Pratt}, G.~W., {Croston}, J.~H., {Arnaud}, M., \& {B{\"o}hringer}, H. 2009,
  \aap, 498, 361

\bibitem[{{Pratt} {et~al.}(2010){Pratt}, {Arnaud}, {Piffaretti},
  {B{\"o}hringer}, {Ponman}, {Croston}, {Voit}, {Borgani}, \&
  {Bower}}]{Pratt10}
{Pratt}, G.~W., {et~al.} 2010, \aap, 511, A85

\bibitem[{{Predehl} {et~al.}(2010){Predehl}, {Andritschke}, {B{\"o}hringer},
  {Bornemann}, {Br{\"a}uninger}, \& et~al.}]{Predehl10}
{Predehl}, P., {Andritschke}, R., {B{\"o}hringer}, H., {Bornemann}, W.,
  {Br{\"a}uninger}, H., \& et~al. 2010, in Society of Photo-Optical
  Instrumentation Engineers (SPIE) Conference Series, Vol. 7732, Society of
  Photo-Optical Instrumentation Engineers (SPIE) Conference Series

\bibitem[{{Puchwein} {et~al.}(2008){Puchwein}, {Sijacki}, \&
  {Springel}}]{Puchwein08}
{Puchwein}, E., {Sijacki}, D., \& {Springel}, V. 2008, \apjl, 687, L53

\bibitem[{{Randall} {et~al.}(2002){Randall}, {Sarazin}, \&
  {Ricker}}]{Randall02}
{Randall}, S.~W., {Sarazin}, C.~L., \& {Ricker}, P.~M. 2002, \apj, 577, 579

\bibitem[{{Rasia} {et~al.}(2006){Rasia}, {Ettori}, {Moscardini}, {Mazzotta},
  {Borgani}, {Dolag}, {Tormen}, {Cheng}, \& {Diaferio}}]{Rasia06}
{Rasia}, E., {et~al.} 2006, \mnras, 369, 2013

\bibitem[{{Rasia} {et~al.}(2012){Rasia}, {Meneghetti}, {Martino}, {Borgani},
  {Bonafede}, {Dolag}, {Ettori}, {Fabjan}, {Giocoli}, {Mazzotta}, {Merten},
  {Radovich}, \& {Tornatore}}]{Rasia12}
---. 2012, New Journal of Physics, 14, 055018

\bibitem[{{Reichardt} {et~al.}(2013){Reichardt}, {Stalder}, {Bleem}, {Montroy},
  {Aird}, {Andersson}, {Armstrong}, {Ashby}, {Bautz}, {Bayliss}, {Bazin},
  {Benson}, {Brodwin}, {Carlstrom}, {Chang}, {Cho}, {Clocchiatti}, {Crawford},
  {Crites}, {de Haan}, {Desai}, {Dobbs}, {Dudley}, {Foley}, {Forman}, {George},
  {Gladders}, {Gonzalez}, {Halverson}, {Harrington}, {High}, {Holder},
  {Holzapfel}, {Hoover}, {Hrubes}, {Jones}, {Joy}, {Keisler}, {Knox}, {Lee},
  {Leitch}, {Liu}, {Lueker}, {Luong-Van}, {Mantz}, {Marrone}, {McDonald},
  {McMahon}, {Mehl}, {Meyer}, {Mocanu}, {Mohr}, {Murray}, {Natoli}, {Padin},
  {Plagge}, {Pryke}, {Rest}, {Ruel}, {Ruhl}, {Saliwanchik}, {Saro}, {Sayre},
  {Schaffer}, {Shaw}, {Shirokoff}, {Song}, {Spieler}, {Staniszewski}, {Stark},
  {Story}, {Stubbs}, {{\v S}uhada}, {van Engelen}, {Vanderlinde}, {Vieira},
  {Vikhlinin}, {Williamson}, {Zahn}, \& {Zenteno}}]{Reichardt13}
{Reichardt}, C.~L., {et~al.} 2013, \apj, 763, 127

\bibitem[{{Reichert} {et~al.}(2011){Reichert}, {B{\"o}hringer}, {Fassbender},
  \& {M{\"u}hlegger}}]{Reichert11}
{Reichert}, A., {B{\"o}hringer}, H., {Fassbender}, R., \& {M{\"u}hlegger}, M.
  2011, \aap, 535, A4

\bibitem[{{Reiprich} {et~al.}(2013){Reiprich}, {Basu}, {Ettori}, {Israel},
  {Lovisari}, {Molendi}, {Pointecouteau}, \& {Roncarelli}}]{Reiprich13}
{Reiprich}, T.~H., {Basu}, K., {Ettori}, S., {Israel}, H., {Lovisari}, L.,
  {Molendi}, S., {Pointecouteau}, E., \& {Roncarelli}, M. 2013, ArXiv e-prints

\bibitem[{{Reiprich} \& {B{\"o}hringer}(2002)}]{Reiprich02}
{Reiprich}, T.~H., \& {B{\"o}hringer}, H. 2002, \apj, 567, 716

\bibitem[{{Rephaeli}(1995)}]{Rephaeli95}
{Rephaeli}, Y. 1995, \araa, 33, 541

\bibitem[{{Rines} {et~al.}(2010){Rines}, {Geller}, \& {Diaferio}}]{Rines10}
{Rines}, K., {Geller}, M.~J., \& {Diaferio}, A. 2010, \apjl, 715, L180

\bibitem[{{Rines} {et~al.}(2013){Rines}, {Geller}, {Diaferio}, \&
  {Kurtz}}]{Rines13}
{Rines}, K., {Geller}, M.~J., {Diaferio}, A., \& {Kurtz}, M.~J. 2013, \apj,
  767, 15

\bibitem[{{Rines} {et~al.}(2003){Rines}, {Geller}, {Kurtz}, \&
  {Diaferio}}]{Rines03}
{Rines}, K., {Geller}, M.~J., {Kurtz}, M.~J., \& {Diaferio}, A. 2003, AJ, 126,
  2152

\bibitem[{{Rosati} {et~al.}(2010){Rosati}, {Borgani}, {Gilli}, {Paolillo}, \&
  {Tozzi}}]{Rosati10}
{Rosati}, P., {Borgani}, S., {Gilli}, R., {Paolillo}, M., \& {Tozzi}, P. 2010,
  ArXiv e-prints

\bibitem[{{Rozo} {et~al.}(2011){Rozo}, {Rykoff}, {Koester}, {Nord}, {Wu},
  {Evrard}, \& {Wechsler}}]{Rozo11}
{Rozo}, E., {Rykoff}, E., {Koester}, B., {Nord}, B., {Wu}, H.-Y., {Evrard}, A.,
  \& {Wechsler}, R. 2011, \apj, 740, 53

\bibitem[{{Rozo} {et~al.}(2009{\natexlab{a}}){Rozo}, {Rykoff}, {Evrard},
  {Becker}, {McKay}, {Wechsler}, {Koester}, {Hao}, {Hansen}, {Sheldon},
  {Johnston}, {Annis}, \& {Frieman}}]{Rozo09}
{Rozo}, E., {et~al.} 2009{\natexlab{a}}, \apj, 699, 768

\bibitem[{{Rozo} {et~al.}(2009{\natexlab{b}}){Rozo}, {Rykoff}, {Koester},
  {McKay}, {Hao}, {Evrard}, {Wechsler}, {Hansen}, {Sheldon}, {Johnston},
  {Becker}, {Annis}, {Bleem}, \& {Scranton}}]{Rozo09b}
---. 2009{\natexlab{b}}, \apj, 703, 601

\bibitem[{{Rykoff} {et~al.}(2008{\natexlab{a}}){Rykoff}, {McKay}, {Becker},
  {Evrard}, {Johnston}, {Koester}, {Rozo}, {Sheldon}, \&
  {Wechsler}}]{Rykoff08b}
{Rykoff}, E.~S., {et~al.} 2008{\natexlab{a}}, \apj, 675, 1106

\bibitem[{{Rykoff} {et~al.}(2008{\natexlab{b}}){Rykoff}, {Evrard}, {McKay},
  {Becker}, {Johnston}, {Koester}, {Nord}, {Rozo}, {Sheldon}, {Stanek}, \&
  {Wechsler}}]{Rykoff08}
---. 2008{\natexlab{b}}, \mnras, 387, L28

\bibitem[{{Rykoff} {et~al.}(2012){Rykoff}, {Koester}, {Rozo}, {Annis},
  {Evrard}, {Hansen}, {Hao}, {Johnston}, {McKay}, \& {Wechsler}}]{Rykoff12}
---. 2012, \apj, 746, 178

\bibitem[{{Sayers} {et~al.}(2012){Sayers}, {Czakon}, {Bridge}, {Golwala},
  {Koch}, {Lin}, {Molnar}, \& {Umetsu}}]{Sayers12}
{Sayers}, J., {Czakon}, N.~G., {Bridge}, C., {Golwala}, S.~R., {Koch}, P.~M.,
  {Lin}, K.-Y., {Molnar}, S.~M., \& {Umetsu}, K. 2012, \apjl, 749, L15

\bibitem[{{Schaye} {et~al.}(2010){Schaye}, {Dalla Vecchia}, {Booth}, {Wiersma},
  {Theuns}, {Haas}, {Bertone}, {Duffy}, {McCarthy}, \& {van de
  Voort}}]{Schaye10}
{Schaye}, J., {et~al.} 2010, \mnras, 402, 1536

\bibitem[{{Schuecker} {et~al.}(2003){Schuecker}, {B{\"o}hringer}, {Collins}, \&
  {Guzzo}}]{Schuecker03}
{Schuecker}, P., {B{\"o}hringer}, H., {Collins}, C.~A., \& {Guzzo}, L. 2003,
  \aap, 398, 867

\bibitem[{{Shaw} {et~al.}(2010){Shaw}, {Nagai}, {Bhattacharya}, \&
  {Lau}}]{Shaw10}
{Shaw}, L.~D., {Nagai}, D., {Bhattacharya}, S., \& {Lau}, E.~T. 2010, \apj,
  725, 1452

\bibitem[{{Sheldon} {et~al.}(2009){Sheldon}, {Johnston}, {Scranton}, {Koester},
  {McKay}, {Oyaizu}, {Cunha}, {Lima}, {Lin}, {Frieman}, {Wechsler}, {Annis},
  {Mandelbaum}, {Bahcall}, \& {Fukugita}}]{Sheldon09}
{Sheldon}, E.~S., {et~al.} 2009, \apj, 703, 2217

\bibitem[{{Short} \& {Thomas}(2009)}]{Short09}
{Short}, C.~J., \& {Thomas}, P.~A. 2009, \apj, 704, 915

\bibitem[{{Short} {et~al.}(2010){Short}, {Thomas}, {Young}, {Pearce},
  {Jenkins}, \& {Muanwong}}]{Short10}
{Short}, C.~J., {Thomas}, P.~A., {Young}, O.~E., {Pearce}, F.~R., {Jenkins},
  A., \& {Muanwong}, O. 2010, \mnras, 408, 2213

\bibitem[{{Sifon} {et~al.}(2012){Sifon}, {Menanteau}, {Hasselfield},
  {Marriage}, {Hughes}, {Barrientos}, {Gonzalez}, {Infante}, {Addison}, \&
  et~al.}]{Sifon12}
{Sifon}, C., {et~al.} 2012, ArXiv e-prints

\bibitem[{{Sijacki} \& {Springel}(2006)}]{Sijacki06}
{Sijacki}, D., \& {Springel}, V. 2006, \mnras, 366, 397

\bibitem[{{Silk} \& {White}(1978)}]{SW78}
{Silk}, J., \& {White}, S.~D.~M. 1978, \apjl, 226, L103

\bibitem[{{Stanek} {et~al.}(2006){Stanek}, {Evrard}, {B{\"o}hringer},
  {Schuecker}, \& {Nord}}]{Stanek06}
{Stanek}, R., {Evrard}, A.~E., {B{\"o}hringer}, H., {Schuecker}, P., \& {Nord},
  B. 2006, \apj, 648, 956

\bibitem[{{Stanek} {et~al.}(2010){Stanek}, {Rasia}, {Evrard}, {Pearce}, \&
  {Gazzola}}]{Stanek10}
{Stanek}, R., {Rasia}, E., {Evrard}, A.~E., {Pearce}, F., \& {Gazzola}, L.
  2010, \apj, 715, 1508

\bibitem[{{Staniszewski} {et~al.}(2009){Staniszewski}, {Ade}, {Aird}, {Benson},
  {Bleem}, {Carlstrom}, {Chang}, {Cho}, {Crawford}, \& et~al.}]{Staniszewski09}
{Staniszewski}, Z., {et~al.} 2009, \apj, 701, 32

\bibitem[{{Story} {et~al.}(2011){Story}, {Aird}, {Andersson}, {Armstrong},
  {Bazin}, {Benson}, {Bleem}, {Bonamente}, {Brodwin}, {Carlstrom}, {Chang},
  {Clocchiatti}, {Crawford}, {Crites}, {de Haan}, {Desai}, {Dobbs}, {Dudley},
  {Foley}, {George}, {Gladders}, {Gonzalez}, {Halverson}, {High}, {Holder},
  {Holzapfel}, {Hoover}, {Hrubes}, {Joy}, {Keisler}, {Knox}, {Lee}, {Leitch},
  {Lueker}, {Luong-Van}, {Marrone}, {McMahon}, {Mehl}, {Meyer}, {Mohr},
  {Montroy}, {Padin}, {Plagge}, {Pryke}, {Reichardt}, {Rest}, {Ruel}, {Ruhl},
  {Saliwanchik}, {Saro}, {Schaffer}, {Shaw}, {Shirokoff}, {Song}, {Spieler},
  {Stalder}, {Staniszewski}, {Stark}, {Stubbs}, {Vanderlinde}, {Vieira},
  {Williamson}, \& {Zenteno}}]{Story11}
{Story}, K., {et~al.} 2011, \apjl, 735, L36

\bibitem[{{Sun} {et~al.}(2009){Sun}, {Voit}, {Donahue}, {Jones}, {Forman}, \&
  {Vikhlinin}}]{Sun09}
{Sun}, M., {Voit}, G.~M., {Donahue}, M., {Jones}, C., {Forman}, W., \&
  {Vikhlinin}, A. 2009, \apj, 693, 1142

\bibitem[{{Sunyaev} \& {Zeldovich}(1980)}]{Sunyaev80}
{Sunyaev}, R.~A., \& {Zeldovich}, I.~B. 1980, \mnras, 190, 413

\bibitem[{{Sunyaev} \& {Zeldovich}(1970)}]{SZ70}
{Sunyaev}, R.~A., \& {Zeldovich}, Y.~B. 1970, Comments on Astrophysics and
  Space Physics, 2,

\bibitem[{{Sunyaev} \& {Zeldovich}(1972)}]{SZ72}
---. 1972, Comments on Astrophysics and Space Physics, 4,

\bibitem[{{Tozzi} \& {Norman}(2001)}]{Tozzi01}
{Tozzi}, P., \& {Norman}, C. 2001, \apj, 546, 63

\bibitem[{{Valageas} \& {Clerc}(2012)}]{Valageas12}
{Valageas}, P., \& {Clerc}, N. 2012, \aap, 547, A100

\bibitem[{{van Weeren} {et~al.}(2012){van Weeren}, {R{\"o}ttgering},
  {Rafferty}, {Pizzo}, {Bonafede}, \& et~al.}]{vanWeeren12}
{van Weeren}, R.~J., {R{\"o}ttgering}, H.~J.~A., {Rafferty}, D.~A., {Pizzo},
  R., {Bonafede}, A., \& et~al. 2012, \aap, 543, A43

\bibitem[{{Vanderlinde} {et~al.}(2010){Vanderlinde}, {Crawford}, {de Haan},
  {Dudley}, {Shaw}, {Ade}, {Aird}, {Benson}, {Bleem}, \&
  et~al.}]{Vanderlinde10}
{Vanderlinde}, K., {et~al.} 2010, \apj, 722, 1180

\bibitem[{{Vikhlinin} {et~al.}(2006){Vikhlinin}, {Kravtsov}, {Forman}, {Jones},
  {Markevitch}, {Murray}, \& {Van Speybroeck}}]{Vikhlinin06}
{Vikhlinin}, A., {Kravtsov}, A., {Forman}, W., {Jones}, C., {Markevitch}, M.,
  {Murray}, S.~S., \& {Van Speybroeck}, L. 2006, \apj, 640, 691

\bibitem[{{Vikhlinin} {et~al.}(2002){Vikhlinin}, {van Speybroeck},
  {Markevitch}, {Forman}, \& {Grego}}]{Vikhlinin02}
{Vikhlinin}, A., {van Speybroeck}, L., {Markevitch}, M., {Forman}, W.~R., \&
  {Grego}, L. 2002, \apjl, 578, L107

\bibitem[{{Vikhlinin} {et~al.}(2003){Vikhlinin}, {Voevodkin}, {Mullis}, {van
  Speybroeck}, {Quintana}, \& et~al.}]{Vikhlinin03}
{Vikhlinin}, A., {Voevodkin}, A., {Mullis}, C.~R., {van Speybroeck}, L.,
  {Quintana}, H., \& et~al. 2003, \apj, 590, 15

\bibitem[{{Vikhlinin} {et~al.}(2009{\natexlab{a}}){Vikhlinin}, {Burenin},
  {Ebeling}, {Forman}, {Hornstrup}, {Jones}, {Kravtsov}, {Murray}, {Nagai},
  {Quintana}, \& {Voevodkin}}]{Vikhlinin09a}
{Vikhlinin}, A., {et~al.} 2009{\natexlab{a}}, \apj, 692, 1033

\bibitem[{{Vikhlinin} {et~al.}(2009{\natexlab{b}}){Vikhlinin}, {Kravtsov},
  {Burenin}, {Ebeling}, {Forman}, {Hornstrup}, {Jones}, {Murray}, {Nagai},
  {Quintana}, \& {Voevodkin}}]{Vikhlinin09b}
---. 2009{\natexlab{b}}, \apj, 692, 1060

\bibitem[{{Voit}(2005)}]{Voit05}
{Voit}, G.~M. 2005, Reviews of Modern Physics, 77, 207

\bibitem[{{Voit} {et~al.}(2002){Voit}, {Bryan}, {Balogh}, \& {Bower}}]{Voit02}
{Voit}, G.~M., {Bryan}, G.~L., {Balogh}, M.~L., \& {Bower}, R.~G. 2002, \apj,
  576, 601

\bibitem[{{White} {et~al.}(2002){White}, {Hernquist}, \& {Springel}}]{White02}
{White}, M., {Hernquist}, L., \& {Springel}, V. 2002, \apj, 579, 16

\bibitem[{{Williamson} {et~al.}(2011){Williamson}, {Benson}, {High},
  {Vanderlinde}, {Ade}, {Aird}, {Andersson}, {Armstrong}, {Ashby}, \&
  et~al.}]{Williamson11}
{Williamson}, R., {et~al.} 2011, ArXiv e-prints, {\tt arXiv:1101.1290}

\bibitem[{{Yang} {et~al.}(2009){Yang}, {Ricker}, \& {Sutter}}]{Yang09}
{Yang}, H.-Y.~K., {Ricker}, P.~M., \& {Sutter}, P.~M. 2009, \apj, 699, 315

\bibitem[{{Yee} \& {Ellingson}(2003)}]{Yee03}
{Yee}, H.~K.~C., \& {Ellingson}, E. 2003, \apj, 585, 215

\bibitem[{{Younger} \& {Bryan}(2007)}]{Younger07}
{Younger}, J.~D., \& {Bryan}, G.~L. 2007, \apj, 666, 647

\bibitem[{{Zhang} {et~al.}(2008){Zhang}, {Finoguenov}, {B{\"o}hringer},
  {Kneib}, {Smith}, {Kneissl}, {Okabe}, \& {Dahle}}]{Zhang08}
{Zhang}, Y.-Y., {Finoguenov}, A., {B{\"o}hringer}, H., {Kneib}, J.-P., {Smith},
  G.~P., {Kneissl}, R., {Okabe}, N., \& {Dahle}, H. 2008, \aap, 482, 451

\bibitem[{{Zwicky}(1933)}]{Zwicky33}
{Zwicky}, F. 1933, Helvetica Physica Acta, 6, 110

\end{thebibliography}
\nocite{*}

\end{document}